\documentclass[a4paper,11pt,twoside]{book} 

\usepackage{epsfig,graphics,subfigure}
\usepackage[centertags]{amsmath}
\usepackage{amssymb}

\usepackage[abbr]{harvard}
\usepackage[sectionbib]{chapterbib}            

\citationstyle{dcu}
\citationmode{abbr}

\usepackage{minitoc}
\usepackage{fancyheadings}

\usepackage{ulem}
\newfont{\dingbat}{dingbat scaled \magstep0}
\newcommand{\dingsymb}[1]{{\dingbat\symbol{#1}}}

\renewcommand{\harvardand}{\&}
\newfont{\vieuxchiffres}{cmmi10 scaled 1100}
\newcommand{\oldstyle}[1]{{\vieuxchiffres #1}}

\newcommand{\tsp}{{\sc tsp}}
\newcommand{\rsb}{{\sc rsb}}
\newcommand{\co}{{\sc co}}
\newcommand{\np}{{\sc np}}
\newcommand{\lk}{{\sc lk}}
\newcommand{\clo}{{\sc clo}}
\newcommand{\spin}{\mathbf{S}}
\newcommand{\bfx}{\mathbf{x}}

\newcommand{\prlpage}[2][-22.8cm]{%
   \raisebox{#1}[0cm][0cm]{
   \hspace{-2cm}
   \includegraphics[scale=0.85]{#2}}
   \newpage\par}

\newcommand{\jppage}[2][-25.6cm]{%
   \raisebox{#1}[0cm][0cm]{
   \hspace{-4.9cm}
   \includegraphics{#2}}
   \newpage\par}

\renewcommand{\thechapter}{\Roman{chapter}}
\renewcommand{\thesection}{\arabic{section}.}

\numberwithin{equation}{chapter}
\renewcommand{\theequation}{\thechapter.\arabic{equation}}
\numberwithin{figure}{chapter}

\numberwithin{table}{chapter}

\numberwithin{footnote}{chapter}
\renewcommand{\thefootnote}{\arabic{footnote}}

\renewcommand{\chaptermark}[1]{\markboth
			       {{\bfseries Chap.~\thechapter :}\bfseries\ #1}
			       {}}
\renewcommand{\sectionmark}[1]{\markright
			       {\bfseries\thesection\ #1}}

\renewcommand{\theenumi}{(\kern -0.15ex{\em\roman{enumi}\/})}

			\makeatletter
\renewcommand{\@seccntformat}[1]{\protect\rule[.6ex]{1.5ex}{.7pt}
				 \csname head#1\endcsname\ 
				 \protect\rule[.6ex]{1.5ex}{.7pt}\ }

\renewcommand{\@makefnmark}{\mbox{$^{\textrm{\thefootnote}}$}}

\renewcommand{\l@chapter}[2]{%
  \ifnum \c@tocdepth >\m@ne
    \addpenalty{-\@highpenalty}%
    \vskip 1.0em \@plus\p@
    \setlength\@tempdima{2em}%
    \begingroup
      \parindent \z@ \rightskip \@pnumwidth
      \parfillskip -\@pnumwidth
      \leavevmode \bfseries
      \advance\leftskip\@tempdima
      \hskip -\leftskip
      #1\nobreak\hfil \nobreak\hb@xt@\@pnumwidth{\hss #2}\par
      \penalty\@highpenalty
    \endgroup
  \fi}
\renewcommand{\l@section}{\@dottedtocline{1}{2em}{2em}}

\renewcommand{\subsection}[1]{\makeatletter
			   \renewcommand{\@seccntformat}[1]
				       {\csname the##1\endcsname\ }
                              \makeatother
			   \@startsection{subsection}{2}{1em}
	      				 {-\baselineskip}{.5\baselineskip}
	      				 {\bfseries\normalsize}
			   {#1} 
			      \makeatletter
			   \renewcommand{\@seccntformat}[1]
			   		{\protect\rule[.6ex]{1.5ex}{.7pt}
				         \csname the##1\endcsname\ 
				 	 \protect\rule[.6ex]{1.5ex}{.7pt}\ 
					}
			      \makeatother
			     }

\renewcommand{\subsubsection}[1]{\refstepcounter{subsubsection}
				 \@startsection{subsubsection}{3}{1.5em}
				 	       {-\baselineskip}
					       {.5\baselineskip}
	      				       {\bfseries\small}
	      			 {\uwave{\thesubsubsection\  #1}}
			          }
				  
			\makeatother

\renewcommand{\thesubsubsection}{\alph{subsection}.\arabic{subsubsection})}

\newcommand{\preliminaires}[1]{{\refstepcounter{section}\setcounter{section}{0}
			   \renewcommand{\thesection}{}
			   \addcontentsline{toc}{section}{
				 	\bfseries\itshape Chapter overview
				                         }
			   \begin{center}
			   \noindent\rule[.6ex]{1.5ex}{.7pt}
			   {\large\bfseries\itshape Chapter overview}
			   \rule[.6ex]{1.5ex}{.7pt}
			   \end{center}
			   \par\vspace{.5cm}
			   \renewcommand{\chaptermark}[1]{\markboth{
			      {\bfseries Chap.~\thechapter :}#1}
			       {}				   
			       					   }
			   \renewcommand{\sectionmark}[1]{
		   		   \markright{\bfseries\itshape Chapter overview
					   	         }
					}
			   \sectionmark{ Chapter overview}
			 }}
	      
\newcommand{\Appendix}[1]{\refstepcounter{chapter}
                          \thispagestyle{empty}
                          {
                           \renewcommand{\thechapter}{\Alph{chapter}}
			   \addcontentsline{toc}{chapter}
			   {\numberline{}\appendixname~\thechapter~--- #1}
			   \begin{center}
			   \rule[.6ex]{1.5ex}{.7pt}\hspace{1ex}
			   {\large\bfseries\appendixname~\thechapter}
			   \rule[.6ex]{1.5ex}{.7pt}\par
                           \bigskip
                           {\bf #1}\par
			   \end{center}\par
                           \bigskip\par
                           \markboth{{\bfseries \appendixname~\thechapter~---
                                      #1}}
                                    {{\bfseries \appendixname~\thechapter~---
                                      #1}}
			 }}

\makeatletter
\renewcommand{\@makechapterhead}[1]{%
  \vspace*{-40\p@}
  {\parindent \z@ \raggedright \normalfont
    \ifnum \c@secnumdepth >\m@ne
      \if@mainmatter
        \huge\bfseries \underline{\@chapapp\space \thechapter} 
        \par\nobreak					       
        \vskip 20\p@
      \fi
    \fi
    \interlinepenalty\@M
    \Huge \bfseries #1\par\nobreak
    \vskip 40\p@
  }}
\makeatother

\newenvironment{abstract}{ \begin{center}
			   \textbf{Abstract}
			   \addcontentsline{toc}{section}
			   {\protect Abstract}%
			   \end{center}
			   \begin{quote}
			 }
			 {\end{quote}}

\newcommand{\clearemptydoublepage}{{\pagestyle{empty}\cleardoublepage}}

\begin{document}


\setcounter{page}{-1}

\thispagestyle{empty}

				\begin{center}

\dingsymb{'141}%
\dingsymb{'142}\dingsymb{'142}\dingsymb{'142}\dingsymb{'142}\dingsymb{'142}%
\dingsymb{'142}\dingsymb{'142}\dingsymb{'142}\dingsymb{'142}\dingsymb{'142}%
\dingsymb{'142}\dingsymb{'142}\dingsymb{'142}\dingsymb{'142}\dingsymb{'142}%
\dingsymb{'142}\dingsymb{'143}%
				\end{center}
				
\hspace{1.075cm}\raisebox{0.4cm}[0cm][0cm]%
{\dingsymb{'144}\sffamily\bfseries\LARGE
Universit\'e Paris~6\hspace{5pt} Doctoral Dissertation\hspace{14.15pt}
\smash{\dingsymb{'145}}}

\vspace{-.84cm}				  
				\begin{center}
\dingsymb{'146}%
\dingsymb{'147}\dingsymb{'147}\dingsymb{'147}\dingsymb{'147}\dingsymb{'147}%
\dingsymb{'147}\dingsymb{'147}\dingsymb{'147}\dingsymb{'147}\dingsymb{'147}%
\dingsymb{'147}\dingsymb{'147}\dingsymb{'147}\dingsymb{'147}\dingsymb{'147}%
\dingsymb{'147}\dingsymb{'150}

\vspace{2cm}

\hspace{-8.5cm}
  {\sffamily\slshape\LARGE\ } 
 
\vspace{.5cm}	 
  {\sffamily\LARGE The Traveling Salesman and Related Stochastic Problems}

\vspace{2.5cm}

	{\sffamily\slshape\Large{by}}
	
\vspace{.5cm}

	{\sffamily\bfseries\LARGE Allon Percus}

\vspace{2.5cm}

        {\sffamily\slshape\Large A dissertation submitted for the degree of}

\vspace{.2cm}

        {\sffamily\slshape\Large {\sc Docteur de l'Universit\'e Paris~6}}

\vspace{.25cm}

        {\sffamily\slshape\Large in Theoretical Physics}

\vspace{2cm}

{\sffamily\slshape\Large{22 September 1997}}

{\sffamily\slshape\Large{\ }}
			     
			     \end{center}

\vspace{1.8cm}

\begin{tabbing}
\hspace{2.5cm}\sffamily\slshape\large PhD Committee:\=\hspace{1cm}	
{\sffamily\bfseries\large Claire Kenyon}\hspace{1.5cm}\=
\sffamily\slshape\large {Reader}\\
\hspace{2.5cm}\sffamily\slshape\large\ \>\hspace{1cm}
{\sffamily\bfseries\large  Olivier Martin} \>\sffamily\slshape\large
{Thesis advisor} \\   
\>\hspace{1cm}
{\sffamily\bfseries\large  R\a'emi Monasson} \>\sffamily\slshape\large {} \\
\>\hspace{1cm}
{\sffamily\bfseries\large  Karol Penson} \>\sffamily\slshape\large
{Committee chair} \\  
\>\hspace{1cm}
{\sffamily\bfseries\large  Nicolas Sourlas} \>\sffamily\slshape\large
{Reader} \\     
\>\hspace{1cm}
{\sffamily\bfseries\large  St\a'ephane Zaleski} \>\sffamily\slshape\large {}\\
\end{tabbing}

\clearemptydoublepage

\thispagestyle{empty}
\renewcommand{\baselinestretch}{1.5}\large

\vglue 1in

\centerline{\bf Abstract}

\bigskip\bigskip

In the traveling salesman problem, one must find the length of the
shortest closed tour visiting given ``cities''.  We study the stochastic
version of the problem, taking the locations of cities and the distances
separating them to be random variables drawn from an ensemble.  We
consider first the ensemble where cities are placed in Euclidean space.
We investigate how the optimum tour length scales with number of cities
and with number of spatial dimensions.  We then examine the analytical
theory behind the random link ensemble, where distances between cities
are independent random variables.  Finally, we look at the related
geometric issue of nearest neighbor distances, and find some remarkable
universalities.

\clearemptydoublepage 

\renewcommand{\baselinestretch}{1.1}\normalsize
\sloppy

\chapter*{Foreword}

Universit\'e Paris 6 authorities require PhD candidates to submit and
defend their dissertations in French.  As I find this policy unhelpful
to the scientific community at large, the manuscript that I have decided
to present here for public distribution is the original (unofficial)
English-language version of the text.
\begin{center}\textbullet\ \textbullet\ \textbullet\end{center}

\bigskip

My graduate student years owe their successful completion --- in both a
personal and a scientific sense --- to a very large number of individuals.
In the limited space I have here, let me highlight the contributions of
those to whom I am most grateful.

\begin{itemize}

\item {\bf Olivier Martin}, my thesis advisor.  I imagine that anyone who
has ever worked with Olivier has found the experience to be an immense
pleasure, from start to finish.  I am delighted to have shared in this
pleasure.  Not only is Olivier an extraordinarily warm, friendly and
patient person; he also has one of the liveliest scientific minds I have
ever encountered.  If my three years of thesis work have been among the
happiest of my life, Olivier deserves a great deal of the credit.

\item {\bf Dominique Vautherin}, former head of our {\it Division de
Physique Th\'eorique\/}, whose efforts to facilitate my stay here
went far above and beyond the call of duty.  His hospitality and constant
encouragement played a large part in making my time at Orsay productive
and enjoyable.

\item Other members of the {\it Division de Physique Th\'eorique\/},
who contributed to the fruitful environment of these past three years:
{\bf Oriol Bohigas}, whose ideas were fundamental to much of our work;
{\bf Xavier Campi}, present head of the Division; {\bf Nadine
Rohart}, Division secretary, whose skill at solving the most complex
administrative problems in under thirty seconds is nothing short of
remarkable; {\bf Mich\`ele Verret}, secretary, who was a constant and
invaluable source of help to me, and who willingly took on the task of
printing and binding this thesis.

\item The students and postdocs in Olivier's group: {\bf Georg
Schreiber}, {\bf J\'er\^ome Houdayer}, {\bf Jacques Boutet de Monvel} and
{\bf Nicolas Cerf}, who could all be relied upon at any moment for
stimulating discussions, friendship and moral support.

\item Other students in the Division, who managed to create an
atmosphere that was both scientifically productive and socially relaxed:
{\bf Alexis Prevost},
{\bf Amaury Mouchet}, {\bf Ulrich Gerland}, {\bf Isabela Porto Cavalcante},
{\bf Christophe Texier}, {\bf Daniel Rouben} and {\bf Guglielmo Iacomelli}.
Amaury deserves special mention --- the formatting style of his own PhD thesis
served as a model for this manuscript, to the point where practically all
of the \TeX\ macros used here are the result of his laborious efforts.

\hyphenation{Isa-belle}
\item My family, and some of my closest friends: {\bf Marie Aronson},
{\bf Marjorie Bertolus}, {\bf Isabelle Billig}, {\bf Paul Boubli},
{\bf Eleonore Dailly}, {\bf Fr\'ed\'eric Huin}, {\bf Isabelle Kraus},
{\bf Karen Meyer-Roux}, {\bf Anouchka Roggeman}, {\bf Tony Seward},
{\bf Eva Sibony} and {\bf Oana Tataru}.  It is difficult for me to express
in words how much their constant support and affection has meant to me
during these past few years in France.

\item My PhD committee, for their dedicated efforts and useful suggestions.

\item And last but not least\dots those whom I have unintentionally
forgotten here, but who surely deserve every bit as much thanks as the
individuals I have mentioned.  I hope they will accept my apologies and,
as this is a first offense, let me off lightly this time.  (Next time
they will not be so lenient.)

\end{itemize}

\clearemptydoublepage

\markboth{}{}

\dominitoc

\tableofcontents

\clearemptydoublepage

\chapter{Introduction}
\markboth{{\bfseries Chap.~\thechapter :}\bfseries\ Introduction}
{{\bfseries Chap.~\thechapter :}\bfseries\ Introduction}

\vspace{1cm}

Consider $N$ sites (``cities'') in a space, and the distances
connecting each pair of cities.  Now find the {\it shortest\/} possible
path (``tour'') visiting each city exactly once, and returning to its
starting point.  This is the traveling salesman problem (\tsp), the most
classic and best known problem in the field of combinatorial optimization (\co).

The \tsp\ is an extremely simple problem to state; deceptively so, because it
is also an extremely difficult one to solve.  The most na\"\i ve algorithm
for finding the optimum tour would have to
consider all $(N-1)!/2$ possible tours.  Working this way, the fastest
computer on the market today would require more time than the current age
of the universe to solve a case with 27 cities.  Of course, far more
sophisticated algorithms exist, but even the best of these is sharply
limited by the inherent complexity of the \tsp.  Computer scientists
classify the problem as being \np-hard: roughly speaking this means that,
most likely, there exists no algorithm that can consistently find the
optimal tour in an amount of time polynomial in $N$.
The statement
provides a technical standard
for ``hardness'', saying that if a polynomial algorithm {\it could\/} ever
be found for the \tsp, one would exist for a whole class of
other \co\ problems as well
\cite[see also discussion in Appendix \ref{app_pnp}]{GareyJohnson}.
For this reason, and because of how straightforward the \tsp\
is to state, it has served as a prototype for the study of difficult
combinatorial optimization problems in general.

\begin{figure}[t]
\begin{center}
\begin{picture}(180,200)
\includegraphics[scale=0.33]{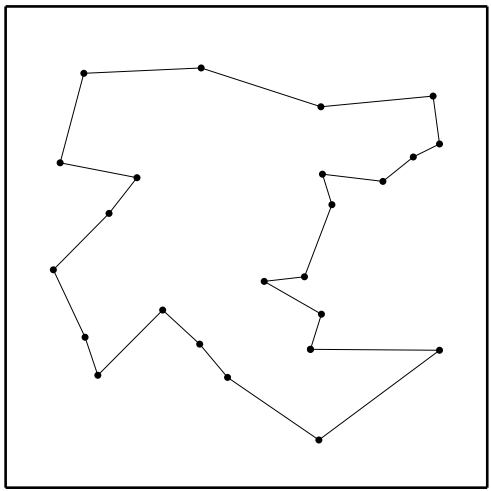}
\end{picture}
\caption{\small\rm An elementary instance of the traveling salesman problem,
and its optimal tour.}
\label{fig_tst}
\end{center}
\end{figure}

The challenge of making progress on the \tsp\ despite its complexity has
carried its appeal well beyond its initial boundaries.  Research on the
\tsp\ and related \co\ issues has become a truly interdisciplinary effort,
uniting computer scientists, pure mathematicians, operations researchers,
and more recently, theoretical
physicists.\footnote{Biology, in the guise of genetic algorithms and
{\sc dna} computing, is but the latest field to arrive on the scene.}
In physics, the \tsp\ has been of particular interest in its {\it stochastic\/}
form, where the configuration (``instance'') of the $N$ cities is
randomly chosen from an ensemble (see Figure \ref{fig_tst} for an
example of an $N=24$ instance and its optimal tour).
Starting in the 1980s, Kirkpatrick,
M\'ezard, Parisi and other theoreticians began applying tools from
statistical mechanics to the stochastic \tsp, noting broad similarities
between \co\ problems and the emerging field of {\it disordered systems\/}.
The hope was, first of all, that numerical methods developed in the
context of disordered systems could be adapted to the \tsp, and second
of all, that the analytical formalism developed for these other problems
could provide a theoretical framework for understanding the \tsp\ --- and
ultimately, could provide exact results as well.

\begin{figure}[!b]
\vspace{0.5cm}
\begin{center}
\begin{picture}(260,200)
\includegraphics[scale=0.6]{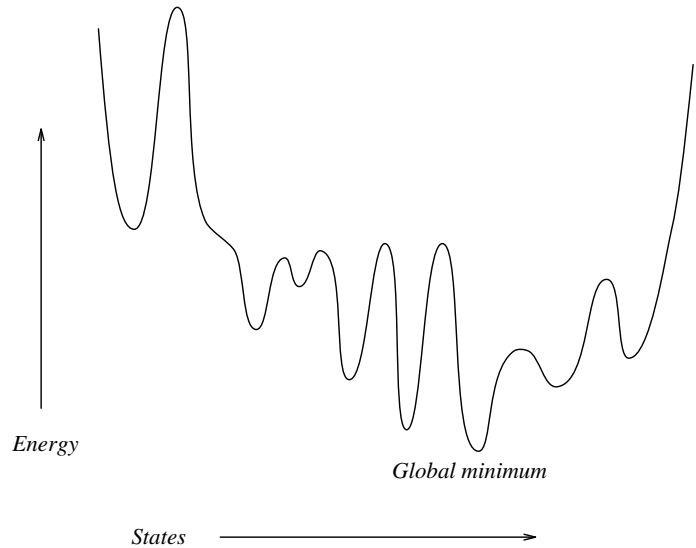}
\end{picture}
\caption{\small\rm Schematic energy ``landscape'', over state space, of a
frustrated disordered system.}
\label{fig_landscape}
\end{center}
\end{figure}

The generic example of a disordered system is the spin glass.  A spin
glass may be thought of as a ferromagnetic Ising model --- the fundamental
lattice model that statistical physicists use to describe magnetism ---
but modified so that the interaction strengths between neighboring
lattice sites are random quantities.  If, as in the case of the spin
glass, they are allowed to be negative as well as positive
(antiferromagnetic as well as ferromagnetic), we have a
{\it frustrated\/} disordered system.  Take a fixed instance of
these random variables (``quenched disorder'').  The total energy of
the system, involving a sum over all pairs of neighboring sites, is
then a complicated function of the particular state of the system (meaning
the binary values of the {\it spins\/} located at each of $N$ sites), and may
be represented schematically by a ``landscape'' of hills and valleys
(see Figure \ref{fig_landscape}).  We shall concern ourselves with what
statistical physicists think of as the {\it zero-temperature\/} case,
where we must find the lowest-energy state, and hence
the global minimum in this landscape.  If we are to
avoid actually enumerating all possible states --- a task which increases
exponentially with the number of sites --- the problem of finding the
global minimum becomes highly nontrivial since
there are numerous local minima in which a minimizing search
procedure could get trapped.

Let us now examine the \tsp\ using this picture.
The system's ``energy'' is the tour length, and is to be minimized.
The quenched ``interaction strengths'' are the distances
between cities.
The energy is once again a complicated function of the state of the system,
where in this case ``state'' refers to the particular tour chosen.
Optimizing the tour is precisely equivalent to finding the system's ground
state, and thus the global minimum in Figure \ref{fig_landscape}.  If we
look at it in this way, the same approximation methods used to
find spin glass ground states may be used to find optimal \tsp\ tours;
simulated annealing, developed independently by \citeasnoun{Kirkpatrick}
and \citeasnoun{Cerny}, is the most famous example.
At the same time, if we look not at a single instance but rather at the
stochastic ensemble of instances, analytical approaches giving exact
results for spin glass problems
can be adapted to provide
predictions for the \tsp\ that avoid algorithmic procedures altogether.

The foregoing analogy can be broadened: disordered systems and \co\ problems
display many similar properties, often making methods
developed in one field applicable to the other.  A common feature in
combinatorial optimization is the existence of a large number of
quantities interacting in a complicated manner, where each quantity
exerts an influence on the others --- and these influences are often
contradictory.  Describing such systems on a microscopic level is
generally so difficult as to be impossible.  A method better suited
to these problems is that of statistical physics, where one contents
oneself with studying the broader picture, reducing the number of
parameters considerably by resorting to a probabilistic approach.
One thus obtains the macroscopic quantities of interest, to reasonable
accuracy, with far less effort.

Let us express these concepts more explicitly.  A \co\ problem
such as the \tsp\ may be formulated as follows.
We have an objective function representing ``cost'' --- the tour length,
for the \tsp\ --- that must be minimized.  This cost may be expressed as
\begin{equation}
L\left(\{n_{ij}\}\right)=\sum_{i,j}l_{ij}n_{ij}\mbox{,}
\end{equation}
where $i$ and $j$ go from 1 up to $N$ (the total system ``size''),
the $\{l_{ij}\}$ are the parameters of the problem, and the $\{n_{ij}\}$
are the variables whose values must be adjusted so as to minimize $L$.
A certain set of constraints determines the allowed values for the
$\{n_{ij}\}$.  In the case of the \tsp, for example, $L$ is simply the tour
length to be minimized, $\{l_{ij}\}$ is the matrix of distances between
city $i$ and city $j$, and $\{n_{ij}\}$ is a binary matrix expressing whether
the link between $i$ and $j$ is used on the tour.  The constraints on
the variables are thus that: (i) each link is either used once or not
at all ($\forall i,j,\ n_{ij}\in\{0,1\}$ and $\forall i,\ n_{ii}=0$);
(ii) each city has one link going into it and one link coming out from it
($\forall j,\ \sum_{i}n_{ij}=1$ and $\forall i,\ \sum_{j}n_{ij}=1$); and
(iii) the $\{n_{ij}\}$ are such that the links form exactly one loop
($\forall i_1,\ \exists i_2,\dots,i_N\ :$
$n_{i_1i_2}=n_{i_2i_3}=\cdots
=n_{i_{N-1}i_N}=n_{i_Ni_1}=1$).\footnote{Alternatively, the final
constraint can be expressed by considering a partition of the
$N$ points into two sets $A$ and $B$, and requiring that $\forall A,B,\
\sum_{i\in A, j\in B}n_{ij}\ge 1$.}

To put this into the language of statistical physics, we think of $L$
as the system's energy; each contribution $l_{ij}n_{ij}$ is then the
interaction energy associated with the pair of sites $i$ and $j$.
We may write down the partition function:
\begin{equation}
Z=\sum_{\{n_{ij}\} }e^{-L(\{n_{ij}\})/T}
\mbox{,}
\end{equation}
where the sum is taken over the ensemble of matrices $\{n_{ij}\}$ that satisfy
the constraints we have posed, and $T$ is the physical ``temperature''
of the system.  In the limit $T\to 0$ the free energy $F=-T\ln Z$ will
be dominated by the ground-state contribution, that is, the particular
state (value of the matrix $\{n_{ij}\}$) that minimizes $L$.  When we speak
of the $T=0$ solution, we are thus speaking of the global optimum.

Rephrasing the problem in this manner does not appear to help us
find the exact state minimizing $L$ for a given set of parameters
$\{l_{ij}\}$ ---
after all, when $N$ is large and we are dealing with matrices of
size $N\times N$, there is still an enormous number of parameters.
However, we may instead consider the $\{l_{ij}\}$ as being chosen
randomly from an ensemble, and ask for properties of the {\it distribution\/}
of $L$, such as its mean over the ensemble.  Analytical tools from
statistical physics, and particularly the approximations that are
typically used, then become relevant and useful.  The stochastic \tsp\
is a prime candidate for this sort of treatment.

Two main versions of the stochastic \tsp\ have been discussed in the
literature, and form the subject of the majority of this thesis.  The
first and more classic version is the Euclidean \tsp\ (also known as
the random point \tsp), where the $N$ cities are placed randomly and
independently in $d$-dimensional Euclidean space, and the distances
between cities are given by the Euclidean metric.  The second is the
random link \tsp, largely developed within the context of disordered
systems.  In the random link \tsp, it is not the positions of the cities
but rather the lengths separating them that are independent and identically
distributed random variables.  We speak of ``lengths'' because there is
no physical ``distance'' here, nor geometry of any sort, apart from a
symmetry requiring that the length $l_{ij}$ be equal to $l_{ji}$ between
any two cities $i$ and $j$.\footnote{Asymmetric versions of the problem do
exist as well, though we do not consider them here.}  The great advantage
of the random link \tsp\ over the Euclidean (random point) \tsp\ is that the
lack of correlations among the lengths $l_{ij}$ brings an
analytical solution within reach.

It was noted by \citeasnoun{VannimenusMezard} that in the limit of large
$N$, one may perfectly well choose the distribution of the
individual $l_{ij}$'s in the random link \tsp\ so as to match that of
the $d$-dimensional Euclidean \tsp.  The probability of finding a point
at Euclidean distance $l$ is simply the surface area of a $d$-dimensional
sphere of radius $l$:
\begin{equation}
\rho_d(l) = \frac{d\,\pi^{d/2}}{\Gamma(d/2+1)}\,l^{d-1}\mbox{.}
\end{equation}
In this way, the random link and Euclidean models can be made to match
one another,
up to correlations between lengths.
We may therefore consider the random link \tsp\ as a {\it random link
approximation\/} to the Euclidean \tsp.  This approximation is far from
perfect, since in reality correlations between Euclidean distances, such
as the triangle inequality, are important.  In terms of the results
obtained, however, the random link approximation turns out to be a
surprisingly good one.

The approach in this thesis is, first of all, to examine the $d$-dimensional
Euclidean \tsp, focusing on the large $N$ behavior of the optimal tour
length $L_E$ (``{\sc e}'' for Euclidean).  Our study is numerical, making
use of good approximate methods for estimating $L_E$ to within small,
well-controlled errors.  This, along with some theoretical insight into
the scaling behavior to expect in arbitrary dimension $d$, enables
us to obtain (in units
where volume equals one) the large $N$ {\it finite size scaling law\/}
for the ensemble average of $L_E$:
\begin{equation}
\langle L_E(N,d)\rangle = \beta_E(d)\,N^{1-1/d}\,\left[ 1+\frac{A(d)}{N}+
O\left(\frac{1}{N^2}\right)\right]\mbox{.}
\end{equation}
At $d=2$ and $d=3$ we find extremely precise numerical values for
$\beta_E(d)$: our results, $\beta_E(2)=0.7120\pm 0.0002$ and
$\beta_E(3)=0.6979\pm 0.0002$, represent a precision
20 times greater than was possible using previously
available
methods.\footnote{These values have subsequently been confirmed by
\citeasnoun{Johnson_HK}, using a modified version of our methods and
more powerful computational resources.  See Appendix \ref{app_dsj} for
further discussion.}
It would of course be nice to have an analytical
estimate of $\beta_E(d)$, and this is where the random link approximation
comes in.  As we have mentioned, this approximation is what makes the
\tsp\ analytically tractable.  Using the ``cavity'' approach of M\'ezard,
Parisi and Krauth \cite{MezardParisi_86b,KrauthMezard,Krauth_Thesis},
we obtain an estimate for $\beta_E(d)$.  For $d=2$ and $d=3$,
the random link approximation turns out to be a remarkably good one,
with an error of less than 2\%.  In the $d\to\infty$ limit, we
argue that in fact the approximation is exact up to $O(1/d^2)$.  This
leads to a conjecture on the behavior of $\beta_E(d)$ at
large $d$, in terms of a series in $1/d$ where leading and subleading
coefficients are given.

It is worth explaining why these Euclidean results are of broader interest
than just to theoretical physicists.  One of the earliest mathematical
results for the Euclidean stochastic \tsp\ \cite{BHH} states that the
random variable $L_E$ obeys a self-averaging property:
\begin{equation}
\label{eq_selfav}
\frac{L_E(N,d)}{N^{1-1/d}}\to\beta_E(d)\mbox{ as }N\to\infty
\end{equation}
with probability 1.  $\beta_E(d)$ is a well-defined quantity and not
a random variable.
This means that if we take a sequence of instances
drawn randomly at large $N$, the relative fluctuations in the optimal
tour length will be small from one instance to the next.  (See
Appendix~\ref{app_selfav} for an outline of the self-averaging proof.)
Using this property,
we may consider some large real-life Euclidean situation, such as
an airplane
serving numerous cities or a delivery truck serving a large
number of retailers, and make the assumption that the instance ``resembles''
one from the random ensemble, {\it i.e.\/}, sites positioned independently.
Knowing $\beta_E(d)$ allows us at least a rough estimate of how far any
prospective tour in this instance is from ``theoretical optimality''.
(A concrete example of this is given in Appendix~\ref{app_nonunif}.)

From a theoretical point of view, the random link \tsp\ is very much of
interest as well, as a model of its own.  The cavity method gives an
analytical prediction for $\beta_{RL}(d)$, the random link counterpart
of $\beta_E(d)$.  However, this prediction is based on some rather
subtle assumptions, reflecting work done in spin glasses.  The central
such assumption is that of {\it replica symmetry\/}, an ergodicity
hypothesis stating that there is a unique thermodynamic limit, in a certain
spin model onto which one maps the random link \tsp.  This assumption
is known to be false in the case of spin glasses.  For the \tsp, however,
numerical analyses performed in the $d=1$ case \cite{Sourlas,KrauthMezard}
suggest that it is in fact correct.  We extend this numerical work to higher
dimensions; our results indicate strongly that the cavity method gives
correct predictions for all $d$.

Finally, there is a related stochastic problem, discussed in this thesis,
that does not directly involve the \tsp\ though it is motivated by the
finite size scaling analysis used for the Euclidean \tsp.  This is a
geometric issue concerning properties of distances between randomly
distributed sites, notably distances from a point to its $k$th-nearest
site.  If the sites are distributed uniformly in Euclidean space
(as cities are in our formulation of the Euclidean \tsp), these $k$th-neighbor
distances display a remarkable universality: they all have the same large
$N$ scaling law.  Letting $\langle D_k(N)\rangle$ be the ensemble average
of distances between $k$th-nearest neighbors, the $N$-dependence and the
$k$-dependence of $\langle D_k(N)\rangle$ separate, and
\begin{equation}
\langle D_k(N)\rangle = \frac{[\Gamma(d/2+1)]^{1/d}}{\sqrt{\pi}}\,
\frac{\Gamma(k+1/d)}{\Gamma(k)}\,\frac{\Gamma(N)}{\Gamma(N+1/d)}
\end{equation}
in $d$-dimensional Euclidean space.  If we work not in Euclidean space
but on an arbitrary geometric manifold, a different universality
emerges: when $\langle D_k(N)\rangle$ is written as a power of $N$
times a correction series in $1/N$, the $O(1/N)$ term of the series is a
topological
invariant and does not depend on the detailed properties of the manifold.
We discuss these universalities and surrounding issues, noting that
some open questions include the application of this work to more complicated
geometric problems such as triangulations.
\begin{center}\textbullet\ \textbullet\ \textbullet\end{center}

In both its stochastic and non-stochastic forms, the \tsp\ has
attracted a large amount of research.  In order to provide the
reader with a better perspective of the context in which our own work
is situated, let us survey some of the most relevant results in the
literature.

The utopian ideal for a \tsp\ algorithmic procedure would be one which
finds the exact optimum, for an arbitrary instance, in a short amount of
time.  The fact that the problem is \np-hard appears to make this
goal unattainable (see Appendix \ref{app_pnp} for a discussion of the
\tsp's complexity).  Nevertheless, computer scientist have for years
tried to develop algorithms that reduce
the amount
of time necessary to find a guaranteed optimal tour.  The most successful
such algorithms have been based on the method of ``branch-and-bound''.
The idea is as follows: a recursive procedure is used to break down an
$N$-city problem into smaller subproblems.  Each subproblem is then
``relaxed'' so as to find a bound.  These bounds are subsequently
used to eliminate further subproblems from consideration.
In this way, the algorithm successively ``branches'' --- divides problems
into subproblems --- and ``bounds'' --- prunes branches that violate
bound restrictions \cite{BalasToth}.  Finding efficient branching and
relaxation procedures are, of course, highly sophisticated problems in
themselves.  A variant on this method known as ``branch-and-cut'' represents
today's state of the art in exact algorithms: problems of up to 100 cities
can typically be solved in minutes on a workstation \cite{PadbergRinaldi}.

However, when instances of larger sizes must be optimized, or a large
sequence of instances must be optimized, exact algorithms are no longer
practical.  The next best choice, then, is to use approximate methods
that find near-optimal tours quickly (typically in $O(N^2)$ operations).
These are known as {\it heuristic\/} algorithms.  For the \tsp, the most
powerful heuristics have been those that start with a non-optimal tour
and successively attempt to improve upon that tour.  The procedure
generally used is known as {\it local search\/}: a certain number of
links are removed from a tour and are reconnected so as to form a new
and better tour; when all possible local changes have been tried without
success, the resulting tour is then a local optimum.
The simplest algorithm of this sort is 2-opt, where two links are deleted
and the paths are reconnected differently.  3-opt does the same but using
3 links in each elementary move \cite{Lin}.  Both heuristics tend to give
tour lengths that are a few percent
higher than the optimum for Euclidean \tsp\ instances \cite{JohnsonMcGeoch}.
Various methods have been popular for improving on this.  The simplest
is to repeat 2-opt or 3-opt from different random starting tours, taking
the best local optimum found over many trials, so as to cover more
of the energy landscape (see Figure \ref{fig_landscape}).  Another
method is simulated annealing, which can outperform repeated 3-opt
(in terms of both running time and percent excess of optimum) only
for sufficiently large instance sizes ($N>1000$).  The most
successful of the local search heuristics, however, is that of
\citeasnoun{LinKernighan}.  This algorithm uses a ``variable-depth
neighborhood search'' method where a large number of links can be
changed at once,
but where
redundancy in searching is reduced to the point where a million-city
Euclidean instance can be solved in less than an hour, with less than
2\% excess over the optimum \citeaffixed{JohnsonMcGeoch}{see Appendix
\ref{app_method} for a description of the heuristic; see also}.

Improving upon Lin-Kernighan has been a goal in algorithmic research
for over 20 years, with relatively limited success.  One particularly
creative attempt has been through the use of {\it genetic algorithms\/}
where, inspired by the process of evolution, one attemps to establish
new starting tours by ``mating'' (combining aspects of two ``parent''
locally optimal tours) and by ``mutation'' (randomly altering a locally
optimal tour).  Although the mating principle has not yet proven to
produce a competitive heuristic, the mutation principle has been used
with success in the ``metaheuristic'' known as Chained Local Optimization
(\clo) by \citeasnoun{MartinOttoFelten_CS}.  \clo\ involves successively
applying a random ``kick'' to a locally optimal tour, and then using
Lin-Kernighan to locally optimize from there.  If the new tour is
better than the old one, the procedure iterates from that point; if
not, the old tour is again used and a new random ``kick'' is attempted.
(A qualitative description may be found in Appendix \ref{app_method}.)
\clo\ outperforms Lin-Kernighan for instances of $N>100$, taking seconds
to get within 1\% of the optimum for a 1000-city instance as opposed
to minutes for the equivalent performance simply repeating Lin-Kernighan
from random starts \cite{MartinOtto_AOR}.

All of what has been mentioned so far applies to optimizing individual
\tsp\ instances.  When we speak of the stochastic \tsp, we must access an
entire ensemble of instances.  One approach is simply to run optimization
heuristics over a large sample from the ensemble, thus obtaining results
to within a quantifiable statistical error.  This approach is indeed
the basis of most numerical studies, including our own.  These numerical
studies, however, have also been accompanied by a certain amount of
theoretical work.

For the Euclidean stochastic \tsp, we have already seen the {\it
self-averaging\/} property shown by Beardwood, {\it et al\/}.  These
authors, in the same article, gave
lower and upper
bounds on the asymptotic limit $\beta_E(d)$.  The lower bound is
simple: it uses the fact that, at best, any city on the tour
has links to its two nearest neighbors.  Nearest neighbor distances
may be calculated over the ensemble (as we discuss in depth in Chapter~IV),
and one finds that $\beta_E(2)\ge 5/8$.  It is straightforward to
generalize this bound to
any dimension $d$.  For the upper bound, a method known as
``strip'' was used, whereby the space is divided into columns (strips)
and a suboptimal tour is constructed by visiting cities alternately
up and down the columns.  In 2 dimensions, this suboptimal tour length
may be calculated exactly over the ensemble; the result is a complicated
expression involving special functions, but may be numerically computed
as $\beta_E(2)\le 0.9204$.

No better exact bounds are known today; the best that has been done is
to use heuristic methods on very large instances (or many very large
instances) to establish estimated bounds.  In this way, an approximation
to the lower bound of \citeasnoun{HeldKarp_A}, involving a linear
programming relaxation of the \tsp, has very recently been used by
\citeasnoun{Johnson_HK}
to estimate $\beta_E(2)\ge 0.708$.  \citeasnoun{LeeChoi} have used a
variant on simulated annealing to obtain the upper bound $\beta_E(2)\le
0.721$.  At $d>2$, virtually no improvements over the exact bounds have
existed prior to our own work.  (Johnson {\it et al.\/} have, however,
in their work on the Held-Karp bound, successfully applied a modified
version of
our numerical methods to higher precision, confirming our own $d=2$ and
$d=3$ results and extending to the $d=4$ case.)
The only other related large $d$ result is
an unproven conjecture by \citeasnoun{BertsimasVanRyzin}, claiming that
when $d\to\infty$, $\beta_E(d)\sim\sqrt{d/2\pi e}$.

For the random link stochastic \tsp, results have been largely due to
the work of theoretical physicists.  \citeasnoun{VannimenusMezard} and
\citeasnoun{KirkpatrickToulouse} first considered the statistical
mechanics of the \tsp, using analytical tools stemming from the study
of spin glasses.  Vannimenus and Mezard showed that the equivalent
of the Bertsimas-van Ryzin conjecture is correct for the random link
case, {\it i.e.\/}, $\beta_{RL}(d)\sim\sqrt{d/2\pi e}$ when $d\to\infty$.
Kirkpatrick and Toulouse studied several low-temperature macroscopic
quantities of the random link model, analyzing the number of links in
common between local optima at finite temperature.  This ``overlap''
leads to the notion of a state-space {\it bond distance\/}, measuring
by how much two tours differ from each other.  In analogous spin glass
problems, such distances between local minima display a property known
as {\it ultrametricity\/}, a stronger form of the triangle inequality,
requiring that for any three distances at least two must be equal
(they form an isoceles triangle) and the third must be less than or equal
to the other two.  Non-trivial ultrametricity is considered a typical
sign of a lack of ergodicity in spin glasses, called {\it replica
symmetry breaking\/}.
\citeasnoun{Sourlas}, numerically extending Kirkpatrick and Toulouse's
low-temperature analysis, found on the basis of the tour overlap that
ultrametricity does not play a role in the $d=1$ random link \tsp, and
that replica symmetry most likely holds.  In the process, he observed
that solutions only contain links between very near neighbors, and
used this property to motivate an improved optimization heuristic that
works by disallowing certain links.

Under the assumption that the replica symmetric solution is the correct
one, \citeasnoun{MezardParisi_86b} and \citeasnoun{KrauthMezard} used
the cavity method to find an analytical expression for $\beta_{RL}(d)$,
in the form of an integral equation.  They performed further numerical
checks at $d=1$ (through direct simulations of the random link \tsp)
that confirmed their analytical cavity results, and briefly discussed
the $d=2$ result.  We are aware of no further
attempts, prior to our own, to extend the analysis of the cavity solutions
to higher dimensions.

\begin{center}\textbullet\ \textbullet\ \textbullet\end{center}

Let us conclude this chapter with a brief guide to the layout of what
follows in the thesis.  Chapter~II deals with scaling laws in the stochastic
\tsp.  In the articles presented in the chapter
we discuss our most important results, notably a numerical study of the
finite size scaling of $L_E$, and an analytical study of the dimensional
scaling of $\beta_E(d)$ by way of the random link approximation.  Our
analysis of $L_E(N,d)$ at large $N$ is, to our knowledge, the first to
use a systematic method for extrapolating to the large $N$ limit.  We
are thus able to improve significantly on previous estimates of $\beta_E(d)$.
Using the random link approximation, we also present evidence supporting
the Bertsimas-van Ryzin conjecture that $\beta_E(d)\sim\sqrt{d/2\pi e}$
at large $d$, and propose a much stronger conjecture: that
\begin{equation}
\beta_E(d) = \sqrt{\frac{d}{2\pi e}}\,(\pi d)^{1/2d}
\left[ 1+\frac{2-\ln 2-2\gamma}{d}
+ O\left(\frac{1}{d^2}\right)\right]\mbox{,}
\end{equation}
where $\gamma$ is Euler's constant ($\gamma\approx 0.57722$).

Chapter~III covers the
random link
\tsp.  We discuss the background of the analytical tools used, and the
physical model serving as the basis of the cavity method.  Through
simulations, we then test the cavity results for $d>1$, and find good
numerical justification for the hypothesis of replica symmetry.

Chapter~IV provides a discussion of the universality in $k$th-nearest
neighbor distances for randomly placed sites in Euclidean space, namely
that the same scaling in $N$ applies to the mean distance $\langle
D_k(N)\rangle$ regardless of $k$.  We generalize this study to arbitrary
geometric manifolds, and find that, although the scaling of
$\langle D_k(N)\rangle$ does depend on $k$ in general, its $O(1/N)$
term is a topological invariant and does not depend on the precise
shape of the manifold.  We note that these properties, though arising
from relatively simple physical arguments, can be applied to far more
complex geometric problems.

In each of these subsequent three chapters, an overview of the subject
is given, following which one or more articles make up the body of
the chapter.  In Chapter~II, these are articles that have already been
published, and are presented in reprint form.
In Chapters~III and IV, the articles have not yet been submitted,
and so it has been possible to take some liberties with their presentation,
following wherever possible the style adopted in the rest of the text.
Not being constrained by editorial restrictions concerning brevity,
we have also attempted to adopt a more pedagogical tone than will
ultimately be permitted in published form.

Several appendices follow the main body of the thesis.  In
Appendix~\ref{app_pnp},
we define the notions {\sc p} and \np\ as used in computational complexity
theory.  We show how to phrase the \tsp\ in such a way that it is an
\np-hard problem, and discuss polynomial time approximation algorithms.
Appendix~\ref{app_selfav} covers self-averaging in $L_E(N,d)$; we
outline a simple proof of this property, establishing that
$L_E(N,d)/N^{1-1/d}$ goes to a constant $\beta_E(d)$ when $N\to\infty$.
In Appendix~\ref{app_nonunif} we explain how our Euclidean stochastic \tsp\
results, although intended for a uniform distribution of cities, may easily
be applied to non-uniform cases.  We give the example of the {\sc at\&t}-532
instance (containing 532 {\sc at\&t} telecommunications sites around the
U.S.), and show
that we come within 1\% of the true optimum by regarding it as a
member of the stochastic ensemble.  In Appendix~\ref{app_method},
we discuss our numerical methodology, giving an explanation of the
workings of the Lin-Kernighan and \clo\ algorithms for optimizing \tsp\
instances.  Finally, in Appendix~\ref{app_dsj}, we describe a more recent
numerical study of $\beta_E(d)$ and $\beta_{RL}(d)$ performed by
\citeasnoun{Johnson_HK}, using a variant on our methods.  They obtain
results in close agreement with ours, and by combining their data and
ours we are in fact able to refine our $\beta_E(2)$ estimate slightly
further.

\newpage
\clearemptydoublepage

\chapter{Scaling laws in the Euclidean TSP}
\thispagestyle{empty}
\minitoc

\vspace{1cm}
\preliminaires{}

Historically, the traveling salesman problem has been of
greatest interest in its Euclidean formulation.  In this form of the
problem, each of the $N$ cities has a position in $d$-dimensional space
(in practice, most often 2-dimensional space), and the lengths separating
pairs of cities are defined to be the associated Euclidean distances.
The traveling salesman tour is thus a tour through a metric space.

Much past work on this problem has concentrated on developing algorithms
to solve certain large instances, or often certain classes of instances.
One goal in operations research has been to find methods of reducing
redundancy in exhaustive search --- methods that, while still exponentially
slow in $N$, at least have a small enough coefficient in the exponential
that they find the exact optimum in a reasonable amount of time.  Another
goal has been to classify special cases of instances.  The idea here
is to identify properties of the layout of cities that could guarantee
finding an optimal tour within, say $O(N^\alpha)$ steps for some fixed
$\alpha$.  Finally, considerable efforts have been devoted to finding
mathematically rigorous bounds on tour lengths.

In the case of the stochastic Euclidean \tsp, however, the emphasis is
somewhat different.  The positions of the cities in the tour are now
independent random variables; in the model that we consider here, the
distribution is uniform over space.\footnote{The non-uniform case is
discussed in Appendix \ref{app_nonunif}.}  It is no longer a matter of
optimizing a single instance.  We are now working with an entire
ensemble of instances, and the problem is to find properties of the optimal
tour length $L_E$ (the actual path that tours follow is of less interest when
the positions of the cities are not fixed quantities) considered in the
ensemble of all possible instances.  The challenge consists of understanding
what happens at large instance sizes.  The fundamental property here is that
of {\it self-averaging\/}, proven by \citeasnoun{BHH}: as $N$ gets large
the relative fluctuations in the optimal tour length go to zero, and the
distribution of $L_E$, up to a scaling factor, becomes more and more
sharply peaked.  The real quantity of interest is then the {\it mean\/}
optimal tour length over the ensemble --- at a given instance size $N$
and for a given dimension $d$ --- which we denote $\langle L_E(N,d)\rangle$,
using units where the volume of our space is 1.

An outline of a more recent and simpler proof of self-averaging, due to
\citeasnoun{KarpSteele}, is given in Appendix~\ref{app_selfav}.  The
precise statement of the property is that with probability 1, for any
sequence of random instances, $L_E(N,d)/N^{1-1/d}$ will approach the
instance-independent quantity $\beta_E(d)$ at large $N$.  Note that
the scaling factor $N^{1-1/d}$ may also be seen by a simple physical argument:
we are working in a fixed volume, so the mean volume {\it per city\/}
scales as $N^{-1}$, and so the mean distance between cities scales as
$N^{-1/d}$; the tour contains $N$ such links, hence its length scales
as $N^{1-1/d}$.

The question then becomes how to determine the limiting 
\begin{equation}
\beta_E(d)\equiv\lim_{N\to\infty} \frac{\langle L_E(N,d)\rangle}{N^{1-1/d}}
\mbox{.}
\end{equation}
For small values
of $d$, this may be answered by combining numerical methods with some
theoretical insights --- namely, {\it how\/} $L_E(N,d)/N^{1-1/d}$ converges
to its
large $N$ limit.  We thus consider the finite size scaling of $L_E(N,d)$.
At large $d$, this no longer feasible numerically.  Fortunately, in the
limit $d\to\infty$, an analytical approach may be developed for obtaining
$\beta_E(d)$ to within an excellent approximation.

What sort of numerical methods are needed for a finite size scaling
analysis?  At a given $d$, we must find $\langle L_E(N,d)\rangle$ over
many values of $N$.  At any particular value of $N$, in order to obtain
this quantity we must, in turn, average $L_E(N,d)$ over a large number
of instances.  (The smaller $N$ is, the more instances we will have to
consider in order to minimize the effects of fluctuations.)  At any
given instance, finally, in order to find the optimum we must have
efficient algorithms.  Unfortunately, exact algorithms are too slow to
be of use here, requiring minutes of CPU time to solve typical instances
of $N=100$;
optimizing a statistically significant sample
of random instances in this way could take weeks of computing time.

There is, fortunately, a better approach.  Powerful {\it heuristic\/}
algorithms exist --- inexact algorithms which find ``good'' tours but
not always the optimal one.  The most suitable method is that of {\it local
search\/}, where the algorithm takes a non-optimal tour and iteratively
improves upon it so as to bring it closer to optimality.  We may think of
this in terms of the landscape in Figure \ref{fig_landscape}; the
algorithm finds the local minimum of the valley in which we start, and
if it is a good algorithm, the landscape will be such that there are
a small number of broad valleys.  As we execute a local search heuristic
increasingly many times on a given instance, each time starting
from a different
random initial tour, we then explore increasing parts of the landscape
and find the true optimum with increasing probability.  Moreover, even
with multiple runs, these heuristics work considerably faster than exact
algorithms --- especially at large $N$ where running time grows only
polynomially and not exponentially in $N$.  This means that in a given
amount of time, we are able to optimize a much larger sample of instances
(and thus have a far smaller statistical error) using heuristics than
using exact algorithms.  The net effect is counterintuitive: because
of computing time considerations, we actually get {\it more\/} precise
results using these inexact algorithms than using exact algorithms.
There is, admittedly, a systematic bias in the results arising from the
use of quasi-optimal tour lengths: no matter how many times we run a
heuristic, we are never {\it sure\/} to have found the global optimum
in the end.  However, we choose the number of runs per instance sufficiently
conservatively that this systematic bias --- estimated for each value of
$N$ on the basis of a number of test instances --- is so small as to be
negligible compared with the statistical error.  (Our numerical methodology
is discussed in Appendix~\ref{app_method}.)

Equipped with this numerical approach and certain theoretical notions
concerning the scaling to expect in $N$ --- described in detail in the
articles that follow --- we find the large $N$ finite size scaling law:
\begin{equation}
\label{eq_pre_fss}
\langle L_E(N,d)\rangle = \beta_E(d)\,N^{1-1/d}\,\left[ 1+\frac{A(d)}{N} +
O\left(\frac{1}{N^2}\right)\right]
\end{equation}
using a $d$-dimensional unit hypercube with periodic boundary
conditions.\footnote{\citeasnoun{Jaillet} has proven that $\beta_E(d)$
is in fact the same regardless of whether periodic or open boundary
conditions are used.}
Understanding the scaling in this manner enables us to obtain $\beta_E(d)$
to high precision at $d=2$ and $d=3$ (the cases that we consider
numerically): $\beta_E(2)=0.7120\pm 0.0002$ and $\beta_E(3)=0.6979\pm 0.0002$.
When $d$ gets larger, however, the feasibility of numerical solutions
rapidly decreases.  In order to get an idea of the dimensional scaling
of $\beta_E(d)$, we therefore turn to the analytical approach used in
the {\it random link approximation\/}.

The random link approximation consists of assuming that lengths $l_{ij}$
between city $i$ and city $j$ ($i<j$) are completely independent of one
another.  Thus under the random link approximation (which could perhaps
more properly be called the ``independent link approximation'') correlations
between distances, such as the triangle inequality, are neglected.  Making
this approximation allows us to take advantage of an analytical approach
developed by
\citeasnoun{MezardParisi_86b}, \citeasnoun{KrauthMezard} and
\citeasnoun{Krauth_Thesis}.
We use this approach to obtain $\beta_{RL}(d)$, the
random link value approximating $\beta_E(d)$.
$\beta_{RL}(d)$ arises from a system of integral equations and does not
appear to have an exact analytical solution, but can on the other hand
be solved numerically to arbitrary precision.  For small values of $d$,
we find that $\beta_{RL}(d)$ is a good approximation to $\beta_E(d)$:
for $d\le 3$, the relative error is within about 2\%.

For large $d$ we are in fact able to express $\beta_{RL}(d)$ analytically,
via a power series in $1/d$.  To leading and subleading order, this gives
\begin{equation}
\label{eq_conjecture1}
\beta_{RL}(d) = \sqrt{\frac{d}{2\pi e}}\,(\pi d)^{1/2d}
\left[ 1+\frac{2-\ln 2-2\gamma}{d}
+ O\left(\frac{1}{d^2}\right)\right]\mbox{,}
\end{equation}
where $\gamma$ represents Euler's constant ($\gamma\approx 0.57722$).
We then present an argument, based on a theoretical analysis of how certain
heuristic algorithms operate, suggesting that the relative error in
the random link approximation itself decreases at least as fast as
$O(1/d^2)$
in the limit $d\to\infty$.  This enables us to conjecture, finally, that
the large $d$ expression (\ref{eq_conjecture1}) applies to $\beta_E(d)$
as well, and thus gives the correct dimensional scaling for the
Euclidean optimal tour length.

\newpage
\clearemptydoublepage

{

\section*{{\bfseries \itshape Reprint: Physical Review Letters}}
\addcontentsline{toc}{section}{Reprint: Physical Review Letters}
\renewcommand{\sectionmark}[1]{\markright{\bfseries\itshape
Reprint: Physical Review Letters}}
\sectionmark{}
\clearemptydoublepage

\prlpage{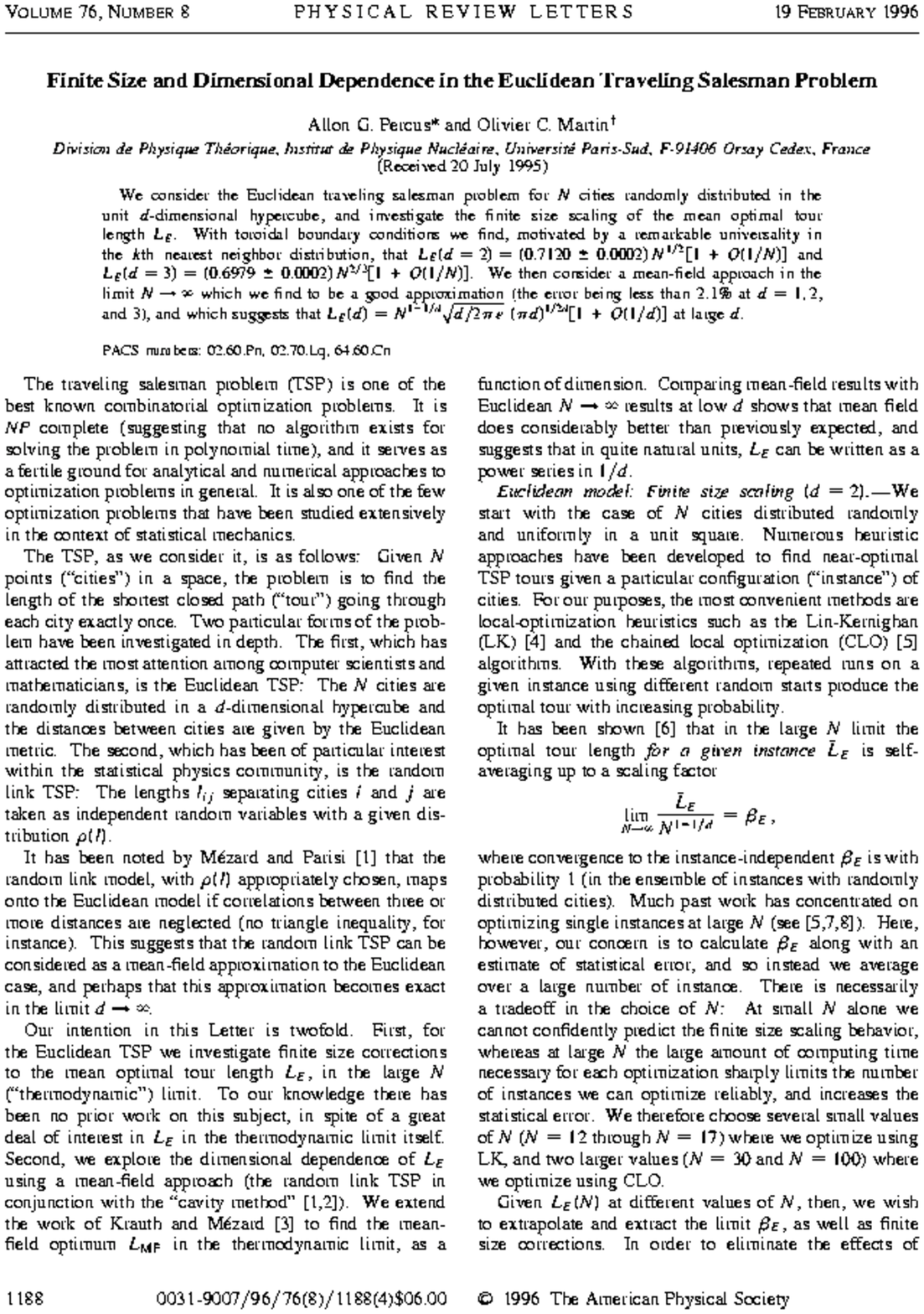}
\prlpage{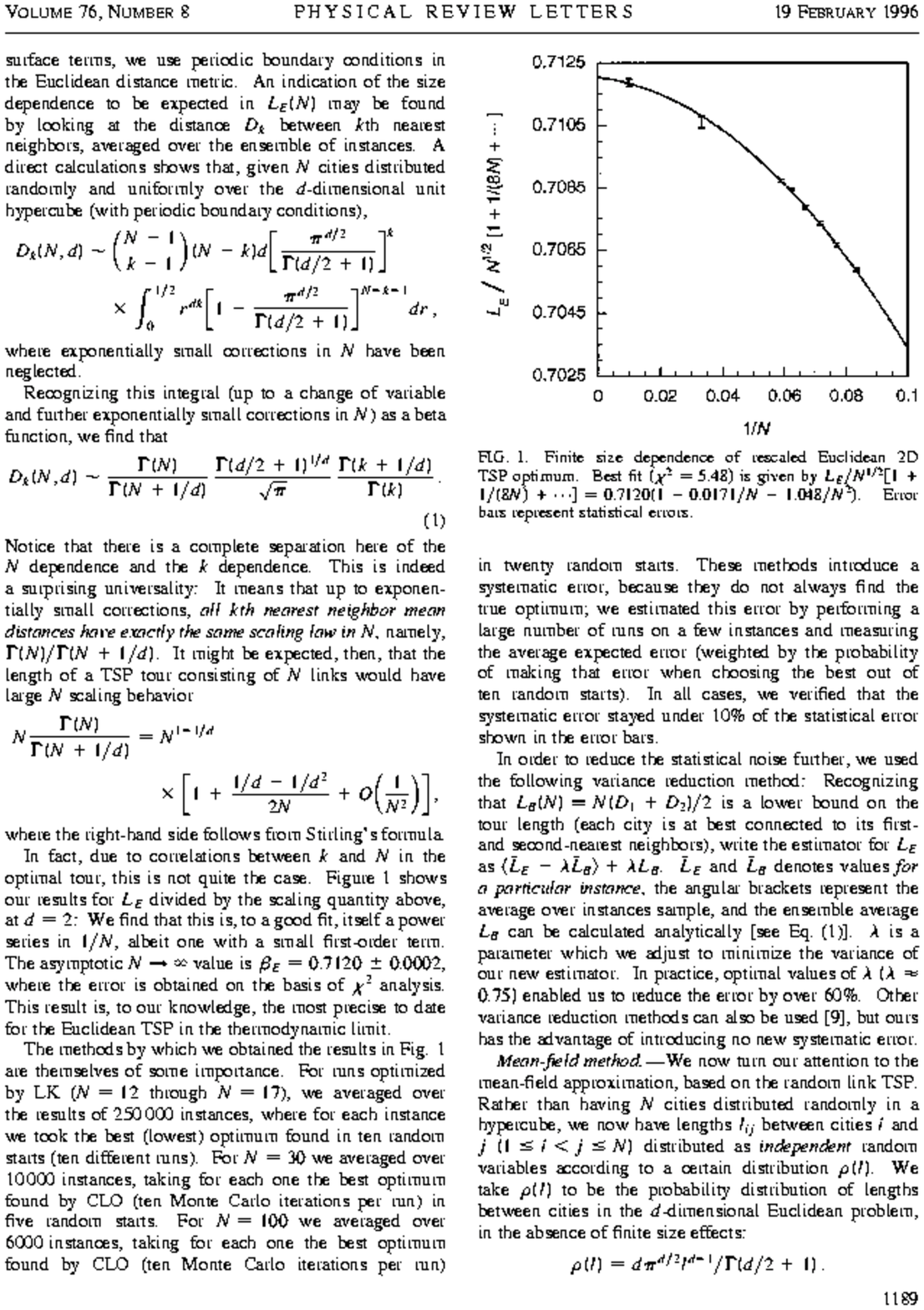}
\prlpage{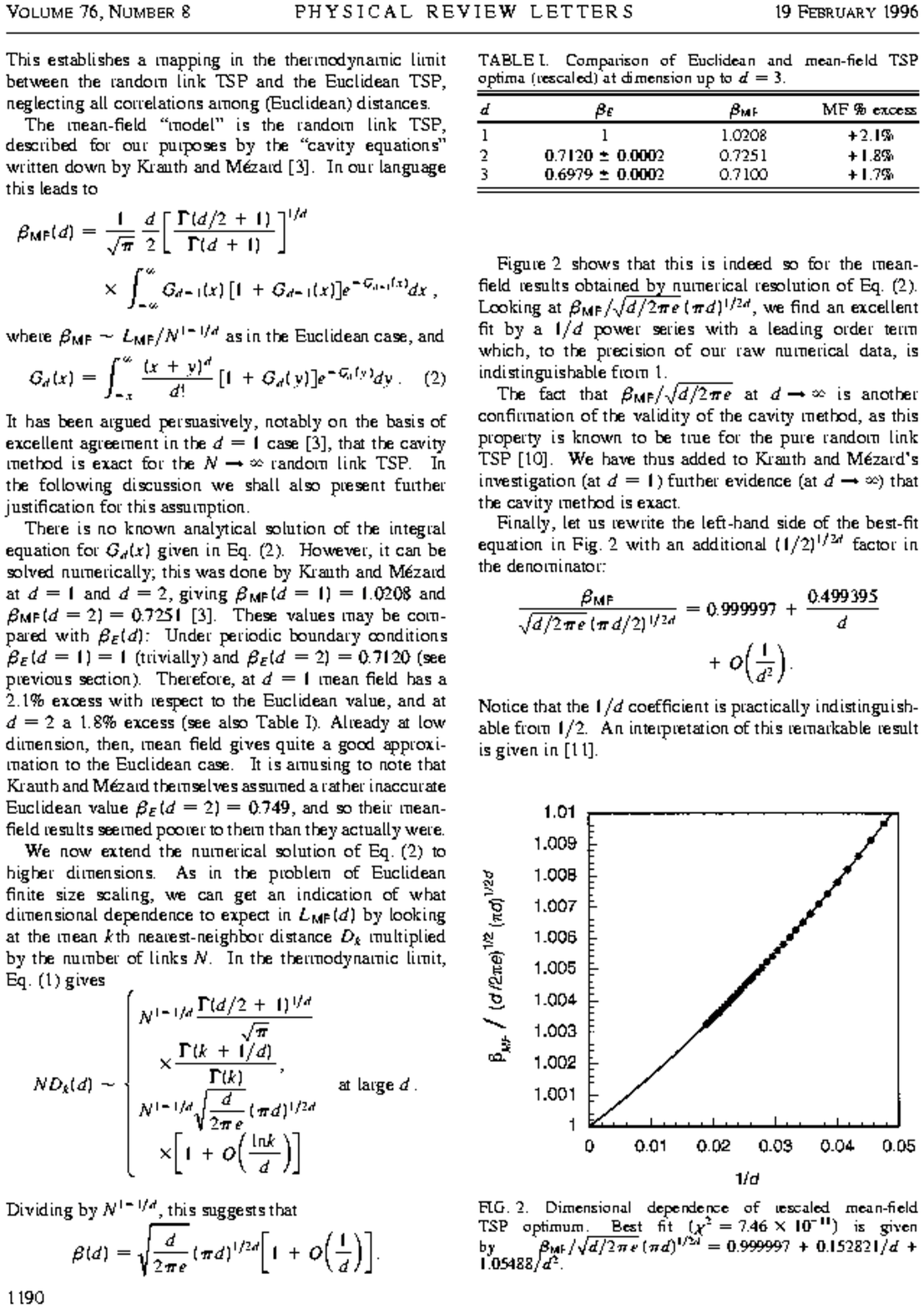}
\prlpage{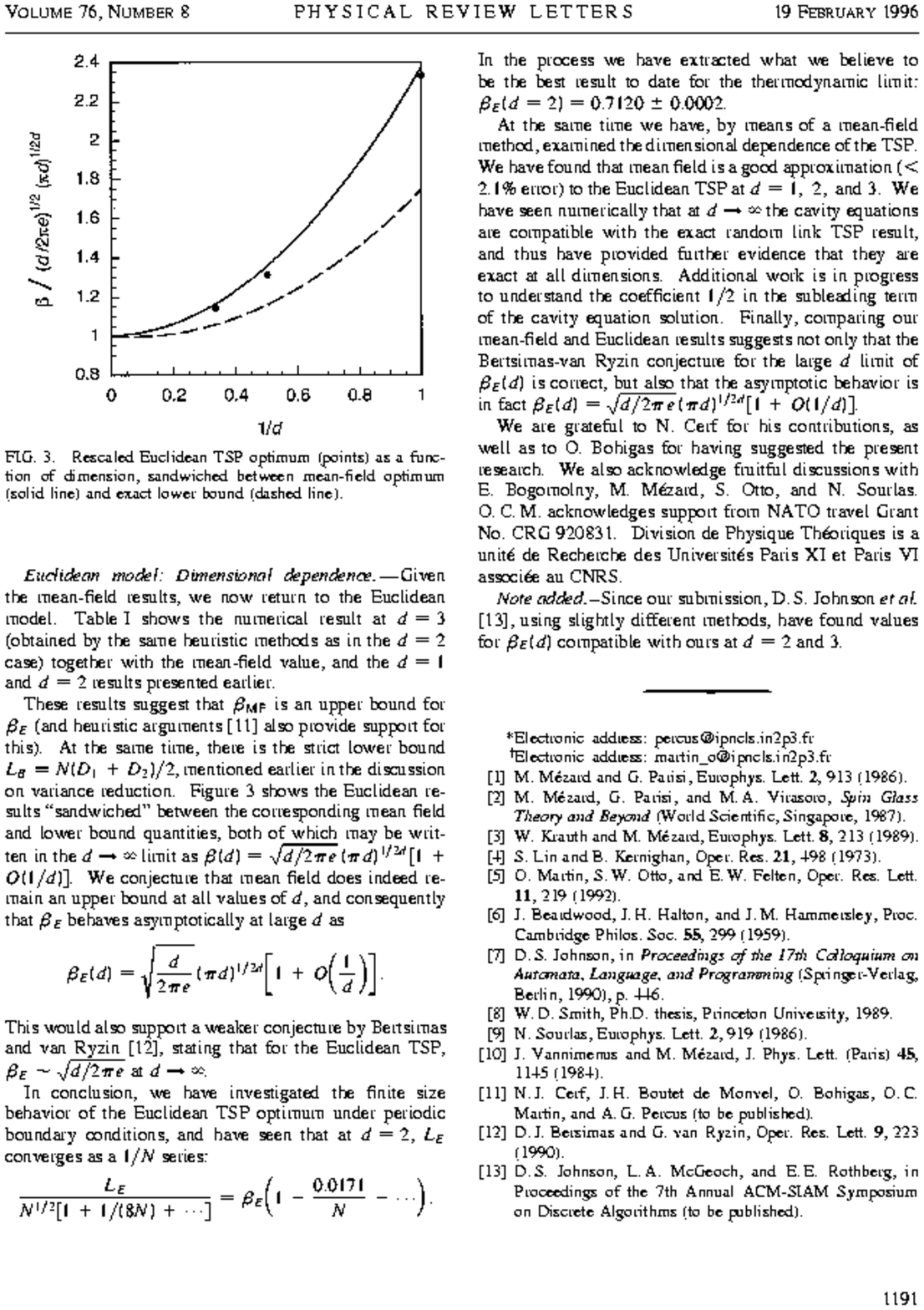}

\section*{{\bfseries \itshape Reprint: Journal de Physique}}
\addcontentsline{toc}{section}{Reprint: Journal de Physique}
\renewcommand{\sectionmark}[1]{\markright{\bfseries\itshape
Reprint: Journal de Physique}}
\sectionmark{}
\clearemptydoublepage

\jppage[-25.2cm]{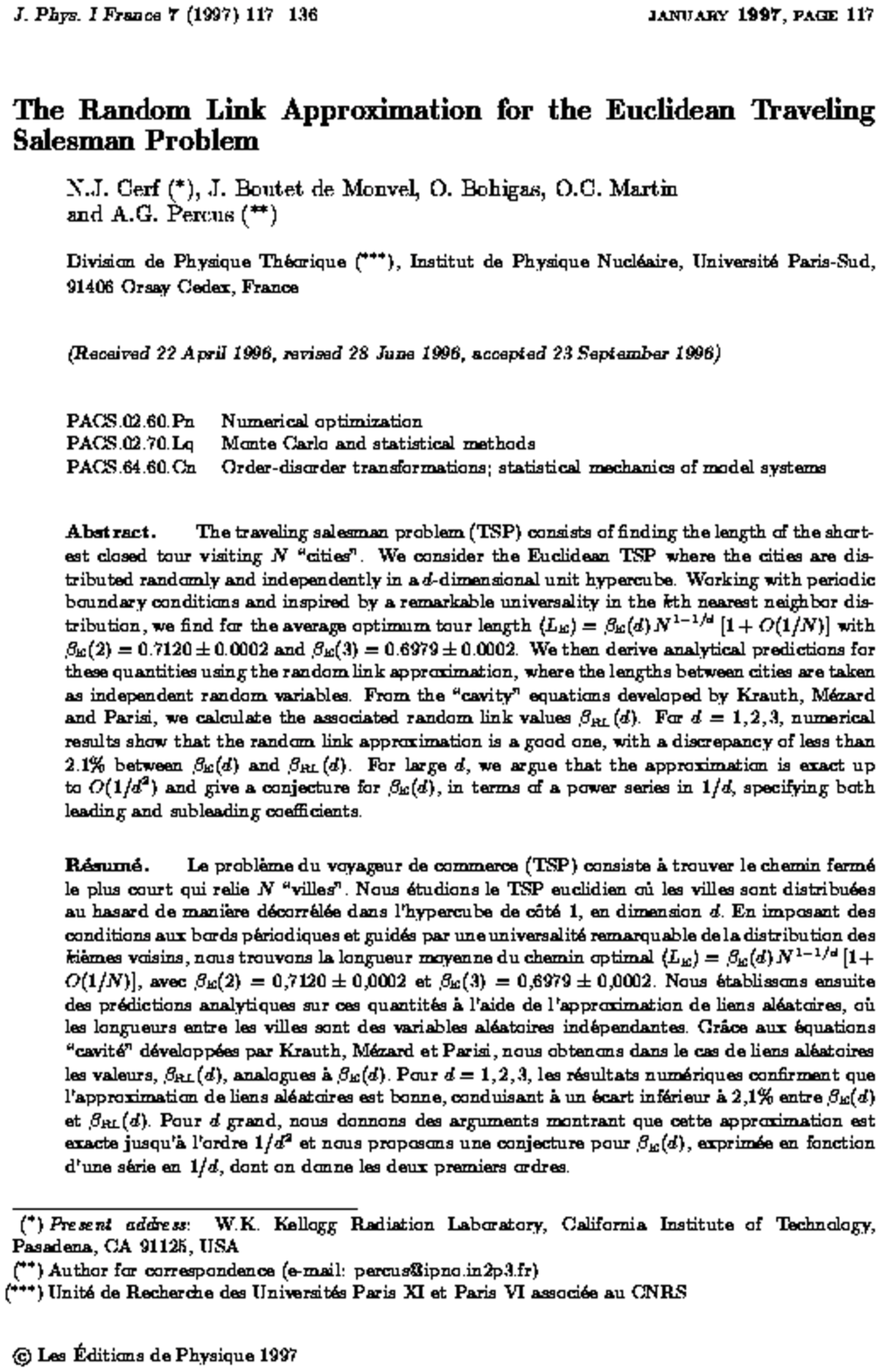}
\jppage{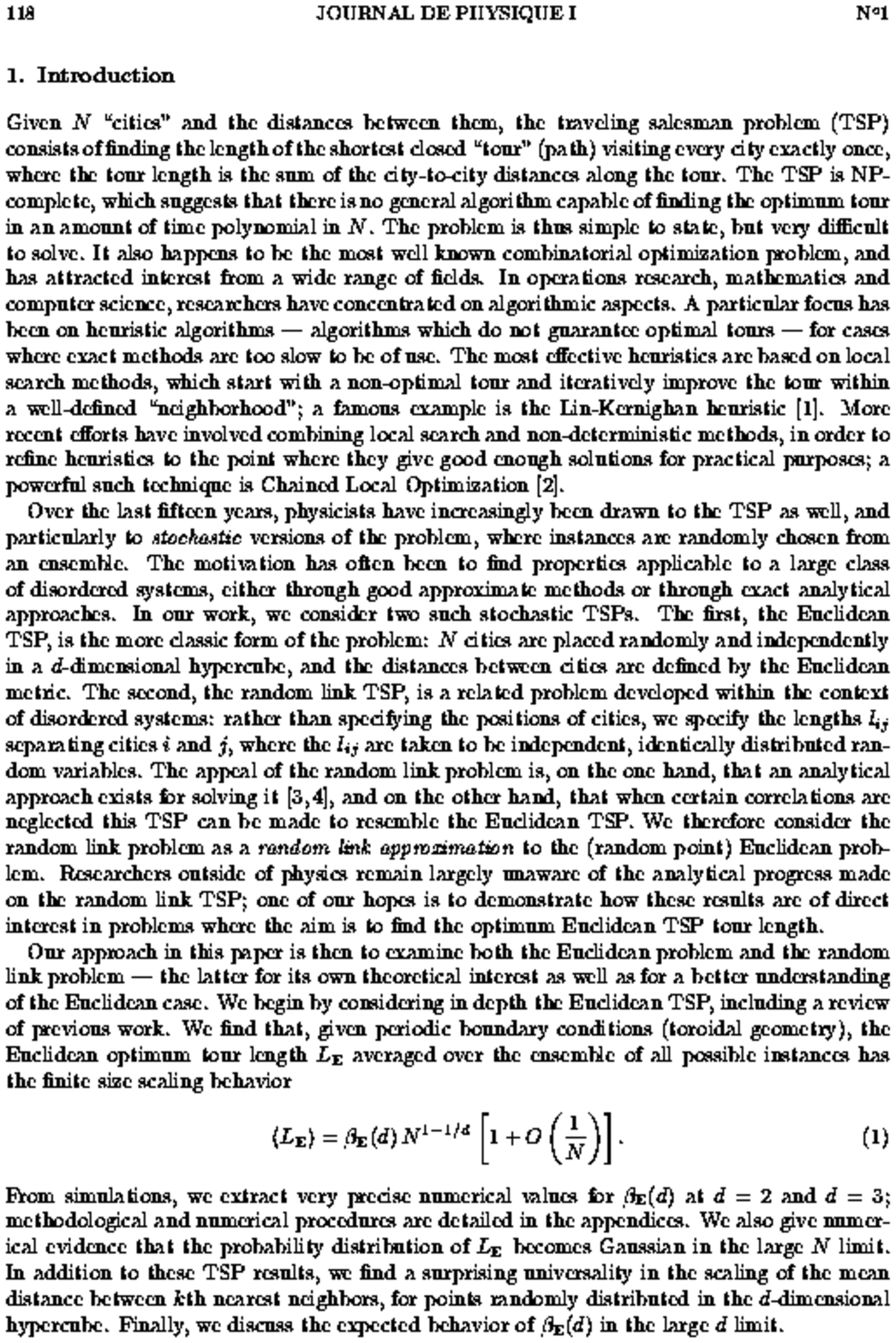}
\jppage{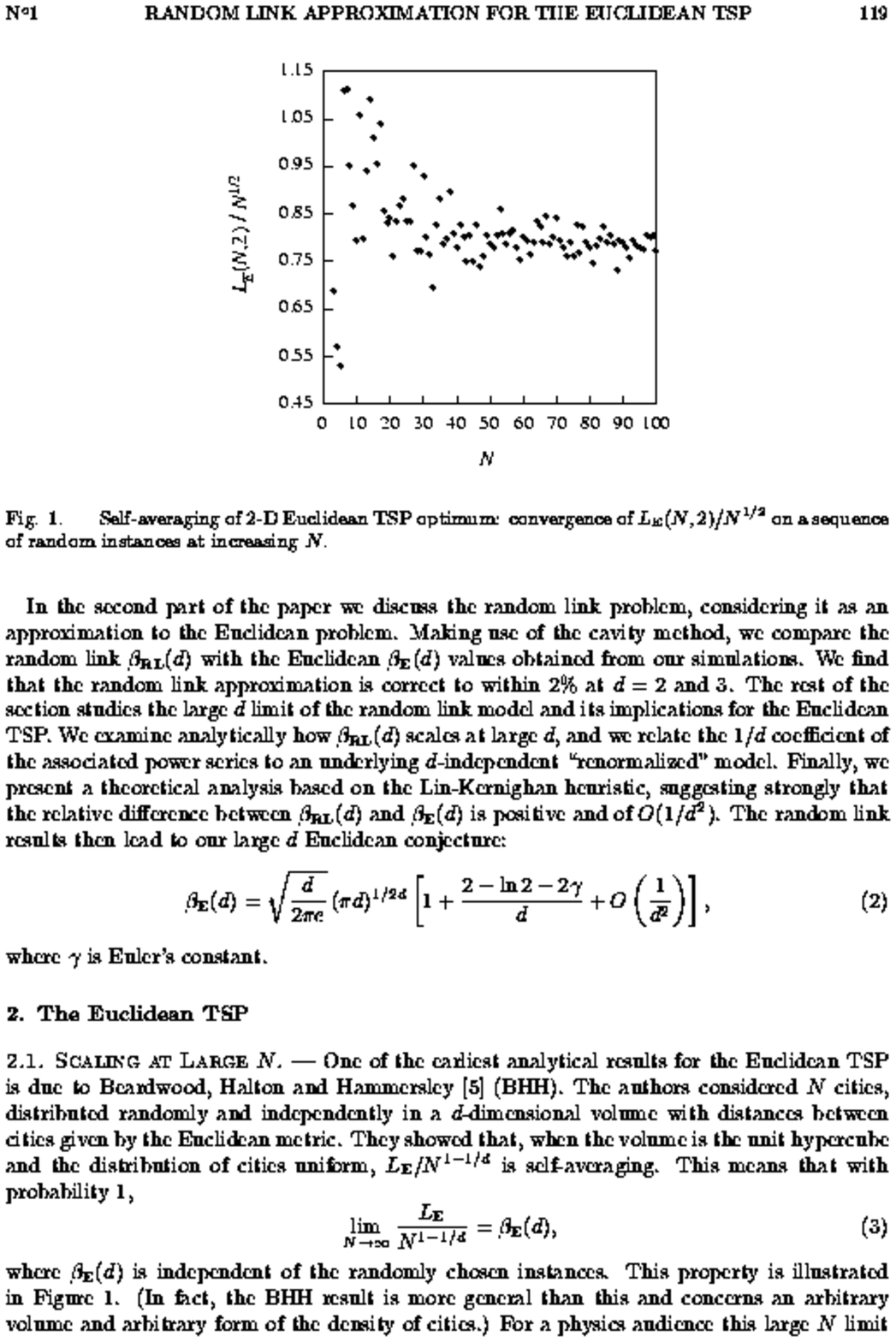}
\jppage{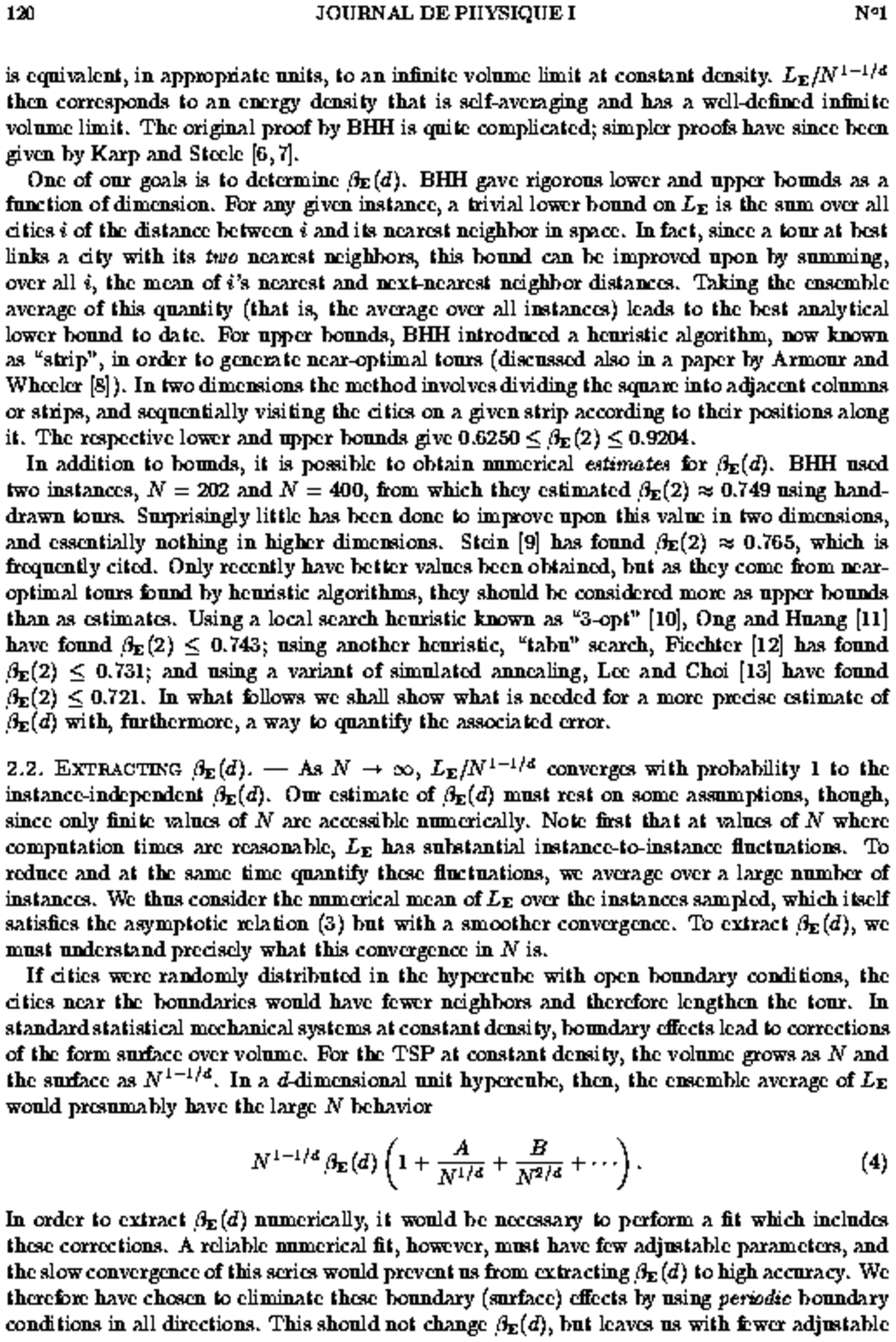}
\jppage{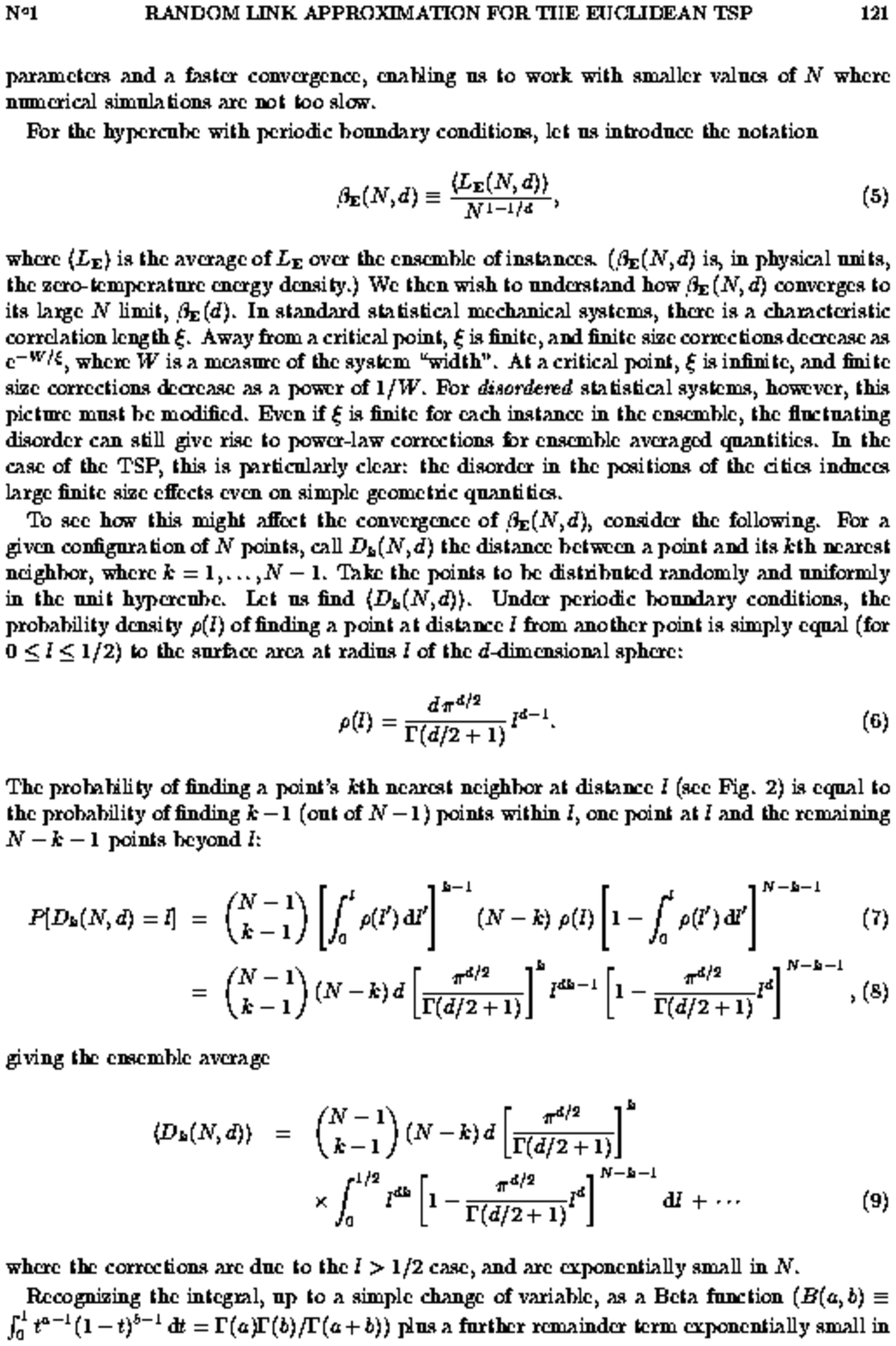}
\jppage{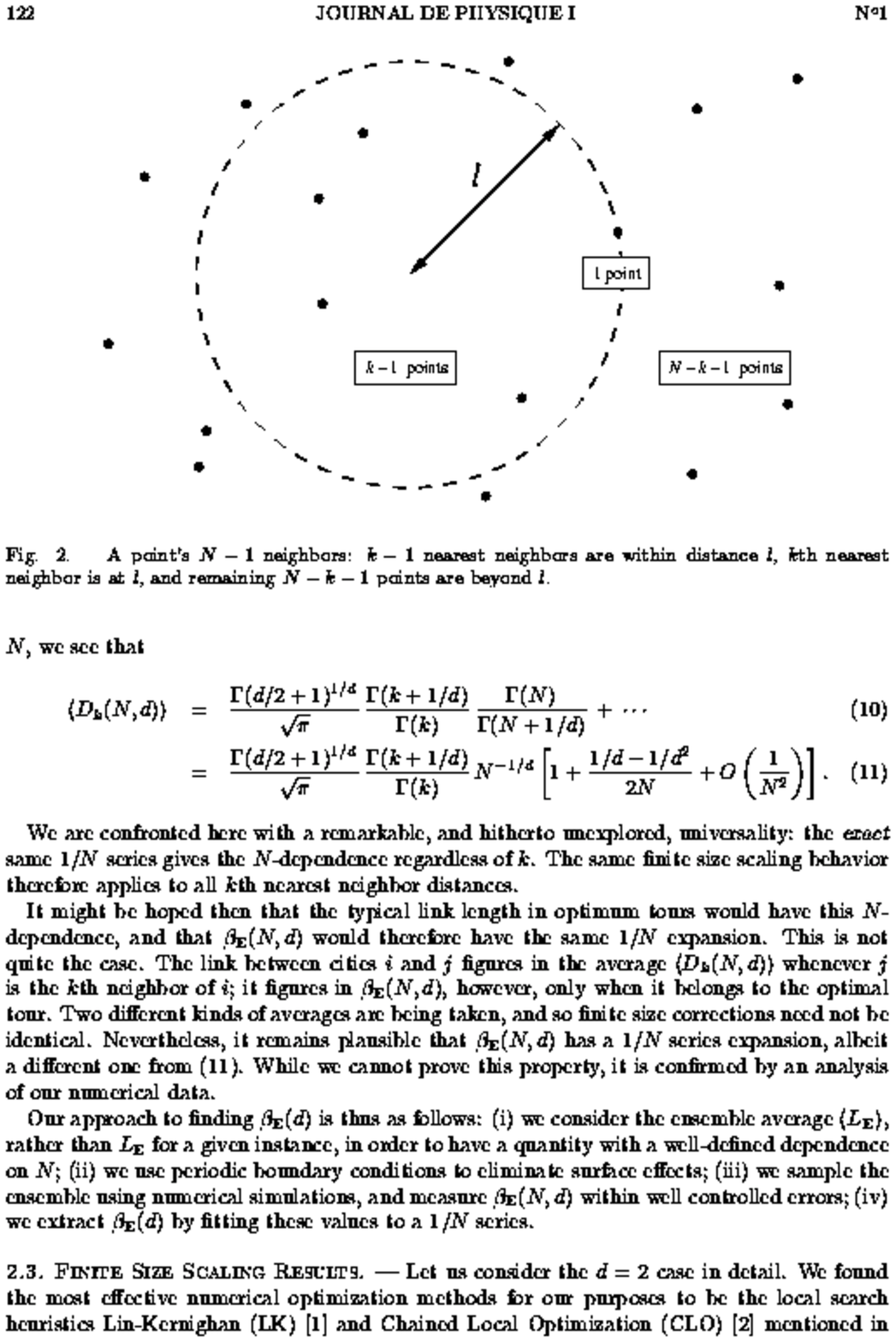}
\jppage{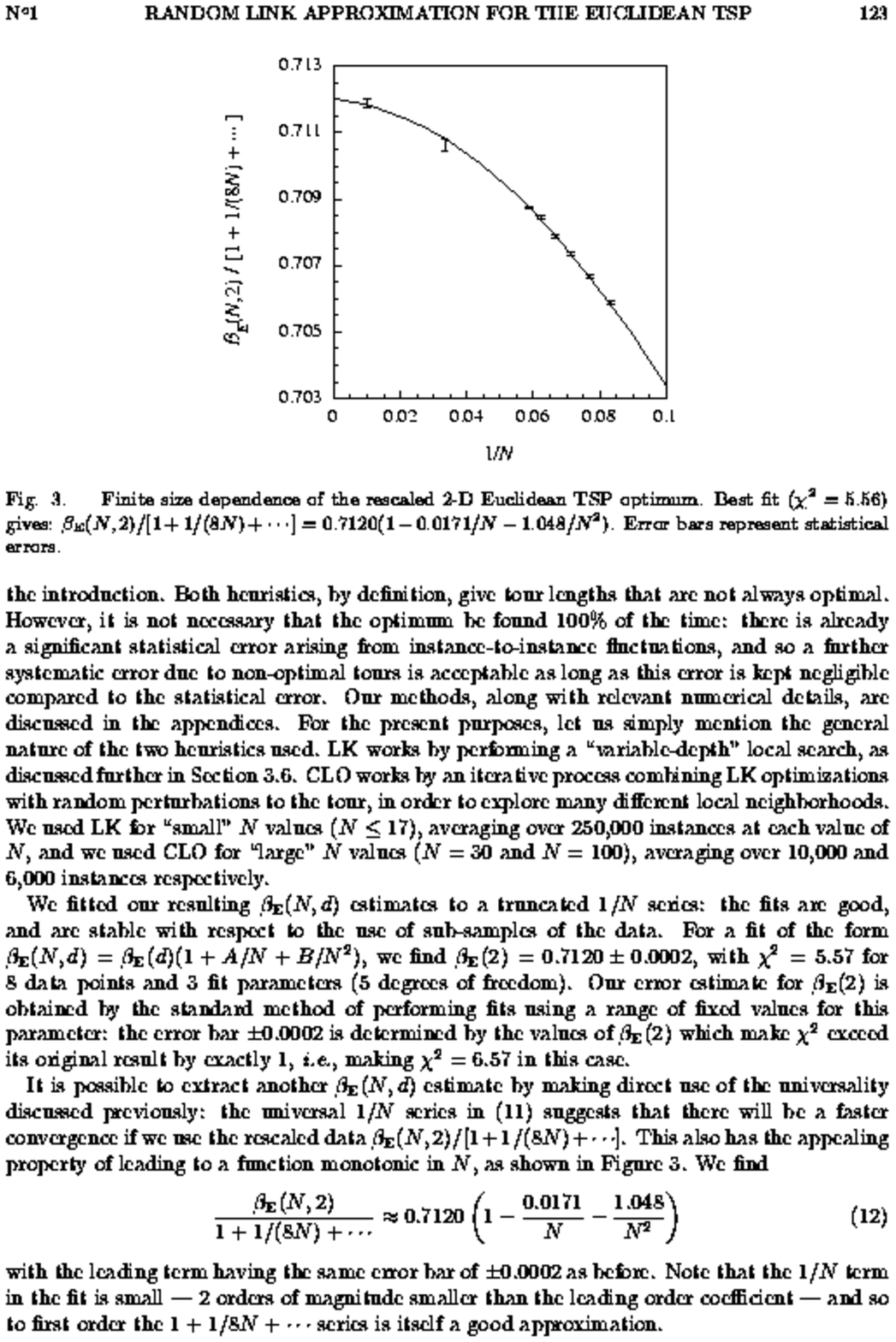}
%
%
\raisebox{-25.6cm}[0cm][0cm]{
   \hspace{-4.9cm}
   \includegraphics{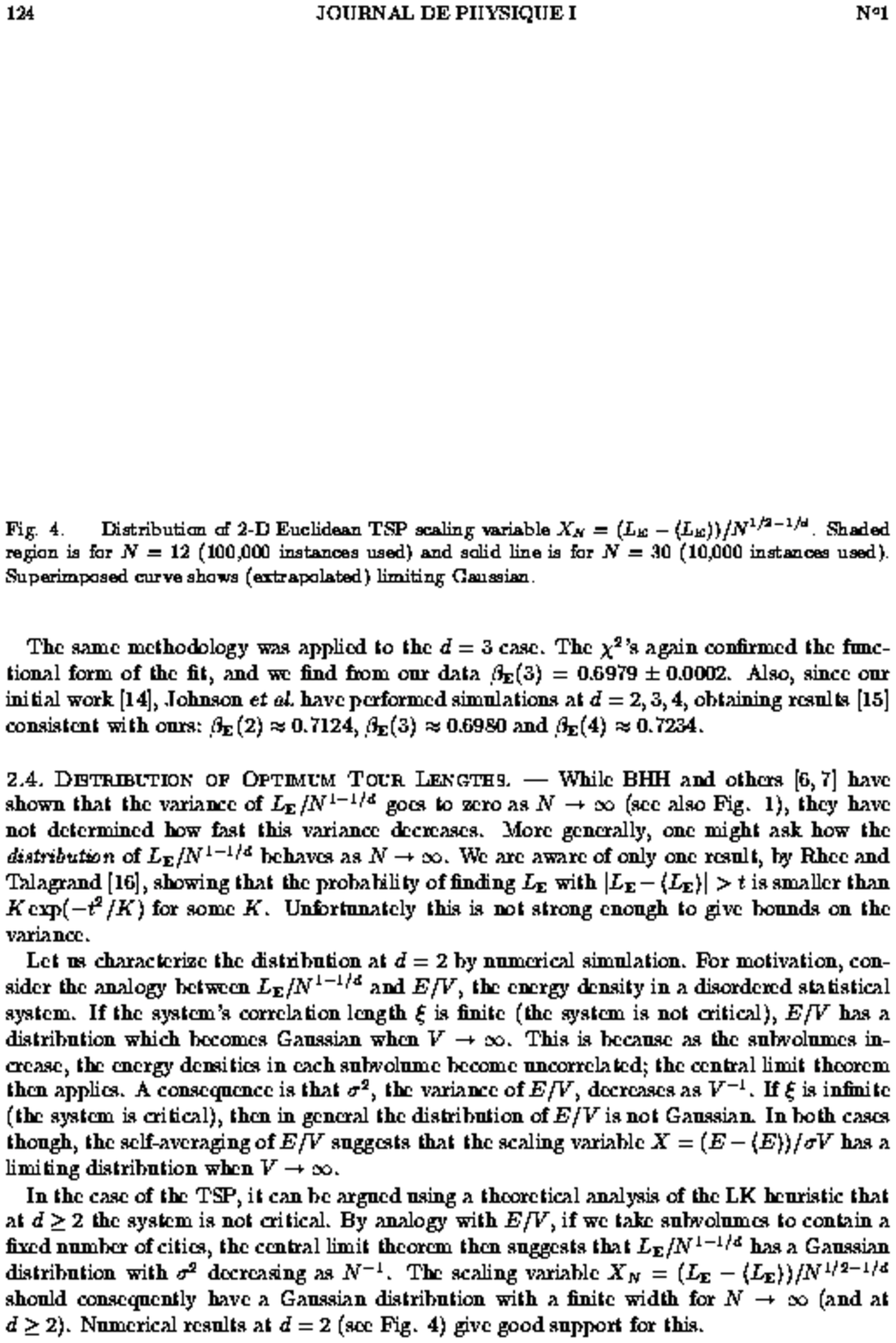}}
\raisebox{-2.7in}{\hspace{1.7in}%
\epsfig{file=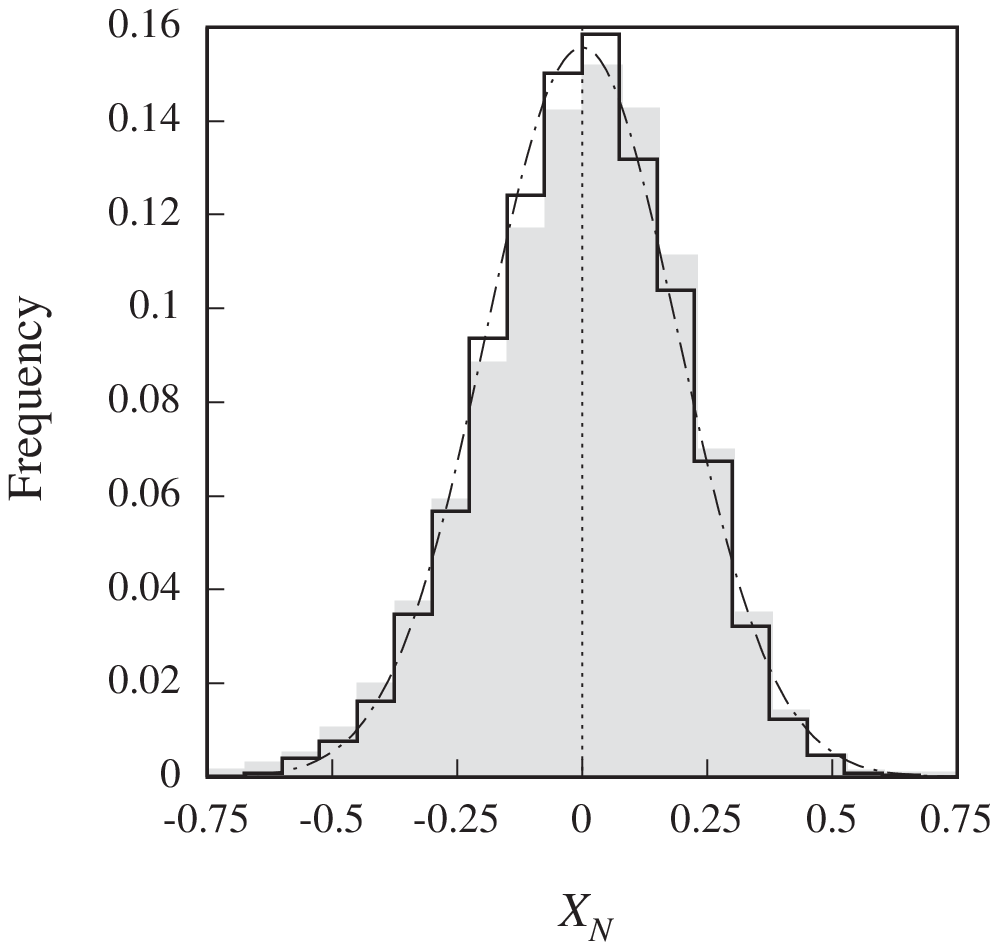,width=7.2cm}}
\newpage

\raisebox{-9.65cm}[0cm][0cm]{
   \hspace{-4.9cm}
   \includegraphics{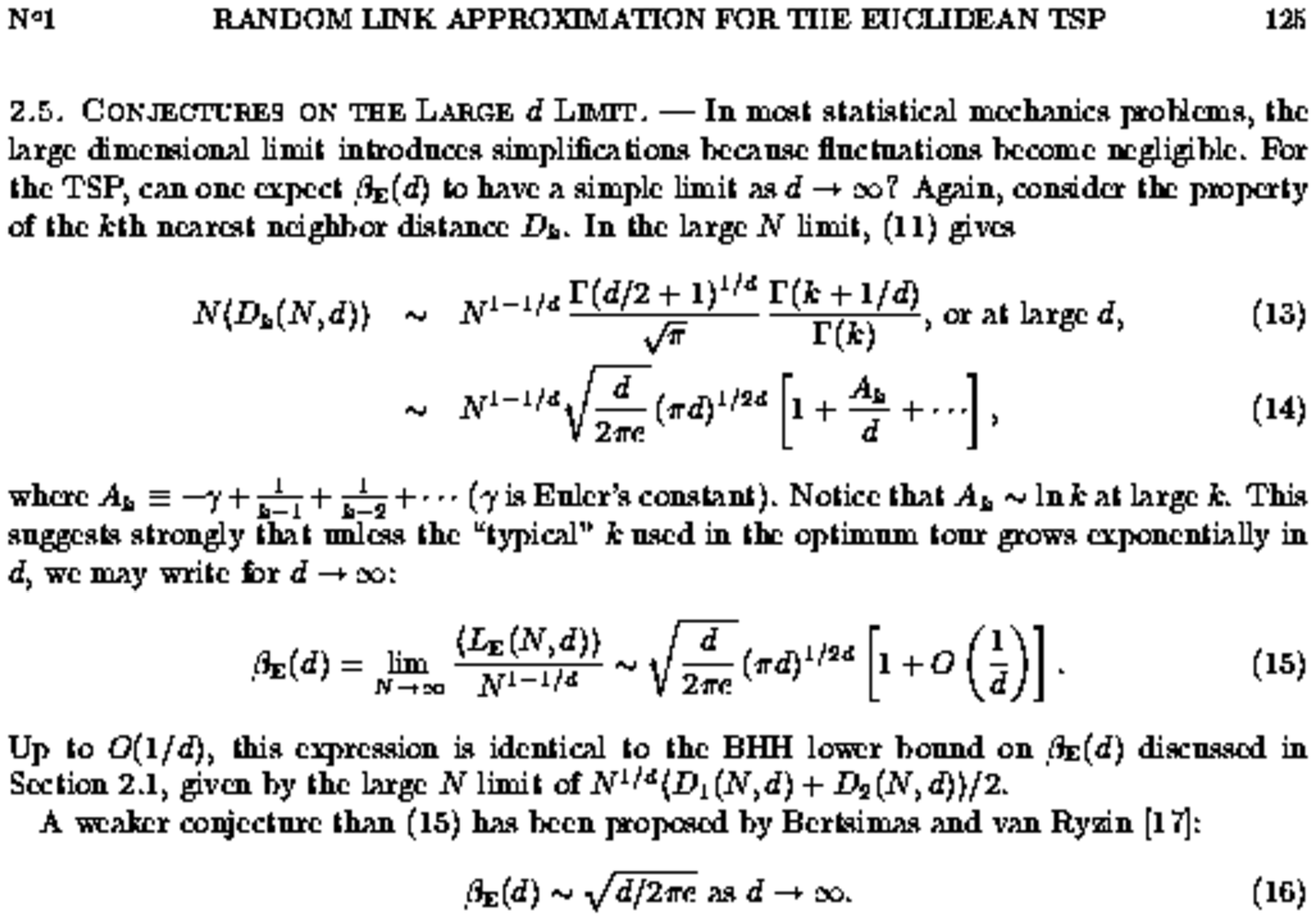}}

\raisebox{-25.05cm}[0cm][0cm]{
   \hspace{-4.9cm}
   \includegraphics{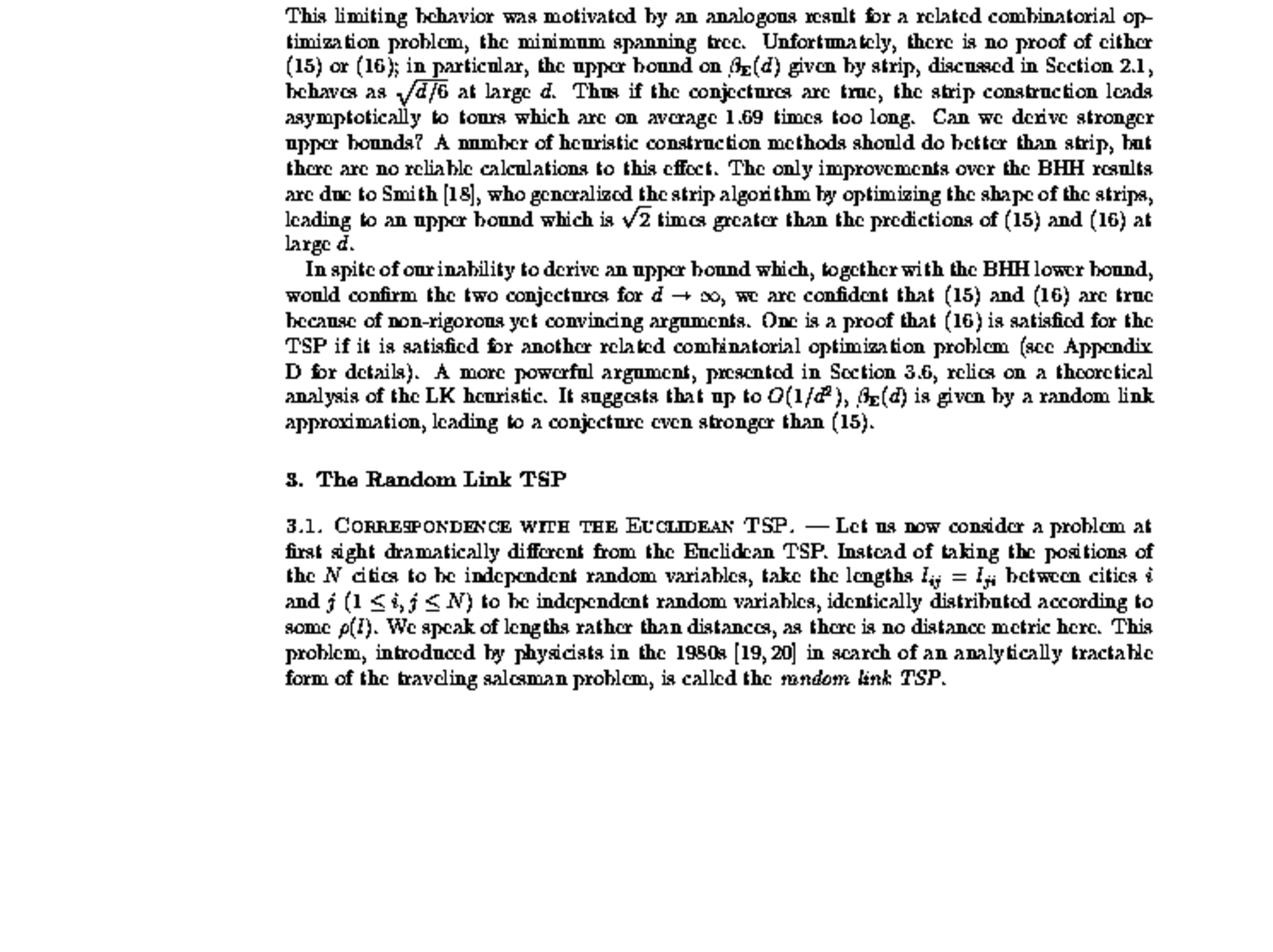}}
\raisebox{-2.7in}{\hspace{1.7in}}
\newpage

\jppage{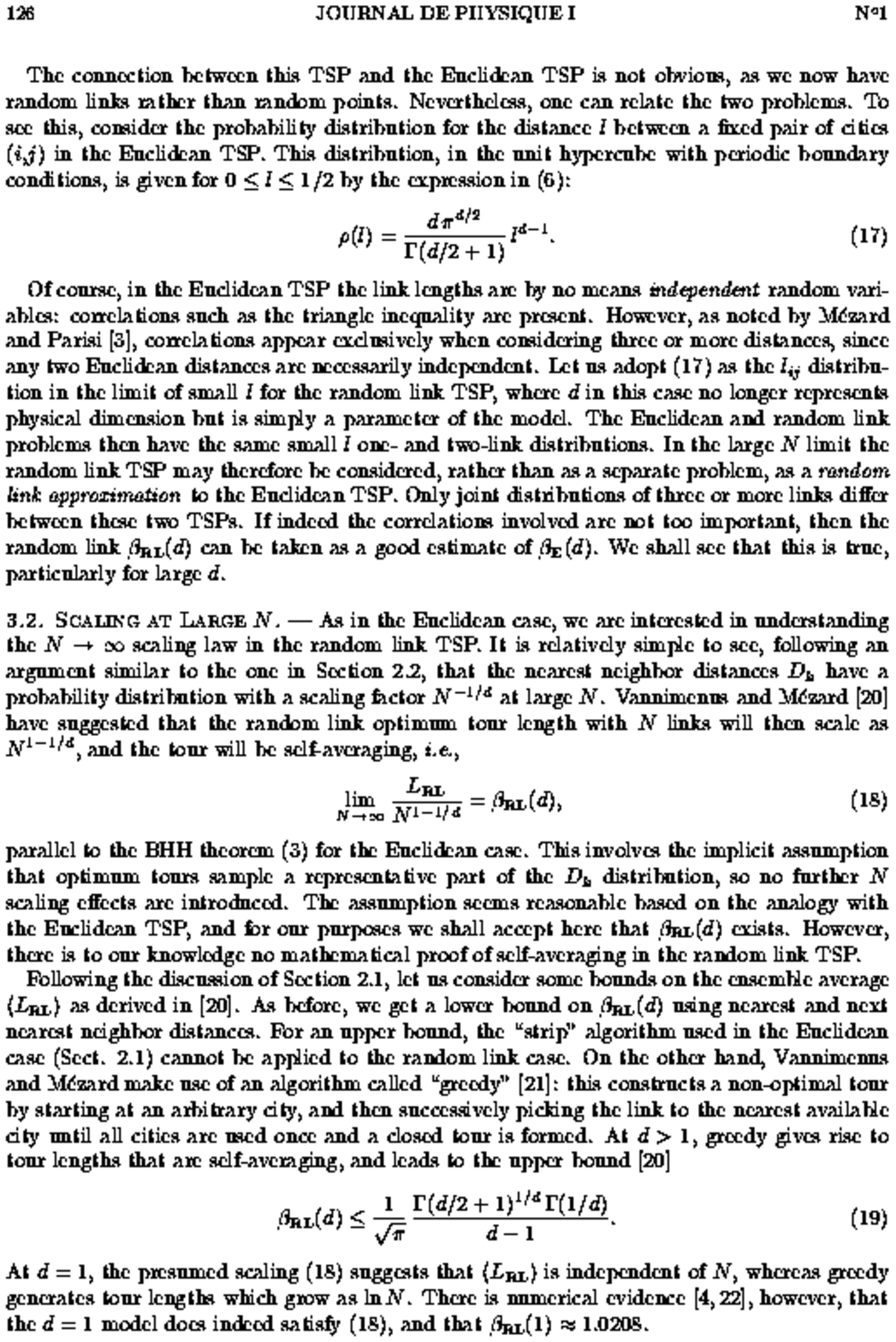}
\jppage{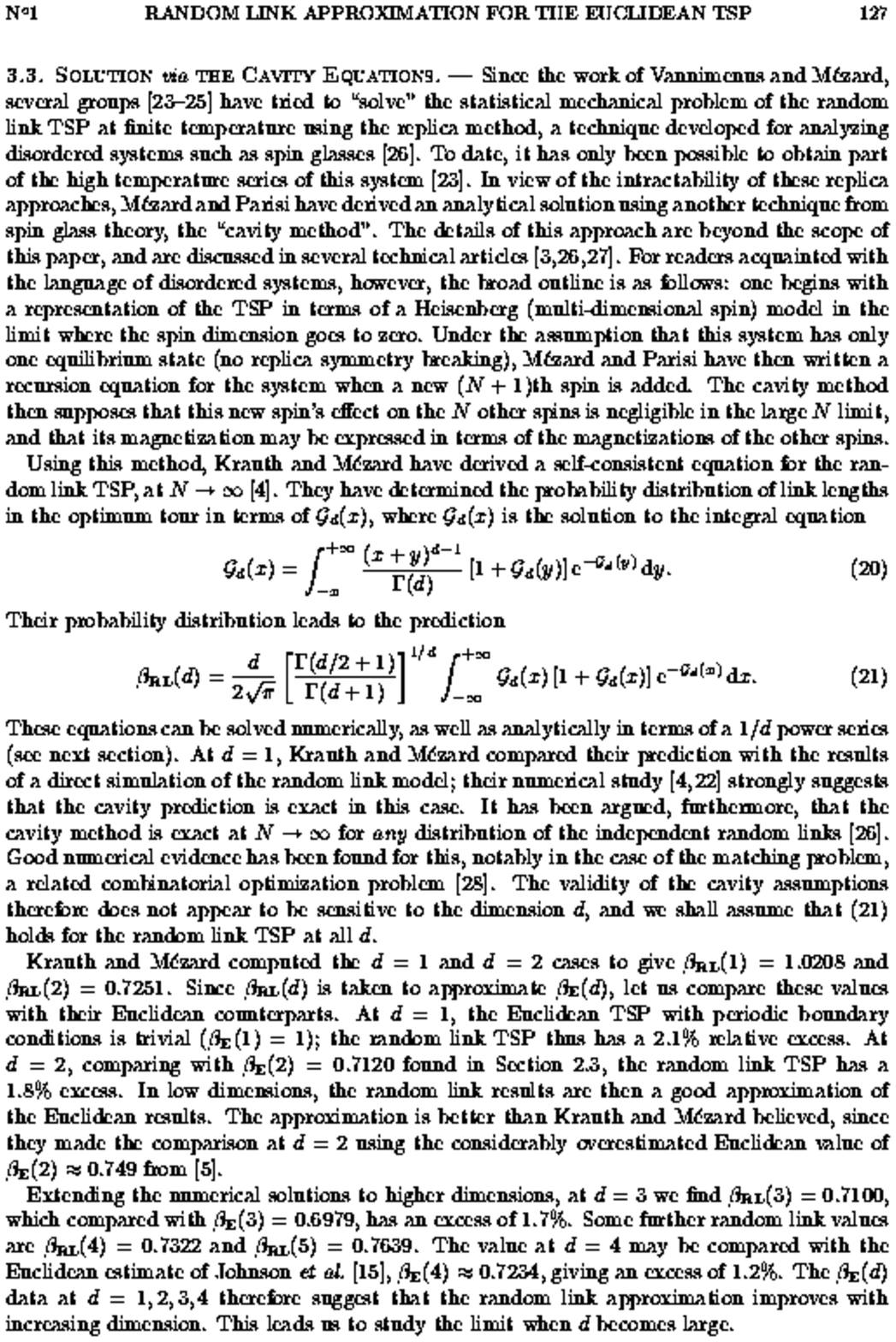}
\jppage{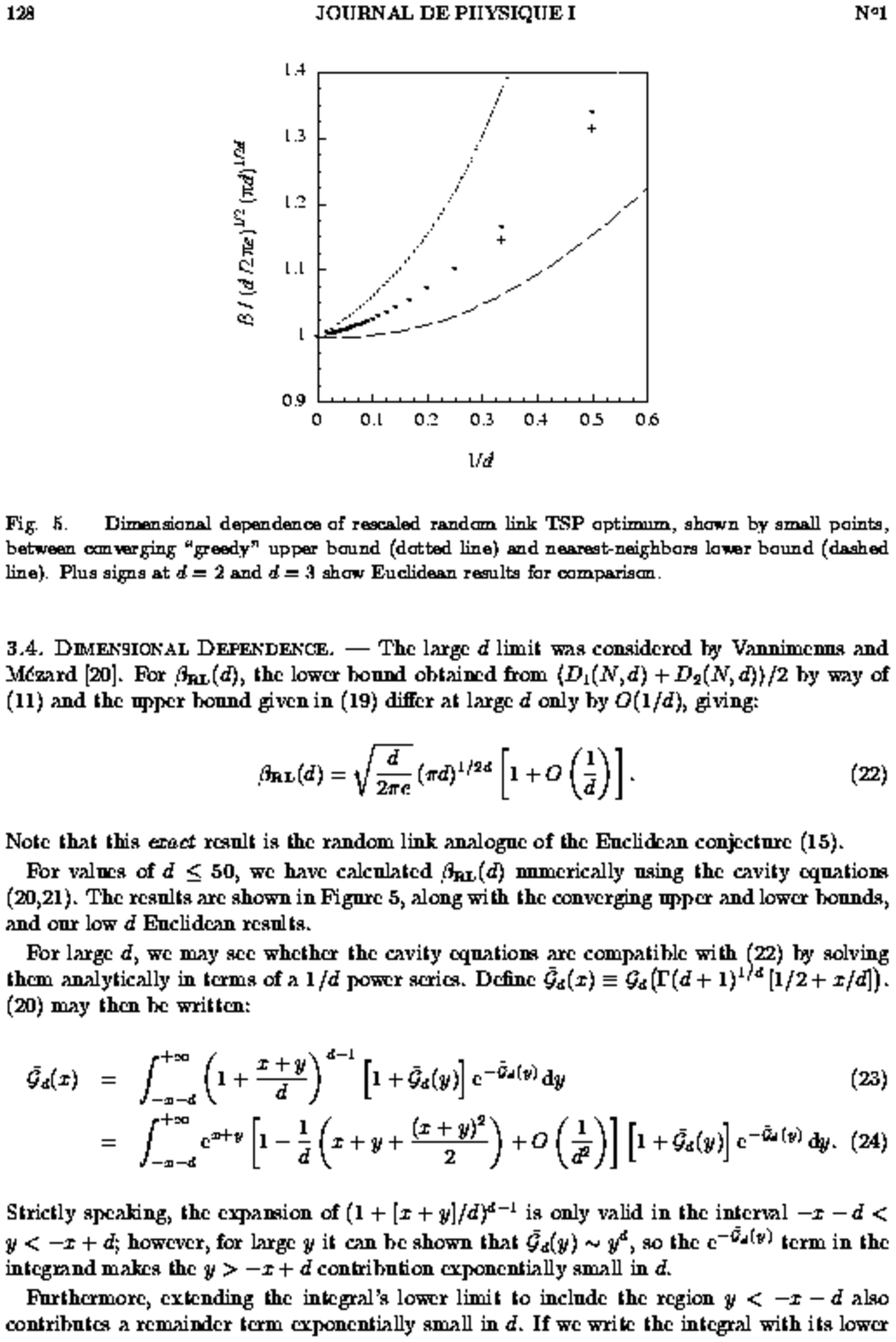}
\jppage{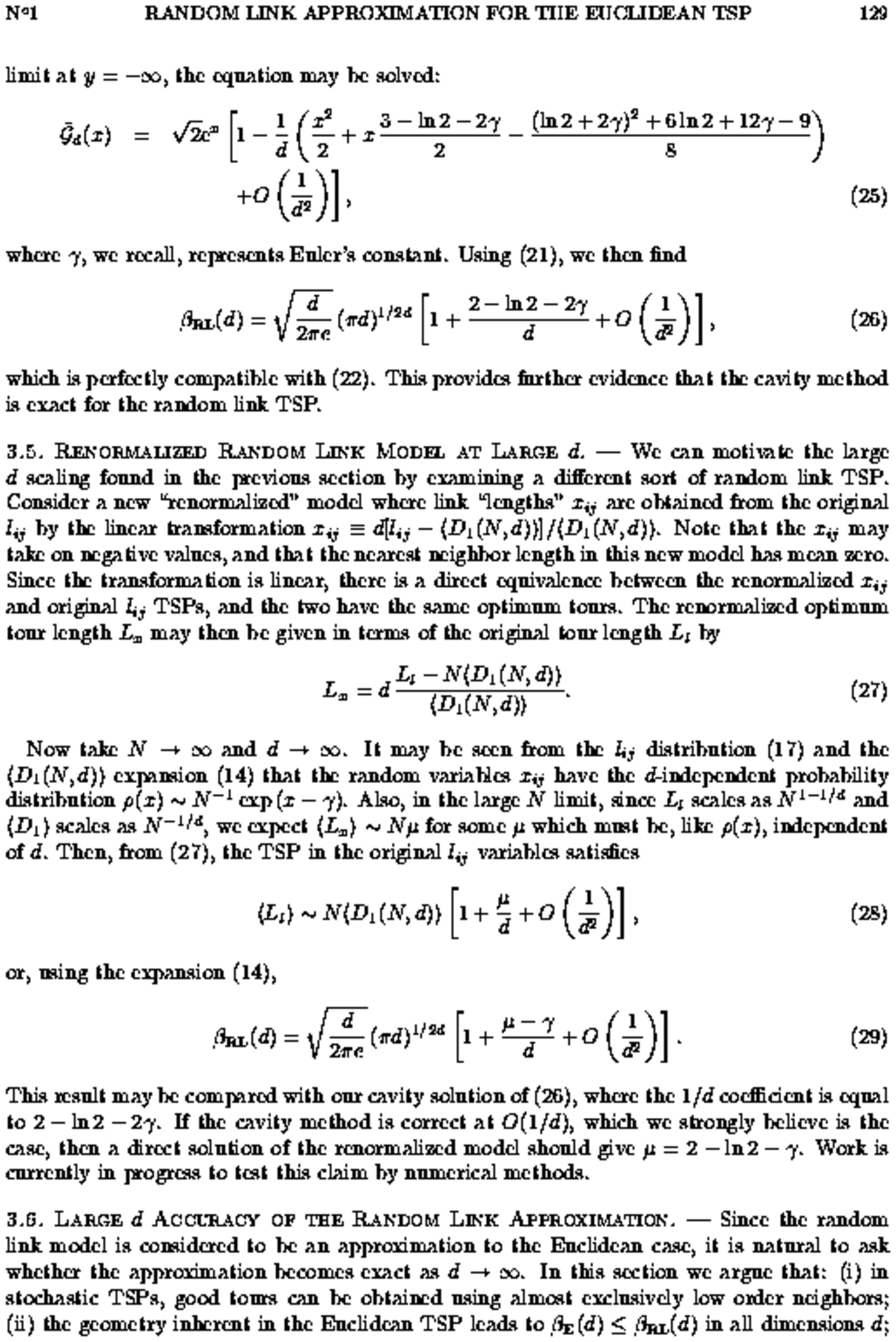}
\jppage{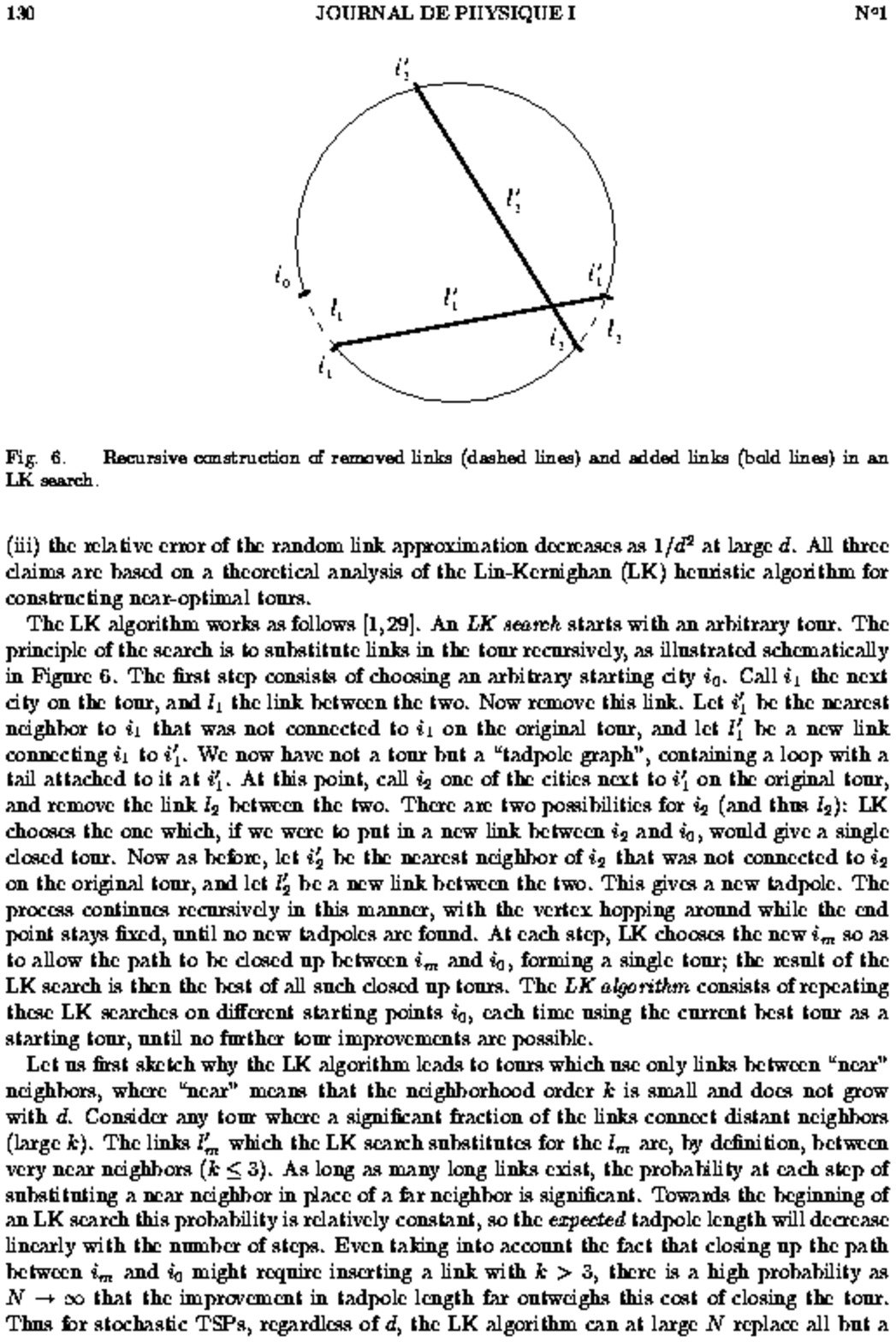}
\jppage{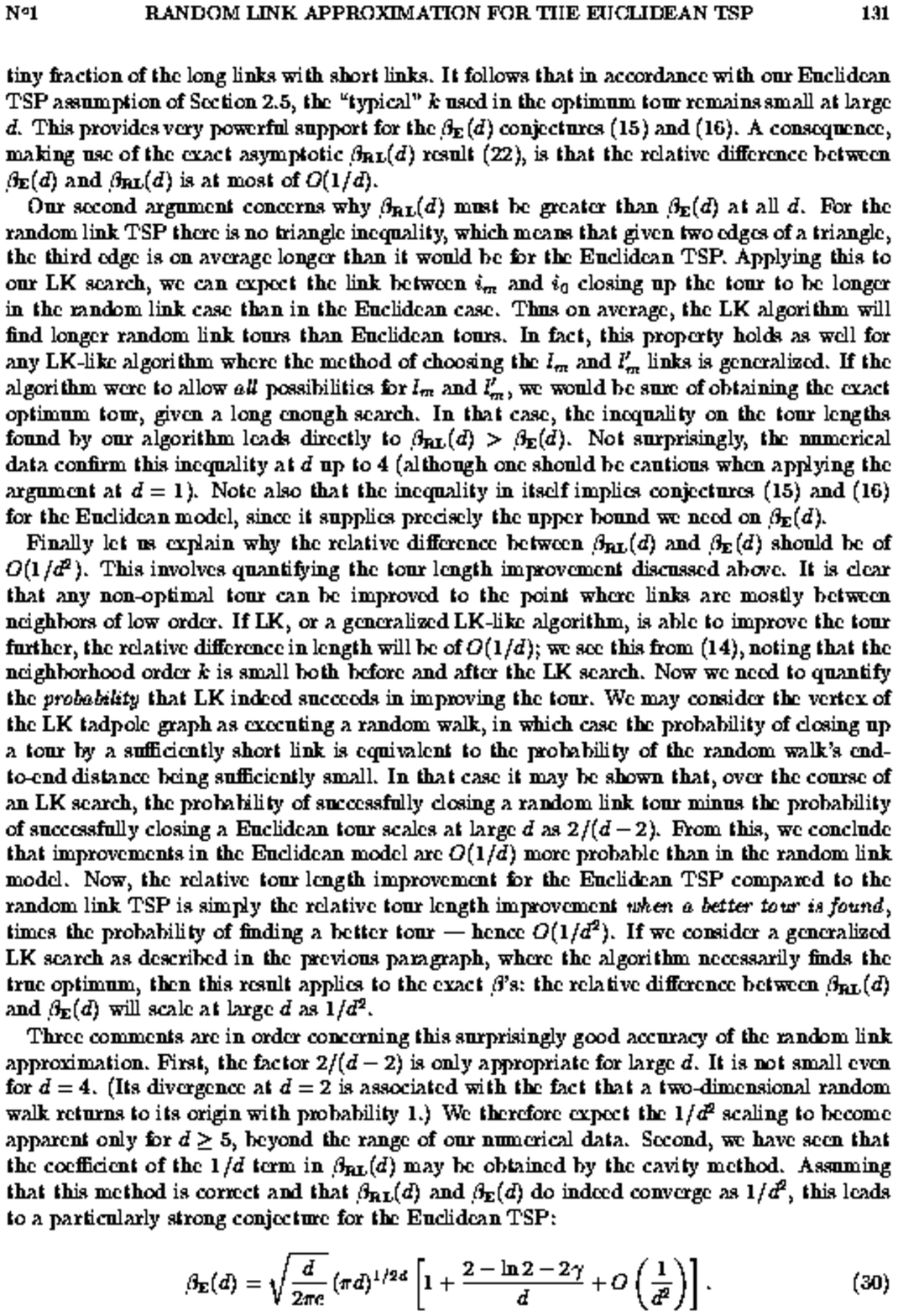}
\jppage{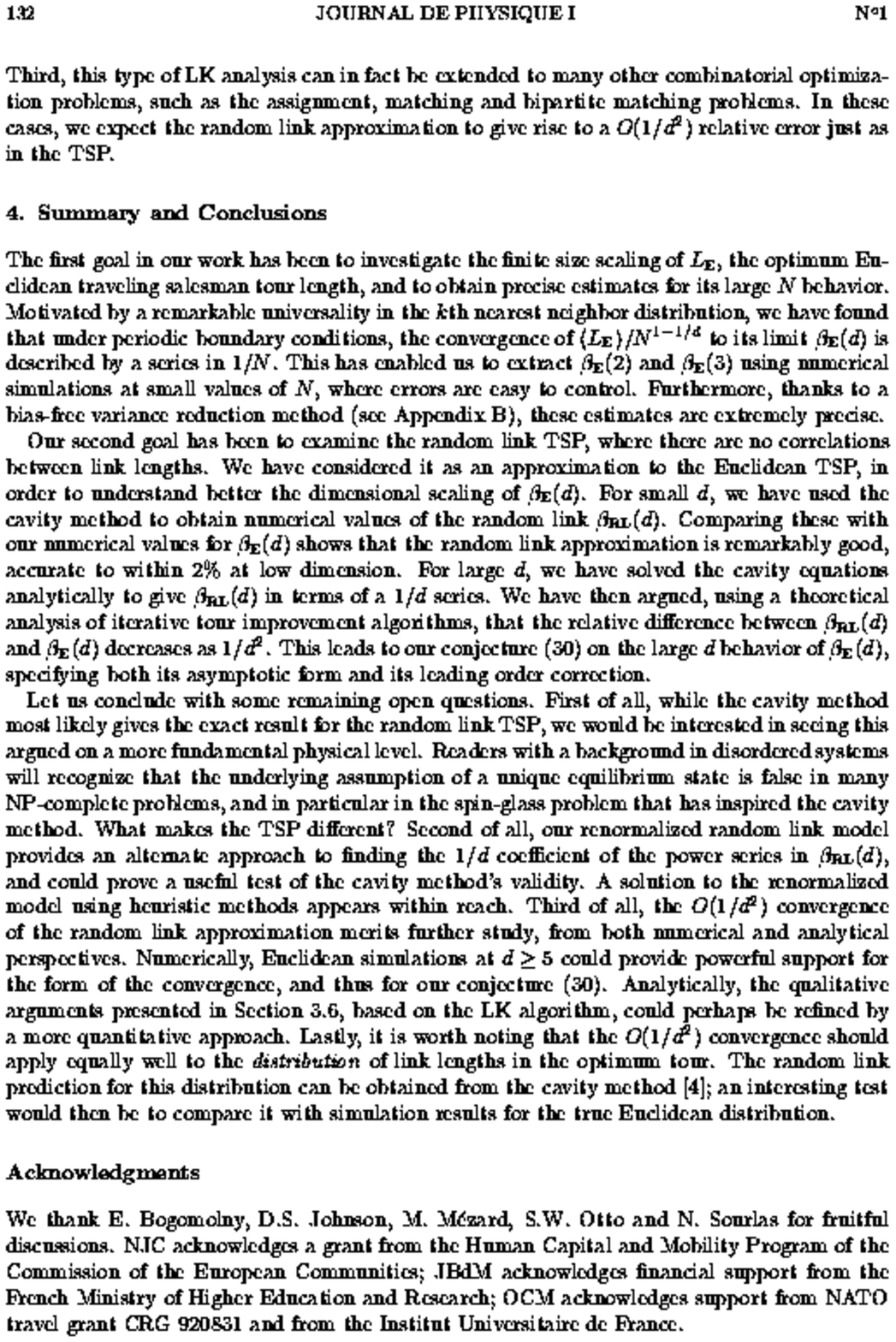}
\jppage{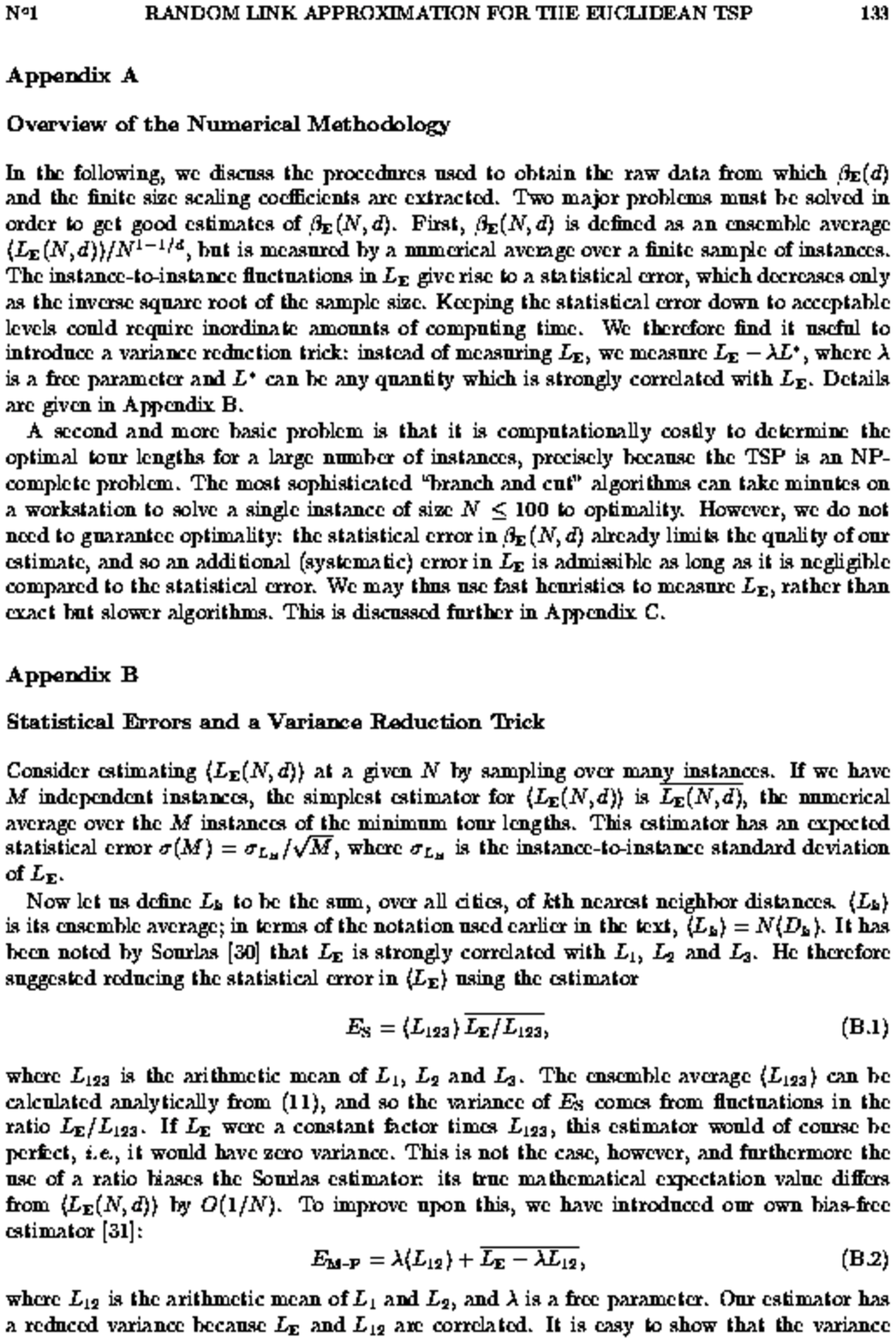}
\jppage{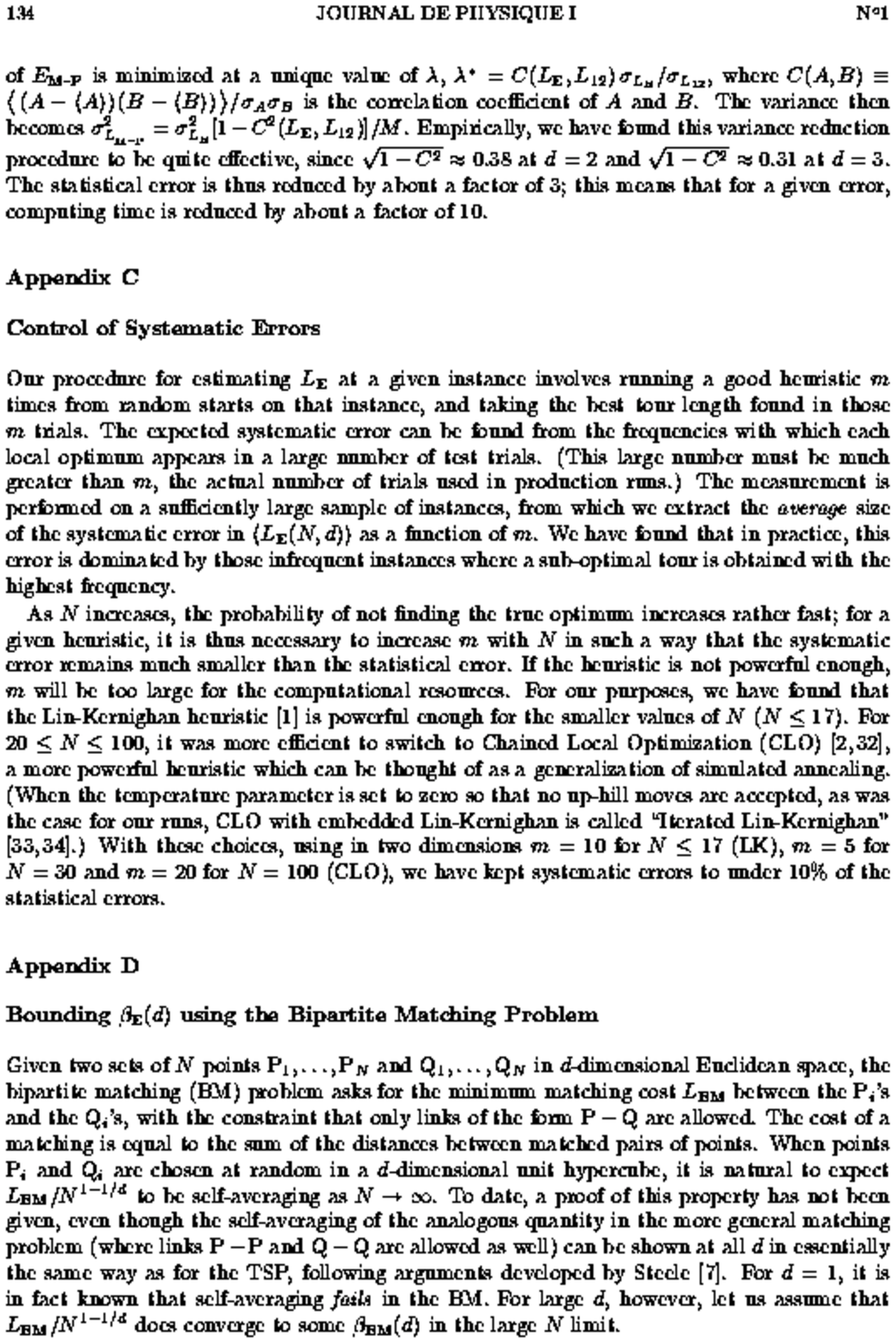}
\jppage{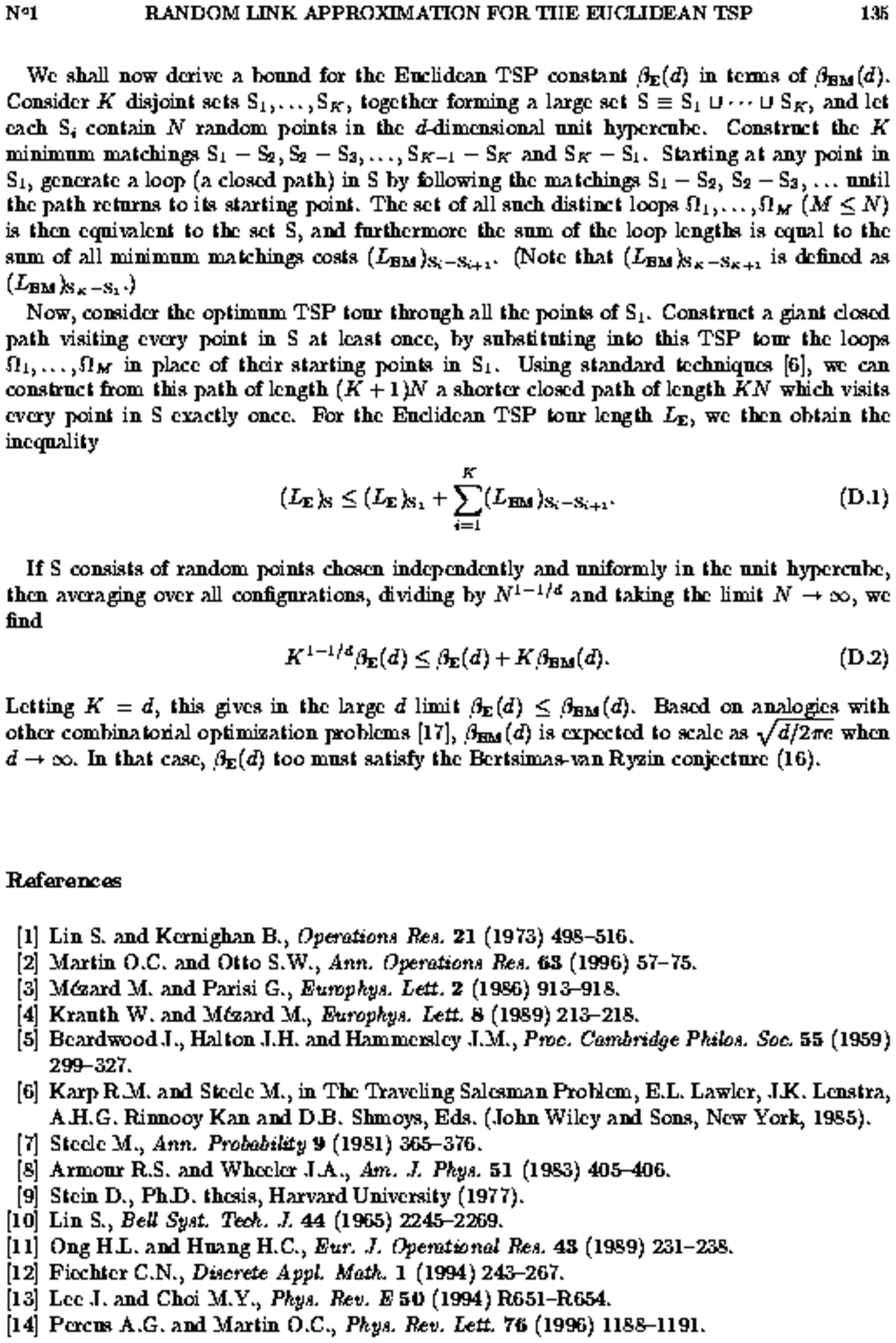}
\jppage{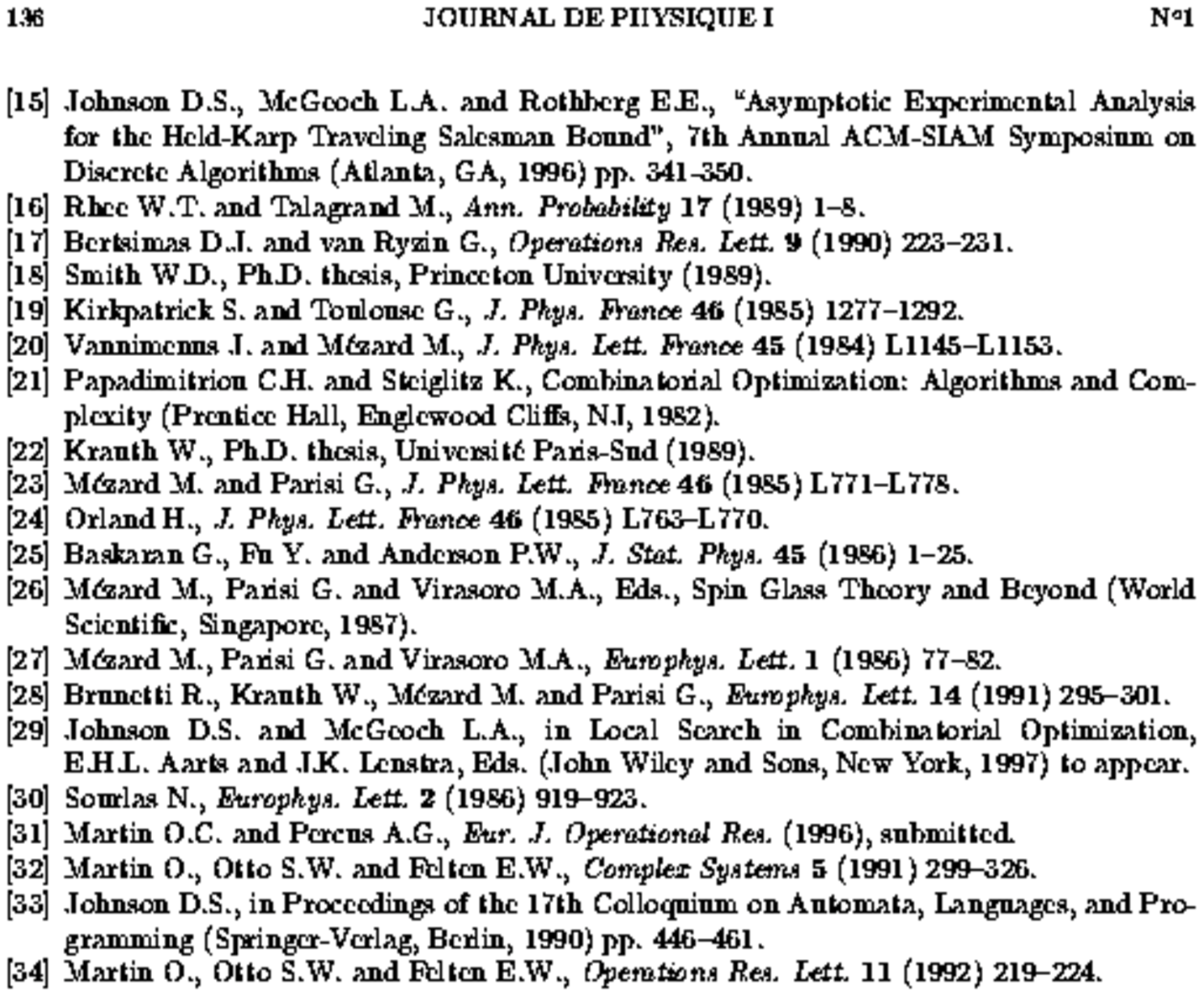}

}
\clearemptydoublepage

\chapter{The random link TSP and its analytical solution}
\thispagestyle{empty}
\minitoc

\vspace{1cm}
\preliminaires{}

Thus far, we have used the Euclidean stochastic \tsp\ as the basis
for discussion.  We have studied the behavior of the optimal tour length
$L_E(N,d)$ at large $N$ using numerical techniques.  We have seen how,
when $N\to\infty$, $L_E(N,d)/N^{1-1/d}$ converges to its instance-independent
limit $\beta_E(d)$.  We have then taken
advantage of the random link approximation to obtain $\beta_{RL}(d)$,
a theoretical prediction for $\beta_E(d)$.  This approximation involves
substituting a different model for the Euclidean model: the {\it random
link \tsp\/}, in which the independent random variables are not the
positions of cities but rather the lengths separating pairs of cities.

The random link \tsp\ was developed by theoreticians in search of an
analytically tractable version of the problem.  A major breakthrough
occurred with the use of the {\it cavity method\/}, by
\citeasnoun{MezardParisi_86b} and later \citeasnoun{KrauthMezard},
to provide an analytical prediction for the (random link) quantity
$\beta_{RL}(d)$.  In this chapter, we discuss the analytical approach
afforded by the cavity method.  The approach relies on certain
assumptions which, while plausible, have thus far been tested numerically
only for a uniform distribution of link lengths --- corresponding to the
$d=1$ case.  We therefore perform numerical simulations to test the
cavity predictions more broadly, and find convincing evidence that these
predictions are correct for all $d$.

The strategy for attacking the random link \tsp\ analytically \cite{Orland}
has been to map the problem onto a model with $m$-component spins;
taking the limit $m\to 0$ leads to a representation of a self-avoiding walk.
This method, originally developed in the context of polymer theory
\cite{DeGennes},
then allows us to recover the partition function for the \tsp\ in terms
of a model that is more easily solvable.  The first success in solving
it was due to \citeasnoun{MezardParisi_86a}, by means of the ``replica''
method developed in spin glasses \citeaffixed{MezardParisiVirasoro}{see}.
The idea behind the replica method is a simple one.  In order to compute
the free energy $F=-T\ln Z$ averaged over the ensemble, we imagine $n$
independent copies (replicas) of the system, and then write
\begin{equation}
\langle\ln Z\rangle = \lim_{n\to 0}\frac{\langle Z^n\rangle -1}{n}\mbox{.}
\end{equation}
Making the assumption that all $n$ replicas are interchangeable, {\it
i.e.\/}, they all reach the same equilibrium state (this ergodicity
hypothesis is known as {\it replica symmetry\/}), M\'ezard and Parisi
obtained a high-temperature expansion for $\beta_{RL}(d=1)$.

Unfortunately, the replica method allowed estimating the solution at
$T=0$ (the global optimum) only by a numerical extrapolation from
higher temperatures.  M\'ezard and Parisi therefore instead tried
using the cavity approach, a mean-field method (also assuming replica
symmetry) whereby one adds an additional spin to the system
and calculates the magnetization at this site in the
absence of correlations between the other spins.  This mean-field
assumption is generally held to be correct in the limit of a large number
of spins.  \citeasnoun{KrauthMezard} then completed the resulting
calculation, finding an integral equation for $\beta_{RL}(d)$.
\citeasnoun{CBBMP} solved the integral equation numerically for small values
of $d$, as well as in a large $d$ expansion:
\begin{equation}
\label{eq_pre_cavity}
\beta_{RL}(d)=\sqrt{\frac{d}{2\pi e}}\,(\pi d)^{1/2d}\,
\left[1+\frac{2-\ln 2-2\gamma}{d}+O\left(\frac{1}{d^2}\right)\right]\mbox{,}
\end{equation}
where $\gamma$ is Euler's constant.

The problematic assumption used here, however, is that of replica
symmetry.  In the spin glass model of \citeasnoun{SherringtonKirkpatrick},
where similar methods were first tried, it was found that the replica
symmetry assumption led to a ground-state energy that was off by about
5\% \cite[pp.~13--14]{MezardParisiVirasoro}.  It was therefore
necessary to resort to a {\it replica symmetry breaking\/} (\rsb)
scheme in order to obtain more realistic results.  This involved
defining an ``overlap'' $q_{ab}$ between two states $a$ and $b$,
so as to provide a measure of ``distance'' between them in state space.
For the \tsp, for example, \citeasnoun{KirkpatrickToulouse} have
defined $q_{ab}$ as the proportion of links in common between two
tours $a$ and $b$.  \rsb\ implies, among other things,
that $q$ is not a self-averaging quantity, {\it i.e.\/}, in the large
$N$ limit the distribution $P(q)$ does not simply approach a delta
function.

In the case of the random link \tsp, \citeasnoun{KrauthMezard} performed
a number of $d=1$ numerical checks indicating that, unlike for the
spin glass, the replica symmetric solution given by the cavity method
is correct.  They solved the cavity equation numerically to obtain
$\beta_{RL}(1)=1.0208$, finding it in close agreement with the earlier
direct simulations of Kirkpatrick and Toulouse --- whose results
suggested $\beta_{RL}(1)\approx 1.045$ --- as well as with M\'ezard
and Parisi's extrapolated value of $\beta_{RL}(1)=1.04\pm 0.015$ from
the replica method.  Recent simulations by \citeasnoun{Johnson_HK} give
$\beta_{RL}(1)=1.0209\pm 0.0002$ (see Appendix~\ref{app_dsj}),
providing
excellent confirmation of the theoretical predictions.  Krauth
and M\'ezard also performed simulations of their own,
and compared the cavity and simulated values for the probability
distribution ${\cal P}_d(l)$ of link lengths $l$ in the optimal tour,
finding close agreement.

More direct evidence of the lack of \rsb\ comes from the analysis by
\citeasnoun{Sourlas}, who considered the low-temperature statistical
mechanics of the $d=1$ random link \tsp.  Performing numerical simulations
at temperatures as low as $T=0.85$, he found that the distribution
for the overlap function, as defined above, does indeed approach a
delta function at large $N$.  This sort of behavior is inconsistent
with \rsb, and suggests that the replica symmetric solution used in the
cavity method is indeed exact.  As Sourlas noted, however, obtaining
good enough statistics for simulations is extremely difficult for
lower $T$, where fluctuations in $q$ become much larger.

We wish to extend these numerical confirmations of the cavity method
to $d>1$, in order to determine its validity for choices of $d$ relevant
to the Euclidean \tsp.  Since we are interested in $T=0$ properties,
direct analysis of the overlap function does not appear within reach.
However, the sort of tests proposed by Krauth and M\'ezard is easily
applicable to cases of interest such as $d=2$, and another method can
be used to test the $O(1/d)$ coefficient in the large $d$ expansion
(\ref{eq_pre_cavity}) for $\beta_{RL}(d)$.  (The leading order term
has already been shown exact by \citeasnoun{VannimenusMezard}.)
This latter method involves defining a ``renormalized'' random link
\tsp\ without the parameter $d$, having tour lengths whose leading
order behavior at large $N$ goes as $L(N)\sim\mu N$.  The quantity
$\mu$ may be expressed in terms of the $O(1/d)$ coefficient in
(\ref{eq_pre_cavity}); by performing numerical simulations on this
renormalized model, we manage to verify the subleading behavior
of the cavity method's $\beta_{RL}(d)$ prediction.

Our simulation methods are summarized in Appendix~\ref{app_method}
and follow those of Chapter~II: we use heuristic algorithms
at values of $N$ up to $N=100$, where both statistical and systematic
errors can be well controlled.  We extrapolate to the large $N$ limit
by fitting to the expected finite size scaling law.  The resulting
fits are excellent, and confirm the cavity predictions both of
$\beta_{RL}(2)$ and of the subleading term in the large $d$ $\beta_{RL}(d)$
expansion to within well under 1\%.  (This may be compared with the
5\% error of the replica symmetric solution in the Sherrington-Kirkpatrick
spin glass model.)  Taken together, then, this provides good arguments
for the claim
that the cavity method is exact for all $d$, and further reason for
conjecturing that there is no \rsb\ down to $T=0$.

\newpage
\clearemptydoublepage

\renewcommand{\baselinestretch}{1.2}\tiny\normalsize
\renewcommand{\thefootnote}{\fnsymbol{footnote}}
\newcommand{\fatops}[2]{\genfrac{}{}{0pt}{1}{#1}{#2}}
\setcounter{footnote}{0}
\thispagestyle{empty}
\vspace{1cm}
			\begin{center}

{\Huge  The stochastic traveling salesman problem:}

\bigskip

{\Huge  Finite size scaling and the cavity prediction}

\vspace{1cm}

{\Large\sc Allon G.~Percus and Olivier C.~Martin}\\
Division de Physique
Th{\'e}orique\footnote{\ Unit{\'e} de recherche des Universit{\'e}s 
			de Paris {\sc xi} et Paris {\sc vi} associ{\'e}e
                        au {\sc cnrs}}
\\ 
Institut de Physique Nucl{\'e}aire\\
Universit{\'e} Paris-Sud\\
F-91406 Orsay Cedex, France. \vspace{0.08in} \\
percus@ipno.in2p3.fr\\
martino@ipno.in2p3.fr\\

			\end{center}

\begin{abstract}
We study the random link \tsp, where lengths $l_{ij}$ between city $i$
and city $j$ ($i<j$) are taken to be independent, identically distributed
random variables.  We discuss a theoretical approach, the cavity method,
that has been proposed for finding the optimal tour length over this random
ensemble,
given the assumption of replica symmetry.  Using finite size scaling and
a renormalized model, we test the cavity predictions against the results
of simulations, and find excellent agreement over a range of
distributions.  In doing so, we provide
numerical evidence that the replica symmetric solution to this problem is
the correct one.
 
\end{abstract} 
 
\clearemptydoublepage
\renewcommand{\baselinestretch}{1.1}\tiny\normalsize
\setcounter{footnote}{0}
\renewcommand{\thefootnote}{\arabic{footnote}}

\section{Introduction}
\label{sec_introduction}

Over the past 15 years, the study of the traveling salesman problem (\tsp)
from the point of view of statistical physics has been gaining added
currency, as theoreticians have improved their understanding of the links
between combinatorial optimization and disordered systems.  One particular
breakthrough occurred with the idea, first formulated by
\citeasnoun{MezardParisi_86b} and later developed by
\citeasnoun{KrauthMezard}, that a version of
the stochastic \tsp\ could be solved using an analytical method inspired
from spin glasses.  This method, known as the {\it cavity method\/},
is based on certain assumptions pertaining to the physical properties
of the system, notably that of replica symmetry.  Although in the case
of the spin glass replica symmetry is known to be broken
\cite{MezardParisiVirasoro}, for the \tsp\ there are various grounds
for suspecting that the assumption is valid \cite{Sourlas,KrauthMezard}.

The traveling salesman problem is as follows: given $N$ sites (or ``cities''),
find the length of the shortest closed path (``tour'') passing through
all cities exactly once.  In the stochastic \tsp, the
matrix of distances separating pairs of cities
is drawn randomly from an ensemble.  The ensemble that has received the
most attention among theoreticians is the {\it random link\/} case, where
the lengths $l_{ij}$ between city $i$ and city $j$ ($i<j$) are taken to
be independent random variables, all identically distributed according
to some $\rho(l)$.  The idea of looking at this random link ensemble,
rather than the more traditional ``random point'' ensemble with cities
distributed uniformly in Euclidean space, originated with an attempt
by Kirkpatrick to find some sort of \tsp\ equivalent to the earlier
\citeasnoun{SherringtonKirkpatrick} model for spin glasses.

The great advantage of the random link \tsp\ over the (random point)
Euclidean \tsp\ is that with the former it is actually possible to make 
analytical progress on the problem.  The cavity solution of
\citeasnoun{MezardParisi_86b} and \citeasnoun{KrauthMezard} leads to
a system of integral equations, which can be solved --- numerically
at least --- to give the optimal tour length in the limit $N\to\infty$.

In a previous paper \cite{CBBMP}, we chose
the random link distribution $\rho(l)$ to match that of the distribution
of individual city-to-city distances in the Euclidean case, 
and used the random link \tsp\
as a {\it random link approximation\/}
to the Euclidean \tsp.  The approximation might seem somewhat crude,
since it neglects all correlations between Euclidean distances, such
as the triangle inequality.  Nevertheless, it gives remarkably good
results.  For the $d$-dimensional Euclidean case in $d=2$ and $d=3$,
a numerical solution of the (random link) cavity equations predicts
asymptotic $N\to\infty$ optimal lengths within 2\% of the values obtained from
direct (Euclidean) simulation.  In the limit $d\to\infty$, this gap
shows all signs of disappearing.  The random link problem is thus more
closely related to the Euclidean problem than most researchers have suspected.

The random link \tsp\ is also interesting in itself, however, particularly
because little numerical work has accompanied the analytical progress
made.  This is all the more important given the questions surrounding
the hypotheses adopted in the cavity method.
In this paper we attempt to redress this imbalance,
providing first of all a numerical study of the finite size scaling of
the random link optimal tour length, and second of all, empirical
arguments suggesting that the cavity solution is in fact the correct one.

\section{Background and the cavity method}
\label{sec_cavity}

In an attempt to apply tools from statistical mechanics to optimization
problems, \citeasnoun{KirkpatrickToulouse} introduced a particularly
simple case of the random link \tsp.  The distribution of link lengths
$l_{ij}$ was taken to be uniform, so that $\rho(l)$ is constant over
a fixed interval.  In light of the random link approximation, one may
think of this as corresponding, at large $N$, to the 1-D Euclidean case.
(When cities are randomly and uniformly distributed on a line segment,
the distribution of lengths between pairs is uniform.)
Although the 1-D Euclidean case is trivial --- particularly
if we adopt periodic boundary conditions, in which case the optimal tour
length is simply the length of the line segment --- the corresponding
random link problem is far from trivial.

The simulations performed by Kirkpatrick and Toulouse
suggested a random link optimal tour
length value of $L_{RL}\approx 1.045$ in the $N\to\infty$ limit.\footnote{Here
we work in units where the line segment is taken to have unit length,
and in order to match the normalized 1-D Euclidean distribution, 
we let $\rho(l)=2$ on $(0,1/2)$.  The 1-D Euclidean value, for comparison,
would thus be $L_E=1$.  Kirkpatrick and Toulouse, among others, choose instead
$\rho(l)=1$ on $(0,1)$, contributing an additional factor of 2 in $L_{RL}$
which we omit when quoting their results.}
\citeasnoun{MezardParisi_86a} attempted to improve both upon this
estimate and upon the theory by investigating the random link \tsp\ using
replica techniques often employed in spin glass problems.  This approach
allowed them to obtain, via a saddle point approximation, many orders of
the high-temperature expansion for the internal energy.  They then
extrapolated down to zero temperature --- corresponding to
the global \tsp\ optimum --- finding $L_{RL}=1.04\pm 0.015$.  This analysis,
like that of Kirkpatrick and Toulouse, was carried out only for the case
of constant $\rho(l)$.

Given the difficulties of pushing the replica method further, M\'ezard
and Parisi then
tried a different but related approach known as the {\it cavity\/} method
\cite{MezardParisi_86b}.
This uses a mean-field approximation which, in the case of spin glasses,
gives the same result as the replica method in the thermodynamic limit
($N\to\infty)$.
Let us sketch what is involved in the cavity method, not least so
that we may enumerate clearly the assumptions made.

\begin{figure}[!b]
\vspace{0.5cm}
\begin{center}
\begin{picture}(260,120)
\includegraphics[scale=0.6]{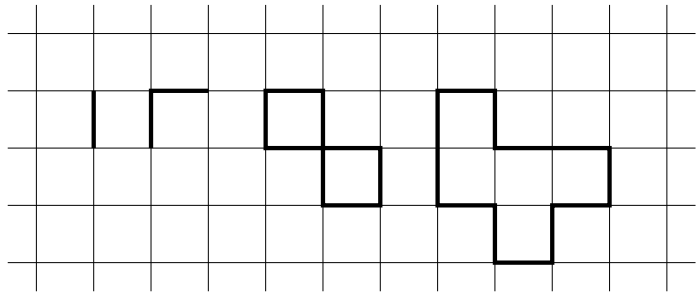}
\end{picture}
\vspace{0.5cm}
\caption{\small A graphical representation of a 1-edge open diagram; a 2-edge
open diagram; an 8-edge closed diagram with two loops; and a 12-edge
closed diagram with one loop.}
\label{fig_diagrams}
\end{center}
\end{figure}

Both the replica and the cavity approaches involve mapping the \tsp\ onto
an $m$-component spin system, writing down the partition function at
temperature $T$,
and then taking the limit $m\to 0$.  More explicity, consider
$N$ spins $\spin_i$, $i=1,\dots,N$ (corresponding to the $N$ cities), where
each spin $\spin_i$ has $m$
components $S_i^\alpha$, $\alpha=1,\dots,m$, and where
$\spin_i\cdot\spin_i=m$ for all $i$.  The partition
function is defined, in terms of a parameter $\omega$, as
\begin{eqnarray}
Z &=& \int d\mu{\{\spin\}}\exp(\omega\,\sum_{i<j}R_{ij}\,\spin_i\cdot
\spin_j)\\
&=& \int d\mu{\{\spin\}} \left[ 1+\omega\,\sum_{i<j}R_{ij}(\spin_i\cdot
\spin_j) + \frac{\omega^2}{2!}\,\sum_{\fatops{i<j}{k<l}}
R_{ij}\,R_{kl}
(\spin_i\cdot\spin_j)(\spin_k\cdot\spin_l)+\cdots
\right]
\label{eq_partfunc}
\end{eqnarray}
where $\int d\mu{\{\spin\}}$ denotes an integral over all possible spin
values (this simply corresponds to the surface of an $m$-dimensional
sphere), and $R_{ij}$ is related to the length $l_{ij}$ between city $i$ and
city $j$ as $R_{ij}\equiv e^{-N^{1/d}l_{ij}/T}$.  Now we
employ a classic diagrammatic argument: let each spin product
$(\spin_p\cdot\spin_q)$ appearing in the series be represented by an
edge in a graph.  The first-order terms ($\omega$)
will consist of one-edge diagrams, the second-order terms ($\omega^2$)
will consist of two-edge diagrams, and so on (see Figure \ref{fig_diagrams}).
What then happens when we integrate over all spin
configurations?  If there is a spin $\spin_p$ which
occurs only once in a given diagram, {\it i.e.\/}, it is an endpoint,
the spherical symmetry of $\spin_p$ will
cause the whole expression to vanish.  We
will therefore be left only with ``closed'' diagrams,
where there is at least one loop.  It may furthermore
be shown that in performing the integration, any one of these
closed diagrams will contribute a factor $m$ for every loop present in the
diagram \cite{DeGennes,Orland}.  If we then consider $(Z-1)/m$ and take
the limit $m\to 0$,
it is clear that only diagrams with a single loop will remain.
Furthermore, since any closed diagram with more than $N$ links must
necessarily contain more than one loop, only diagrams up to order $\omega^N$
will remain.  Finally, take the limit $\omega\to\infty$.  The term that
will then dominate in (\ref{eq_partfunc}) is the order $\omega^N$ term
which, being a single loop diagram, represents precisely a closed tour
passing through all $N$ sites.  We may write it without the combinatorial
factor $N!$ by expressing it as a sum over ordered pairs in the tour,
and we thus find:
\begin{eqnarray}
\label{eq_ztsp}
\lim_{\fatops{m\to 0}{\omega\to\infty}}\frac{Z-1}{m\omega^N} &=&
\sum_{\fatops{N\mbox{\scriptsize -link single loops}}{(i_1,i_2,\dots,i_n)}}
R_{i_1i_2}\,R_{i_2i_3}\cdots R_{i_{N-1}i_N}\,R_{i_1i_N}\\
&=& \sum_{N\mbox{\scriptsize -city tours}} e^{-N^{1/d}L/T}
\end{eqnarray}
where $L$ is the total tour length.  We thus obtain exactly the partition
function for the traveling salesman problem, with the correct canonical
ensemble Boltzmann weights, using the tour length as the energy to be
minimized (up to a factor $N^{1/d}$, necessary for the energy to be
extensive).

The cavity method now consists of adding an additional
$(N+1)$th spin to the system in the presence of an infinitesimally
small external magnetic field oriented along component 1,
leading to a spontaneous magnetization $\langle S_i^1\rangle$.  The
magnetization $\langle S_{N+1}^1\rangle$ is then expressed in
terms of what the other $\langle S_i^1\rangle$'s would be in the
absence of site $N+1$, hence the notion of a ``cavity''.  In order
to do this, two important assumptions are made.
First of all, we assume that there is one
single equilibrium state.  That is to say, at all temperatures a unique
thermodynamic limit exists.  A similar assumption of ergodicity is made
to obtain the replica solution given earlier.  The assumption is known
as replica symmetry: if a large number of ``copies''
of the system are allowed to develop, they all reach the same
equilibrium state.  The notion of
replica symmetry is central to the study of disordered systems.  Replica
symmetry is known to be broken in spin glasses, though there is no
reason why this should necessarily be so in other \np-hard problems.
Showing
that the cavity solution correctly predicts macroscopic quantities for
the random link \tsp\ suggests that indeed the \tsp\ exhibits replica symmetry.

The second important assumption on which the cavity method is based
involves neglecting the correlations between spins $\spin_i, i=1,\dots,N$,
when calculating $\langle S_{N+1}^1\rangle$.
The assumption is that at large $N$, the contributions due to these
correlations are $O(1/N)$ or less, so that when $N\to\infty$ this
sort of mean-field approximation is justified.
Proceeding in this way, one obtains from the partition function a recursion
relation for $\langle S_{N+1}^1\rangle$ in terms of the other
$\langle S_i^1\rangle$'s:
\begin{equation}
\label{eq_recur1}
\langle S_{N+1}^1\rangle = \frac{\sum_{i=1}^N R_{N+1,i}\langle S_i^1\rangle}
{\sum_{1\le i<j\le N}R_{N+1,i}\langle S_i^1\rangle R_{N+1,j}
\langle S_j^1\rangle}
\mbox{.}
\end{equation}
Furthermore, it may be seen from (\ref{eq_partfunc})
that the diagrams contributing to the thermal average
$\langle\spin_p\cdot\spin_q\rangle$ are those that have link
$p$---$q$ occupied; $\langle\spin_p\cdot\spin_q\rangle$ is then
the occupation number $n_{pq}$ of
the link.  Since the spontaneous magnetization is along component 1,
$n_{pq}$ is simply $\langle S_p^1 S_q^1\rangle$.
From the partition function, a recursion relation for $n_{N+1,i}$
in terms of the $\langle S_i^1\rangle$'s may
be obtained much as in (\ref{eq_recur1}):
\begin{equation}
\label{eq_recur2}
n_{N+1,i} = R_{N+1,i}\langle S_{N+1}^1\rangle\langle S_i^1\rangle
\frac{\sum_{j\ne i}R_{N+1,j}\langle S_j^1\rangle}
{\sum_{j=i}^N R_{N+1,j}\langle S_j^1\rangle}
\mbox{.}
\end{equation}

Now consider the effect of our second (mean-field) assumption over the
ensemble of instances (distribution over the disorder).  As far as
(\ref{eq_recur1})
is concerned, we may treat the magnetizations $\langle S_i^1\rangle$ as
independent identically distributed random variables.  Requiring
$\langle S_{N+1}^1\rangle$ to have the same distribution as the others
imposes, for a given link length distribution $\rho(l)$, a unique
self-consistent probability distribution of the magnetizations.  Via
(\ref{eq_recur2}), then, the distribution
${\cal P}_d(l)$ of link lengths $l$ {\it in the optimal tour\/}
(at $N\to\infty$) may be found.  \citeasnoun{KrauthMezard} carried
out this calculation in the $T\to 0$ limit, for $\rho(l)$ corresponding
to that of the $d$-dimensional Euclidean case, namely
\begin{equation}
\rho(l)=\frac{d\,\pi^{d/2}}{\Gamma(d/2+1)}\,l^{d-1}\mbox{.}
\end{equation}
They found
\begin{eqnarray}
\label{eq_lldist}
{\cal P}_d(l)&=&N^{-1/d}\pi^{d/2}\,\frac{\Gamma(d/2+1)}{\Gamma(d+1)}\,
\frac{l^{d-1}}{2\Gamma(d)}\\
&&\times\left(-\frac{\partial}{\partial l}\right)
\int_{-\infty}^{+\infty}\left[ 1+H_d(x)\right] e^{-H_d(x)}
\left[ 1+H_d(l-x))\right] e^{-H_d(l-x)}\,dx\mbox{,}
\end{eqnarray}
where $H_d(x)$ is the solution to the integral equation
\begin{equation}
H_d(x)=\pi^{d/2}\,\frac{\Gamma(d/2+1)}{\Gamma(d+1)}
\int_{-x}^{+\infty}\frac{(x+y)^{d-1}}{\Gamma(d)}\,
\left[ 1+H_d(y)\right]\, e^{-H_d(y)}\,dy\mbox{.}
\end{equation}
From ${\cal P}_d(l)$, one may obtain the mean link length in the tour,
and thus the total length of the tour:
\begin{eqnarray}
L_{RL}(N,d)&=&N\int_0^{+\infty} l\,{\cal P}_d(l)\,dl\\
&=&N^{1-1/d}\,\frac{d}{2}\int_{-\infty}^{+\infty}
H_d(x)\left[ 1+H_d(x)\right] e^{-H_d(x)}\,dx
\end{eqnarray}
for large $N$.
At $d=1$, Krauth and M\'ezard solved these equations numerically,
obtaining $L_{RL} = 1.0208$.  It is difficult to compare this
with Kirkpatrick's value of 1.045 from direct simulations (as no
error estimate exists for the latter quantity), however \citeasnoun{Johnson_HK}
have recently obtained the numerical result $L_{RL}(N=10$,$000)=
1.0211\pm 0.0003$, giving strong credence to the cavity value.
Krauth and M\'ezard also performed
a numerical study of ${\cal P}_1(l)$.  They found the
cavity predictions to be in good agreement
with what they found in their own direct simulations.  Further numerical
evidence supporting the assumption of replica symmetry was found in an
analysis, by \citeasnoun{Sourlas}, of the low temperature statistical
mechanics of the system.  Thus, for the $l_{ij}$ distribution at
$d=1$, there is good reason to believe that the cavity
assumptions are valid and that the resulting predictions are exact
at large $N$.

The cavity solution was extended to higher dimensions by the
present authors \cite{PercusMartin_PRL,CBBMP}, and a large $d$
power series solution was given for the asymptotic value
$\beta_{RL}(d)\equiv\lim_{N\to\infty}L_{RL}(N,d)/N^{1-1/d}$:
\begin{equation}
\label{eq_cavity}
\beta_{RL}(d)=\sqrt{\frac{d}{2\pi e}}\,(\pi d)^{1/2d}\,
\left[1+\frac{2-\ln 2-2\gamma}{d}+O\left(\frac{1}{d^2}\right)\right]\mbox{,}
\end{equation}
where $\gamma$ represents Euler's constant ($\gamma\approx 0.57722$).
It would be nice, then, to have evidence that the cavity method is
exact for {\it all\/} $d$ and not just at $d=1$.  While certain
arguments have been advanced in favor of this claim
\cite{MezardParisiVirasoro,Sourlas}, they have yet to be backed up
by numerical evidence --- as they have been, for instance, in a
related combinatorial optimization problem known as the matching
problem \cite{BKMP,Boutet_Thesis}.
We now turn to this task, considering first the $d=2$ case, and then a
``renormalized'' random link model enabling us to verify numerically
the $O(1/d)$ coefficient in (\ref{eq_cavity}).

\section{Numerical analysis: $d=2$ case}
\label{sec_numerd2}

We have implicitly been making the assumption so far, via our notation,
that as $N\to\infty$ the random variable $L_{RL}(N,d)/N^{1-1/d}$
approaches a unique value $\beta_{RL}(d)$ with probability 1.
This is a property known as self-averaging.  The analogous
property has been shown for the Euclidean \tsp\ at all
dimensions \cite{BHH}.  For the random link \tsp, however, the only
case where a proof of self-averaging is known is in the $d\to\infty$ limit,
where a converging upper and lower bound give in fact the {\it exact\/}
result \cite{VannimenusMezard}:
\begin{equation}
\label{eq_rlexact}
\beta_{RL}(d)=\sqrt{\frac{d}{2\pi e}}\,(\pi d)^{1/2d}
\left[ 1 + O\left( \frac{1}{d}\right) \right]\mbox{.}
\end{equation}
Note, incidentally, that this already shows that the cavity solution
(\ref{eq_cavity}) is correct in the infinite dimensional limit.

\begin{figure}[!b]
\begin{center}
\begin{picture}(340,270)
\includegraphics[scale=0.6]{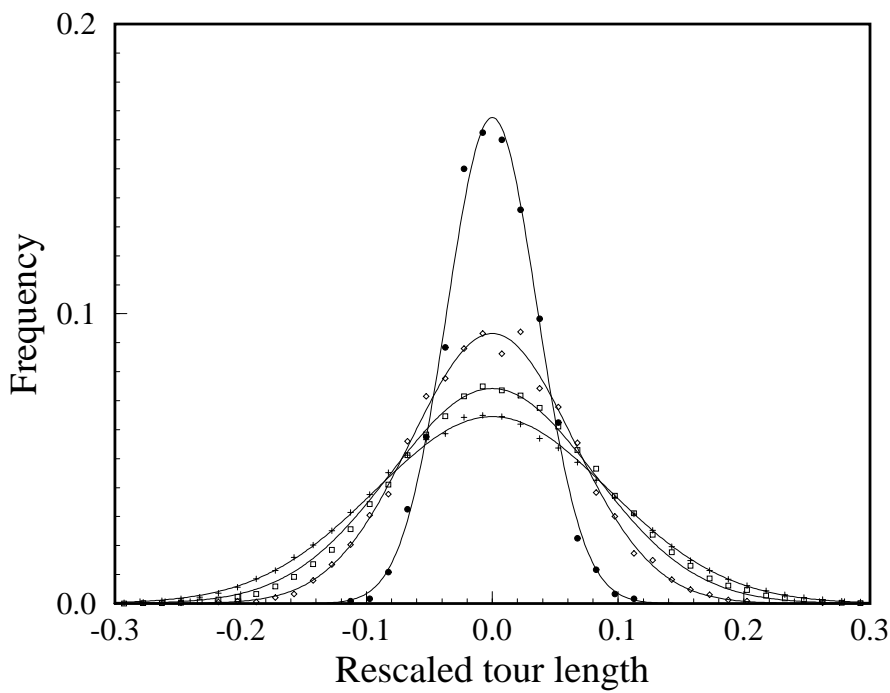}
\end{picture}
\caption{\small Distribution of 2-D random link rescaled tour length
$(L_{RL}-\langle
L_{RL}\rangle)/\sqrt{N}$ for increasing values of N.  Plus signs show
$N=12$ (100,000 instances used), squares show $N=17$ (100,000 instances
used), diamonds show $N=30$ (4,000 instances used), and dots show $N=100$
(1,200 instances used).
Solid lines represent Gaussian fits for each value of $N$ plotted.}
\label{fig_rl_selfavd2}
\end{center}
\end{figure}

For finite $d$, however, it has not been shown analytically that
$\beta_{RL}(d)$
even exists.  To some extent, the difficulty in proving this can be
traced to the non-satisfaction of the triangle inequality.  The reader
acquainted with the proof by Beardwood, {\it et al.\/}, may see that the ideas
used there are not applicable to the random link case; for instance,
combining good subtours using simple insertions will not lead to
near-optimal global tours, making the problem particularly challenging.
Let us therefore examine the distribution of $d=2$ optimum tour lengths
using numerical simulations, in order to give empirical support for the
assertion that the $N\to\infty$ limit is well-defined.

In Figure \ref{fig_rl_selfavd2}, we see that the $L_{RL}(N,2)/\sqrt{N}$
distribution indeed becomes increasingly sharply peaked for increasing
$N$.  Furthermore the variance of $L_{RL}(N,2)$ remains relatively
constant in $N$ (see Table \ref{table_rl_vard2}), indicating that the
width $\sigma$ for the distribution shown in the figure decreases as
$1/\sqrt{N}$, strongly suggesting a Gaussian distribution.  Similar
results were found
in an earlier study of ours concerning the Euclidean \tsp\ \cite{CBBMP}
(albeit in that case with $\sigma$ approximately half of its random
link value).  This is precisely the sort of behavior one would expect
were the central limit theorem to be applicable.

\begin{table}[h]
\caption{\small Variance of the non-rescaled optimum tour length $L_{RL}(N,2)$
with increasing $N$.}
\label{table_rl_vard2}
\begin{center}
\begin{tabular}{ccc}
$N$&$\sigma^2$&\# instances used\\
\hline
12&$0.3200$&$100$,$000$\\
17&$0.3578$&$100$,$000$\\
30&$0.3492$&$4$,$000$\\
100&$0.3490$&$1$,$200$\\
\end{tabular}
\end{center}
\end{table}

It is worth noting that self-averaging in a different quantity, the
Parisi overlap $q$ \cite{MPSTV,MezardParisiVirasoro}, would in fact be
a direct indication of replica symmetry.  The overlap
$q_{ab}$ between two states (tours) $a$ and $b$ may be defined in the
\tsp, following Kirkpatrick and Toulouse, as the fraction of links that
are common to tours $a$ and $b$.  If we then consider suboptimal tours
produced by a finite-temperature algorithm, and measure the overlaps
between these tours, replica symmetry requires that the overlap
distribution become more and more sharply peaked as
$N\to\infty$.  Results by \citeasnoun{Sourlas} confirm that this is
so for $d=1$, at temperatures going down to $T=0.85$.  Unfortunately,
such simulations at lower temperatures do not appear feasible: Sourlas'
results suggest that fluctuations in $q$ become uncontrollable as
$T\to 0$.  As we are interested primarily in the $T=0$ case, we do not
consider $q$ in the present analysis, resorting instead to direct tests
of the cavity predictions.

The algorithmic procedures we use for simulations are identical
to those we used in our Euclidean study; for details, the interested
reader is referred to that article.
Let us, however, note here that our optimization procedure involves using
the \lk\ and \clo\ local search heuristic algorithms
\cite{LinKernighan,MartinOtto_AOR} where for each instance of the
ensemble we run the
heuristic over multiple random starts.  \lk\ is used for smaller values
of $N$ ($N\le 17$) and \clo, a more sophisticated method combining \lk\
optimization with random jumps, for larger values of $N$ ($N=30$ and
$N=100$).  Practitioners may note at this point that given the problems
we have mentioned in obtaining near-optimal tours from near-optimal
subtours in the random link \tsp, local search methods would be expected
to perform very poorly, leading to relative excess lengths which diverge
as $N\to\infty$ and thus next to useless for a large $N$ analysis.
In the case of the \lk\ heuristic, interestingly enough, the divergence
appears to be no worse than logarithmic in $N$
\cite{JohnsonMcGeoch}, presumably because of the
``variable depth'' search that \lk\ conducts.  There is, nevertheless, a
certain probability that even over the course of multiple random starts,
our heuristics will not find the true optimum of an instance.  We
estimate the associated systematic bias using a number of test instances,
and adjust the number of random starts to keep this bias at least an
order of magnitude below other sources of error discussed below.
(At its maximum --- occurring in the $N=100$ case --- the systematic
bias is estimated as under 1 part in 20,000.)

\begin{figure}[t]
\begin{center}
\begin{picture}(340,300)
\epsfig{file=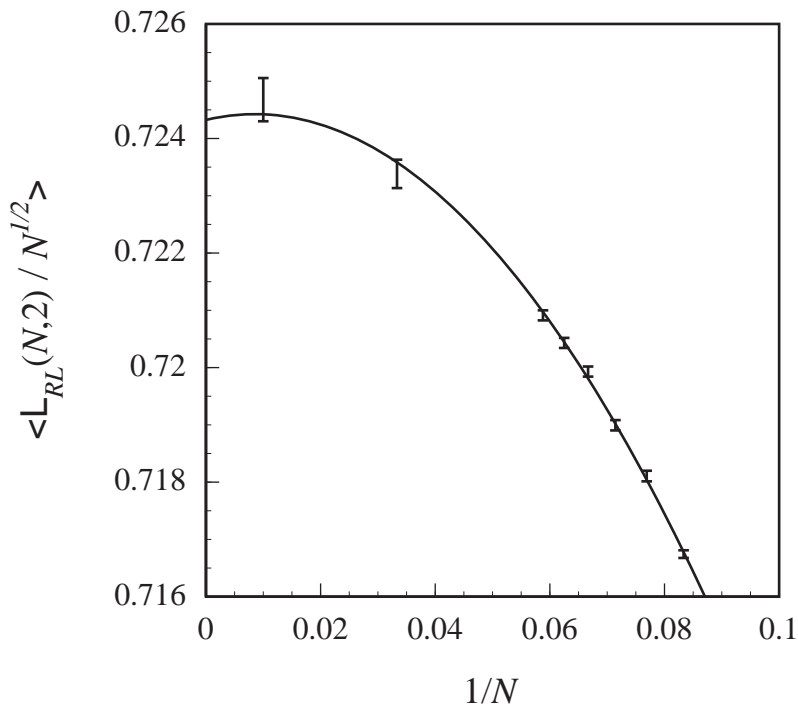}
\end{picture}
\caption{\small Finite size scaling of $d=2$ optimum.  Best fit ($\chi^2=4.46$)
is given by:
$\langle L_{RL}(N,2)\rangle/N^{1/2} = 0.7243
(1 + 0.0322/N - 1.886/N^2)$.  Error bars show one standard deviation
(statistical error).}
\label{fig_rl_fssd2}
\end{center}
\end{figure}

Following this numerical method, let us now consider the large $N$ limit
of $L_{RL}(N,2)$.  In the Euclidean case, it has been observed that the
finite size scaling law can be written in terms of a power series in
$1/N$.  We expect this same behavior in the random link case, namely
that the ensemble average $\langle L_{RL}(N,2)\rangle$ satisfies
\begin{equation}
\langle L_{RL}(N,d)\rangle=\beta_{RL}(d)\,N^{1-1/d}\,
\left[ 1+\frac{A(d)}{N} + \cdots
\right]\mbox{.}
\end{equation}
In order to obtain $\langle L_{RL}(N,2)\rangle$ at a given value
of $N$ from simulations,
we average over a large number of instances to reduce the statistical
error arising from instance-to-instance fluctuations.  Figure
\ref{fig_rl_fssd2}
shows the results of this, with accompanying error bars, fitted to the
expected finite size scaling law.  The fit is a good one ($\chi^2=4.46$
for 5 degrees of freedom) and gives an extrapolated value of
$\beta_{RL}(2)=0.7243\pm 0.0004$.  This is in superb agreement with the
cavity result of 0.7251.  The relative discrepancy between the cavity
prediction and our numerical estimate is approximately $0.1\%$, 
which in consistent with our statistical error bars; by
comparison, the error in the replica symmetric solution to the
Sherrington-Kirkpatrick spin glass ground state energy is estimated to
be of the order of $5\%$ \cite[pp.~13--14]{MezardParisiVirasoro}.

\begin{figure}[!b]
\begin{center}
\begin{picture}(260,270)
\includegraphics[scale=0.6]{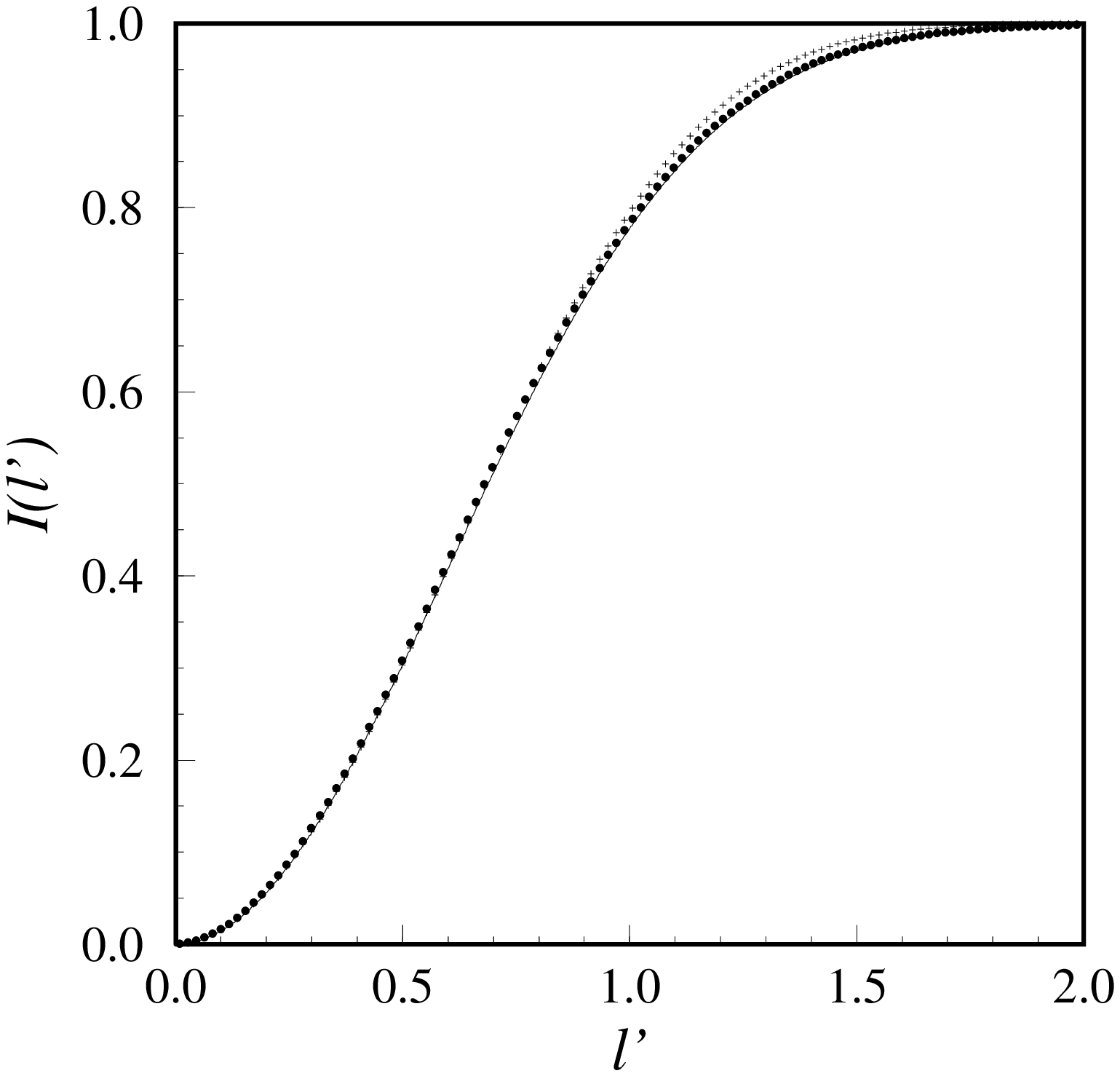}
\end{picture}
\caption{\small Integrated probability distribution of link lengths in the
optimal tour, for $d=2$, using rescaled length $l'=l\sqrt{N}$.
Plus signs represent $N=12$ simulation results, dots represent $N=100$
simulation results, and solid line represents cavity prediction.}
\label{fig_rl_distscale}
\end{center}
\end{figure}

Another quantity that was considered in the $d=1$ numerical study of
\citeasnoun{KrauthMezard} is the optimal tour link length distribution
${\cal P}_d(l)$ given in (\ref{eq_lldist}).  Let us consider ${\cal P}_2(l)$,
and following their example, let us look specifically at the integrated
distribution $I_d(l)\equiv\int_0^l {\cal P}_d(\tilde{l})\,d\tilde{l}$.
The cavity
result for $I_d(l)$ can, like $\beta_{RL}(d)$, be computed numerically
to arbitrary precision.
In Figure \ref{fig_rl_distscale} we compare this with the results of direct
simulations, at increasing values of $N$.
The improving agreement for
increasing $N$ (already within 2\% at $N=100$) strongly suggests that the
cavity solution gives the exact $N\to\infty$ result.

\begin{figure}[t]
\begin{center}
\begin{picture}(340,270)
\includegraphics[scale=0.6]{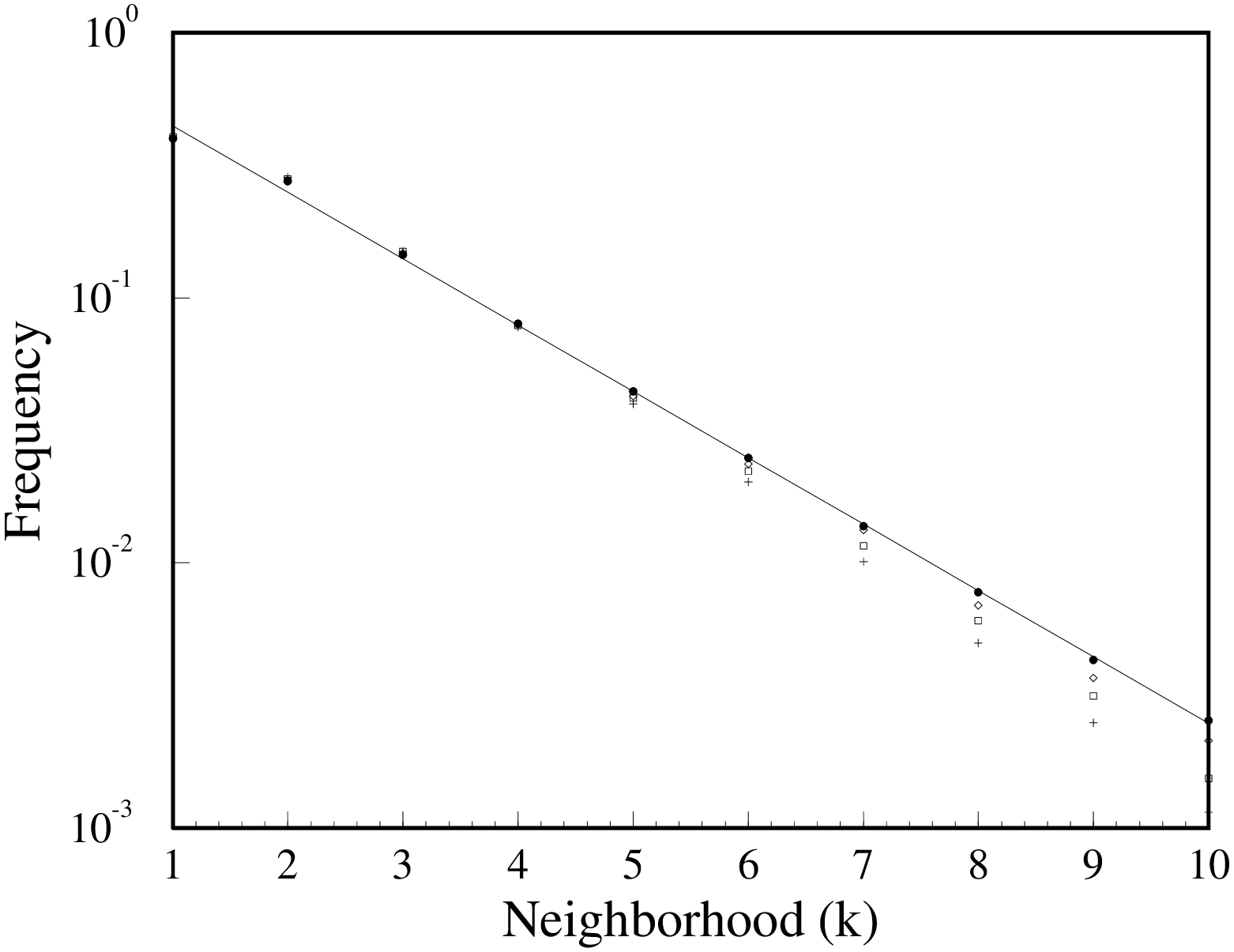}
\end{picture}
\caption{\small Frequencies with which $k$th-nearest neighbors are used in
optimal 2-D random link tours.  Plus signs show values for $N=12$,
squares for $N=17$, diamonds for $N=30$, and dots for $N=100$.  Best
exponential fit (appearing linear on log plot) is shown for $N=100$ data.}
\label{fig_rl_neighbord2}
\end{center}
\end{figure}

Finally, it may be of interest to consider one further quantity in
the $d=2$ random link simulations: the frequencies of ``neighborhoods''
used in the optimal tour, that is, the proportion of links connecting
nearest neighbors, 2nd-nearest neighbors, etc.  While there is no
cavity prediction for this quantity, \citeasnoun{Sourlas} has noted
that in practice in the $d=1$ case, the frequency falls off rapidly
with increasing neighborhood --- suggesting that optimization heuristics
could be improved by preferentially choosing links between very near
neighbors.  Results of our simulations, shown in Figure
\ref{fig_rl_neighbord2}, suggest that at $d=2$ the decrease is in fact
very close to exponential.  We have no theoretical explanation for this
behavior at present, and consider it a fascinating open question.

\section{Numerical analysis: renormalized model}
\label{sec_numerrn}

Let us now consider a different sort of random link \tsp, proposed in
\citeasnoun{CBBMP}, allowing us to test numerically  the $1/d$ coefficient
in the cavity result (\ref{eq_cavity}).  Define $\langle D_1(N,d)\rangle$
to be the distance between a city and its nearest neighbor, averaged over
all cities in the instance and over all instances in the
ensemble.\footnote{Note
that $\langle D_1(N,d)\rangle$ itself does not involve the notion of optimal
tours, or tours of any sort for that matter.}  For large $d$, it may be
shown that
\begin{equation}
\lim_{N\to\infty}N^{1/d}\langle D_1(N,d)\rangle =
\sqrt{\frac{d}{2\pi e}}\,(\pi d)^{1/2d} \left[ 1 - \frac{\gamma}{d} +
O\left(\frac{1}{d^2}\right)\right]
\end{equation}
where $\gamma$ is Euler's constant.  It is not surprising that
this quantity is reminiscent of (\ref{eq_rlexact}), since
$N^{1/d}\langle D_1(N,d)\rangle$ represents precisely a lower bound
on $\beta_{RL}(d)$.

In the renormalized random link model, we define a new link length
$x_{ij}\equiv d [l_{ij} -\langle D_1(N,d)\rangle ] /
\langle D_1(N,d)\rangle$, where the $l_{ij}$ have the usual distribution
corresponding to $d$ dimensions.  The $x_{ij}$ are ``lengths'' only in the
loosest sense, as they can be both positive and negative.  The optimal
tour in the $x_{ij}$ model will, however, follow the same ``path'' as
the optimal tour in the associated $l_{ij}$ model since the transformation
is linear.  Its length $L_x(N,d)$ will simply be given in terms of
$L_l(N,d)$ by:
\begin{eqnarray}
L_x(N,d)&=&d\,\frac{L_l(N,d)-N\langle D_1(N,d)\rangle}{\langle D_1(N,d)\rangle}
\mbox{, so}\\
L_l(N,d)&=&N\langle D_1(N,d)\rangle\,\left[1+\frac{L_x(N,d)}{dN}\right]\mbox{.}
\end{eqnarray}
By definition, then,
\begin{eqnarray}
\beta_{RL}(d)&=&\lim_{N\to\infty}N^{1/d}\langle D_1(N,d)\rangle
\mbox{, or at large $d$,}\\
&=&\sqrt{\frac{d}{2\pi e}}\,(\pi d)^{1/2d} \left[ 1 - \frac{\gamma}{d} +
O\left(\frac{1}{d^2}\right)\right]\,\lim_{N\to\infty}\left[1+
\frac{L_x(N,d)}{dN}\right]\mbox{.}
\label{eq_rnbeta1}
\end{eqnarray}
If $\beta_{RL}(d)$ is to be well-defined, then there must exist
a value $\mu(d)$ such that $\lim_{N\to\infty}
L_x(N,d)/N=\mu(d)$.

What will be the distribution of lengths $\rho(x)$ corresponding to
$\rho(l)$?  From the definition of $\rho(l)$ and that of $x$,
\begin{eqnarray}
\rho(x)&=& \frac{d\,\pi^{d/2}\,l^{d-1}}{\Gamma(d/2+1)}\,
\frac{\langle D_1(N,d)\rangle}{d}\mbox{, and substituting for $l$,}\\
&=& \frac{\pi^{d/2}}{\Gamma(d/2+1)}\,\left(1+\frac{x}{d}\right)^{d-1}\,
\langle D_1(N,d)\rangle^d\mbox{.}
\end{eqnarray}
In the limit $N\to\infty$, this gives
\begin{eqnarray}
\rho(x)&\sim&\frac{\pi^{d/2}}{\Gamma(d/2+1)}\,\left(1+\frac{x}{d}\right)^{d-1}
N^{-1}\,\left(\frac{d}{2\pi e}\right)^{d/2}\sqrt{\pi d}\,
\left[1-\frac{\gamma}{d}+\cdots\right]^d\\
&\sim&N^{-1}\left(1-\frac{\gamma}{d}\right)^d
\left(1+\frac{x}{d}\right)^{d-1}\left[1+O\left(\frac{1}{d}\right)\right]
\mbox{ by Stirling's formula}\\
&\sim&N^{-1}e^{x-\gamma}\,\left[1+O\left(\frac{1}{d}\right)\right]
\label{eq_rndist}
\end{eqnarray}
At large $d$, we see that $\rho(x)$ is to leading order independent of $d$;
the same must then be true for $L_x(N,d)$, and hence for $\mu(d)$, which we
now write as $\mu_0 [1+O(1/d)]$.  From (\ref{eq_rnbeta1}), we obtain
\begin{equation}
\beta_{RL}(d)=\sqrt{\frac{d}{2\pi e}}\,(\pi d)^{1/2d} \left[ 1 +
\frac{\mu_0 - \gamma}{d} + O\left(\frac{1}{d^2}\right)\right]\mbox{.}
\label{eq_rnbeta2}
\end{equation}

\begin{figure}[!b]
\begin{center}
\begin{picture}(340,270)
\includegraphics[scale=0.6]{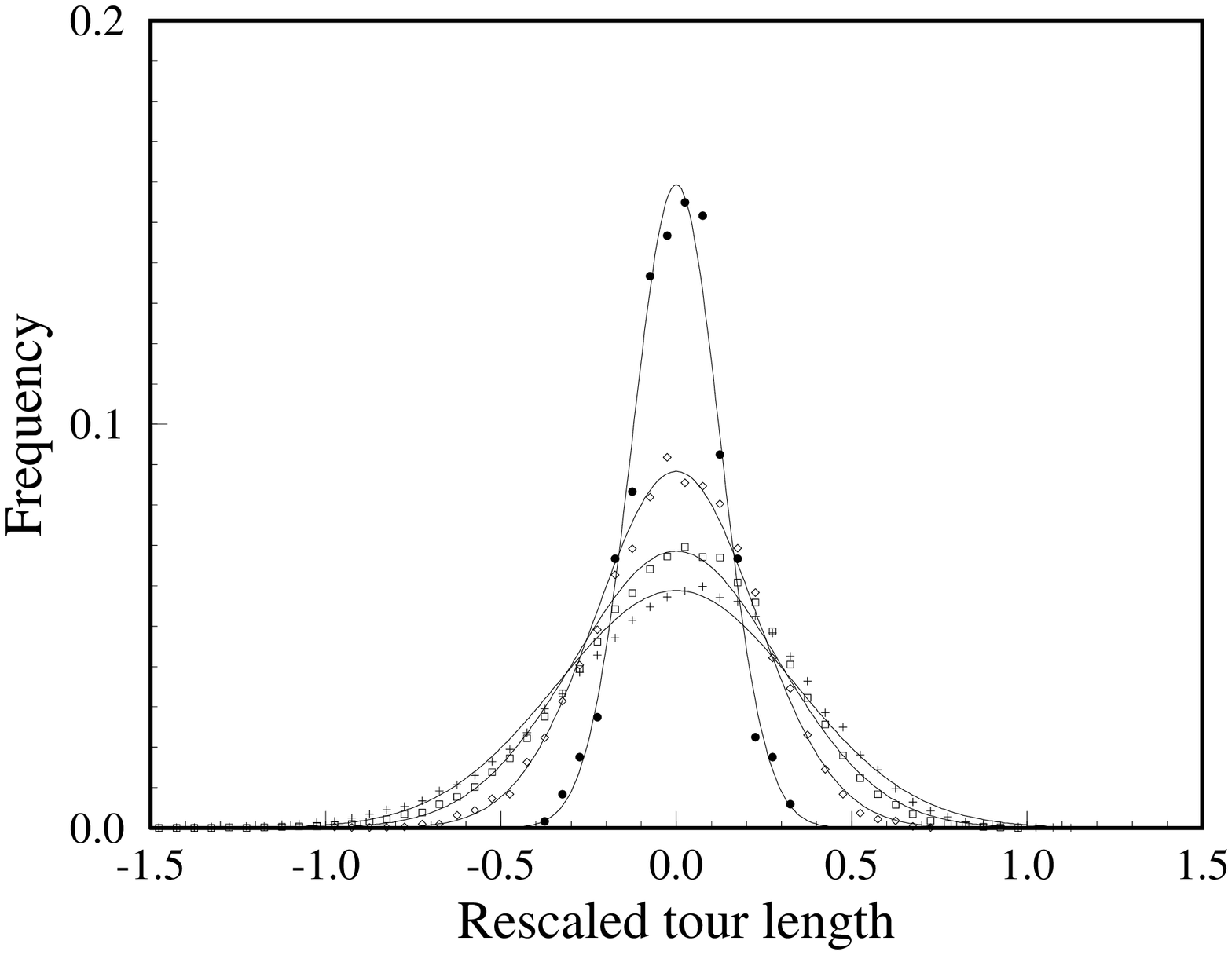}
\end{picture}
\caption{\small Distribution of renormalized random link rescaled tour length
$(L_{x}-\langle
L_{x}\rangle)/N$ for increasing values of N.  Plus signs show
$N=12$ (100,000 instances used), squares show $N=17$ (100,000 instances
used), diamonds show $N=30$ (4,000 instances used), and dots show $N=100$
(1,200 instances used).
Solid lines represent Gaussian fits for each value of $N$ plotted.}
\label{fig_rl_selfavrn}
\end{center}
\end{figure}
\par\noindent
Finally, we may perfectly well define $\rho(x)$ in the $d\to\infty$ limit,
in which case (\ref{eq_rndist}) gives $\rho(x)=N^{-1}\exp (x-\gamma)$.
Doing so will give us a renormalized model with $\lim_{N\to\infty} L_x(N)/N=
\mu_0$.  By performing direct simulations on a random link model with
this exponential distribution, we may find the value of $\mu_0$
numerically, and thus from (\ref{eq_rnbeta2}) the $1/d$ coefficient for the
$d$-dimensional (non-renormalized) case.

\begin{figure}[t]
\begin{center}
\begin{picture}(340,300)
\epsfig{file=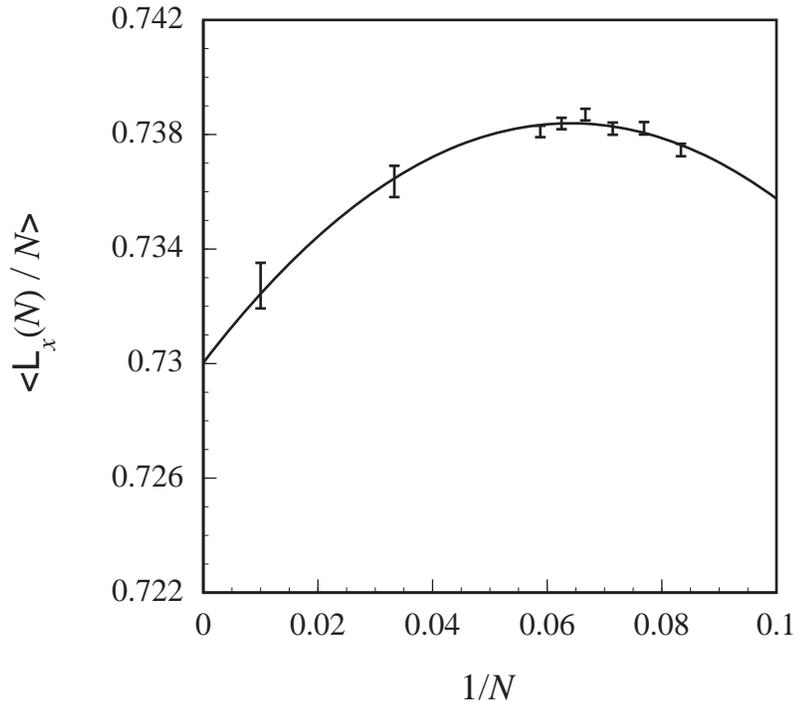}
\end{picture}
\caption{\small Finite size scaling of ``renormalized model'' optimum.
Best fit ($\chi^2=5.23$) is given by: $\langle L_x(N)\rangle/N = 0.7300
(1 + 0.3575/N - 2.791/N^2)$.  Error bars show one standard deviation
(statistical error).}
\label{fig_rl_fssrn}
\end{center}
\end{figure}

Figures \ref{fig_rl_selfavrn} and \ref{fig_rl_fssrn} show the numerical
results for the renormalized model.  In Figure \ref{fig_rl_selfavrn},
we see that just as in the $d=2$ case, the distribution of the optimal
tour length becomes sharply peaked at large $N$ and the asymptotic
limit $\mu_0$ is well-defined.  Via (\ref{eq_rnbeta1}), this provides
very good reason for believing that $\beta_{RL}(d)$ is well-defined
{\it for all\/} $d$, and self-averaging holds for the random link
\tsp\ in general.
In Figure \ref{fig_rl_fssrn}, we show the finite size
scaling of $\langle L_x(N)/N\rangle$.  The fit is again quite satisfactory
(with $\chi^2$=5.23 for 5 degrees of freedom), giving the asymptotic result
$\mu_0=0.7300\pm 0.0010$.  The simulated value for the $1/d$ coefficient
in $\beta_{RL}(d)$ is then $\mu_0-\gamma=0.1528\pm 0.0010$, in excellent
agreement (error under $0.3\%$) with the cavity value
$2-\ln 2-2\gamma\approx 0.1524$ given in (\ref{eq_cavity}).

\begin{figure}[t]
\begin{center}
\begin{picture}(340,270)
\includegraphics[scale=0.6]{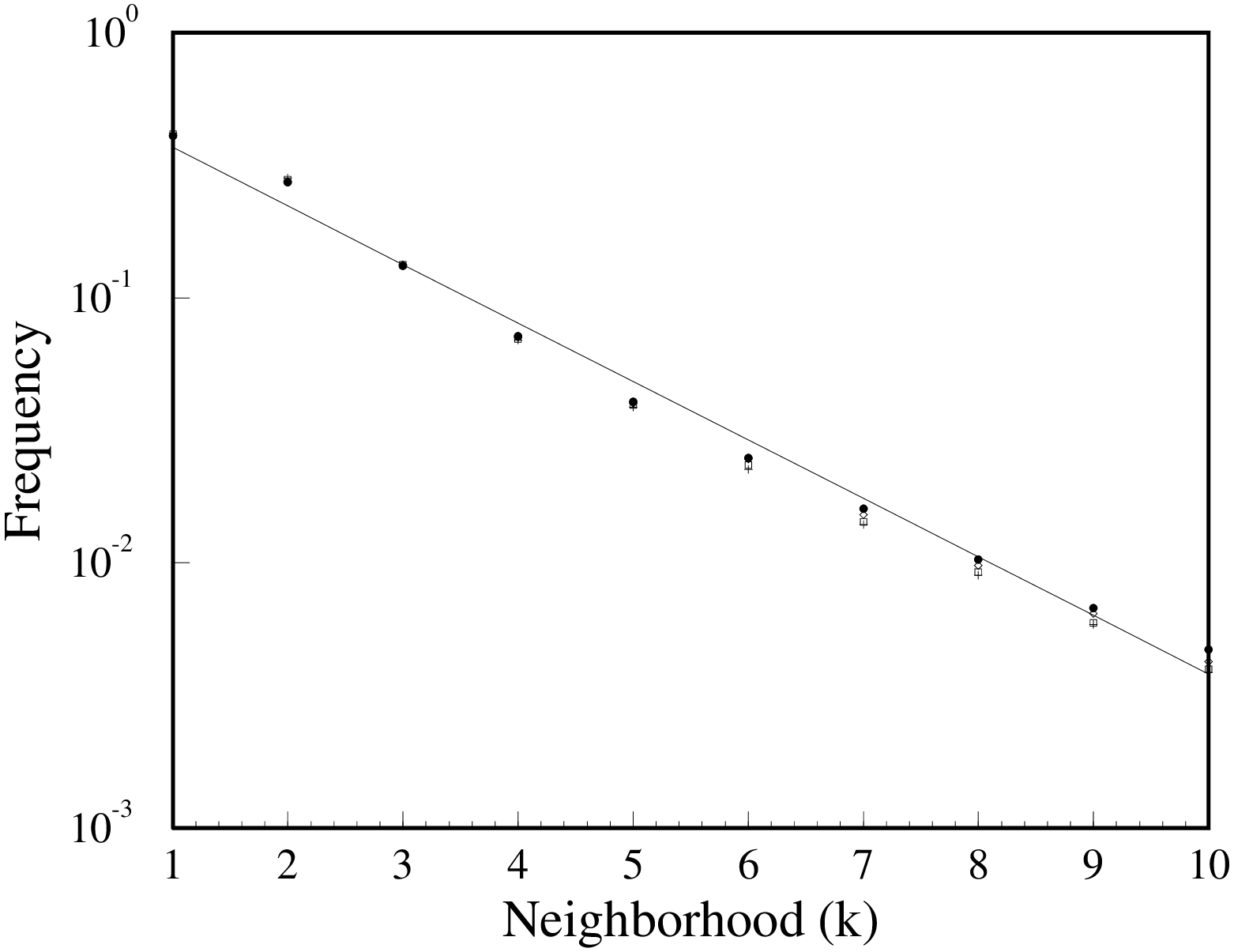}
\end{picture}
\caption{\small Frequencies with which $k$th-nearest neighbors are used in
optimal renormalized random link tours.  Plus signs show values for $N=12$,
squares for $N=17$, diamonds for $N=30$, and dots for $N=100$.  Best
exponential fit (appearing linear on log plot) is shown for $N=100$ data.}
\label{fig_rl_neighborrn}
\end{center}
\end{figure}

Also as in the $d=2$ case, let us briefly consider the frequencies
of $k$th-nearest neighbors used in optimal tours.  This is given in
Figure \ref{fig_rl_neighborrn}.  Even though the exponential fit is
not as excellent as in the $d=2$ case, it is still striking here.
We have seen that the renormalized random link model is analogous to
the $d\to\infty$ limit of the standard random model, as far as tour
paths are concerned.  These $k$th-neighbor frequency results then give
empirical evidence that even at infinite dimension, the ``typical''
$k$ used remains bounded.

\section{Conclusion}
\label{sec_conclusion}

The random link \tsp\ has interested theoreticians primarily because of its
analytical tractability, allowing presumably exact results that are not
possible in the more traditional Euclidean \tsp.  Other than in the $d=1$ case,
however, it has attracted little attention from the angle of numerics.
In this paper we have provided a numerical study of the random link \tsp\
that was lacking up to this point, and which addresses some long-unanswered
questions.  Through simulations, we have tested the validity of the
theoretical cavity method predictions.
While in other disordered systems, such as spin glasses, the replica symmetric
solution gives values of macroscopic quantities that are inexact (typically
by at least $5\%$), in the random link \tsp\ it shows all signs of
being exact.  
We have studied various macroscopic quantities at $d=2$ and found
that the numerical results confirm the cavity predictions to within
$0.1\%$.  Furthermore,
we have confirmed, by way of a renormalized random link model,
that the analytical cavity solution gives a large $d$ expansion for the
optimal tour length whose $1/d$ coefficient is correct to within well
under $1\%$.  The excellent agreement found at $d=1$
\cite{KrauthMezard,Sourlas},
$d=2$, and to $O(1/d)$ at large $d$, then suggest strongly
that the cavity predictions are exact.  This, finally, provides indirect
evidence that the assumption of replica symmetry --- on which the
cavity approach is based --- is indeed justified.

\ifx\undefined\BySame
\newcommand{\BySame}{\leavevmode\rule[.5ex]{3em}{.5pt}\ }
\fi
\ifx\undefined\textsc
\newcommand{\textsc}[1]{{\sc #1}}
\fi
\ifx\undefined\emph
\newcommand{\emph}[1]{{\em #1\/}}
\fi

\newpage
\clearemptydoublepage

\chapter{Universalities in nearest neighbor distances}
\thispagestyle{empty}
\minitoc

\vspace{1cm}
\preliminaires{}

In this final chapter we turn to a problem that does not involve
the \tsp\ as such, but is closely inspired by our \tsp\ work and mirrors much
of the language we have employed so far.  We have seen in Chapter~II
how the $N\to\infty$ Euclidean optimal tour length, where $N$ represents
the number of cities, could be extrapolated from reasonably small sized
instances via an understanding of the finite size scaling behavior.
We were thus able to obtain precise numerical estimates for
this asymptotic value without undue computational effort.

In order to understand the finite size scaling, we decided to set aside
temporarily the notion of optimal tours --- or, for that matter, tours of
any sort.
Instead we simply looked at cities placed randomly and independently,
with a uniform distribution, and considered the distances between nearest
neighboring cities, second-nearest neighboring cities, and so on.
What we found was an unexpected sort of
universality.  If one considers the ensemble average $\langle D_k(N)\rangle$
of the distances between $k$th-nearest neighbors, the large $N$ scaling
law of this quantity is independent of $k$.  Putting it another way,
the $N$-dependence and the $k$-dependence of $\langle D_k(N)\rangle$
separate; in $d$ dimensions we are left with, up to corrections
exponentially small in N,
\begin{equation}
\langle D_k(N)\rangle = \frac{[\Gamma(d/2+1)]^{1/d}}{\sqrt{\pi}}\,
\frac{\Gamma(k+1/d)}{\Gamma(k)}\,\frac{\Gamma(N)}{\Gamma(N+1/d)}\mbox{.}
\end{equation}

We now wish to investigate this universality further.  How, first of all,
does the property depend on the nature of the physical space used?  The
universality only applies in flat ({\it i.e.\/}, Euclidean) spaces, but leads
us to notice another interesting and more subtle property on curved surfaces.
Let us look at the large $N$ behavior of $\langle D_k(N)\rangle$ (averaged
over the whole surface, if the surface is not homogeneous).  We find that
$\langle D_k(N)\rangle$ scales as a power of $N$ times a series in $1/N$
whose $O(1/N)$ term, while now explicitly depending on $k$, has the form of
a topological invariant: $(\chi(2k+1)-9)/24$, where $\chi$ is the Euler
characteristic.  To $O(1/N)$, then, it is the topology of the
surface and not its detailed shape that plays a role in the finite size
scaling behavior.

The $1/N$ series describing $\langle D_k(N)\rangle$ may, for a general
manifold, be obtained directly from the area $A(l)$ of a geodesic disc of
radius $l$ on the manifold.  A geodesic disc about some point $\bfx$
is simply defined as the locus of points whose distance from $\bfx$
{\it along the manifold\/} is smaller than the radius.  For a flat surface,
of course, $A(l)=\pi l^2$.  For a curved surface, $A(l)$ may be written
as a series expansion in $l$; higher-order terms in this expansion will
then correspond to higher-order terms in the $1/N$ series.
We must therefore find the series expansion
of $A(l)$, whose coefficients
involve gradients of the surface's Gaussian curvature $K$ at the point
$\bfx$ about which the expansion is being performed.  Unfortunately, we
know of no general expression for these coefficients.  Calculating them
is a somewhat laborious procedure, which we have carried out up to the
$O(l^8)$ term in $A(l)$:
\begin{eqnarray}
A(l) &=& \pi l^2-\frac{\pi l^4}{12}\,K + \frac{\pi l^6}{720}\bigl(
2K^2-3\nabla^2K\bigr)\nonumber\\
&& - \frac{\pi l^8}{161280}\bigl( 8K^3 -
3[10(\nabla K)^2 + 14K\nabla^2K - 5\nabla^4K]\bigr)
+ O(l^{10})\mbox{.}
\end{eqnarray}

From this, terms through $O(1/N^3)$ can be found in the $1/N$ series
for $\langle D_k(N)\rangle$.  We discuss these higher-order terms, although
it appears that no further universalities exist beyond the topological
invariant at $O(1/N)$.  We then turn briefly to higher-dimensional manifolds
--- where $A(l)$ is far less obvious --- and examine the cases in which a
topological invariant could exist there.  Finally, we provide an
interpretation of our results via a Regge calculus approach, where we
consider sites distributed on a polyhedral surface rather than a smooth
manifold; we find that there too, we recover the topological invariant.

Let us alert the reader to some notational differences in this
final chapter that we unfortunately cannot avoid.  We find it clearer
in the present context to speak of the distance from an arbitrary point
$\bfx$ on the manifold to its $k$th-nearest site, rather than of the
distance between a site and its $k$th-nearest neighbor.  $D_k(N)$ is
thus a random variable defined everywhere in (continuous) space, and not
only at
the positions of one of the $N$ sites.  $\langle D_k(N)\rangle$ is its
ensemble average, which for inhomogeneous surfaces will vary with $\bfx$.
Our previous notation amounted to restricting $\bfx$ to be the position
of one of the cities, thus reducing the number of possible neighbors by 1.
Our results for $N$ cities, in our previous notation, are therefore
equivalent to our results for $N-1$ cities, in our new notation.  Note,
also, that while the terms ``cities'' and ``sites'' are interchangeable,
they are to be distinguished from the term ``point'', which is used here
in its most precise geometric meaning to represent {\it any\/} position
$\bfx$ on the manifold.

\newpage
\clearemptydoublepage

\renewcommand{\baselinestretch}{1.2}\tiny\normalsize
\renewcommand{\thefootnote}{\fnsymbol{footnote}}
\setcounter{footnote}{0}
\thispagestyle{empty}
\vspace{1cm}
                        \begin{center}

{\Huge  Finite size scaling universalities of $k$th-nearest}

\bigskip

{\Huge  neighbor distances on closed manifolds}

\vspace{1cm}

{\Large\sc Allon G.~Percus and Olivier C.~Martin}\\
Division de Physique
Th{\'e}orique \footnote{\ Unit{\'e} de recherche des Universit{\'e}s
                        de Paris {\sc xi} et Paris {\sc vi} associ{\'e}e
                        au {\sc cnrs}}
\\
Institut de Physique Nucl{\'e}aire\\
Universit{\'e} Paris-Sud\\
F-91406 Orsay Cedex, France. \vspace{0.08in} \\
percus@ipno.in2p3.fr\\
martino@ipno.in2p3.fr\\

                        \end{center}

\begin{abstract}
Take $N$ sites distributed randomly and uniformly on a closed surface.
We express
the expected distance $\langle D_k(N)\rangle$ from an arbitrary point on the
surface to its $k$th-nearest neighboring site, in terms of the surface
area $A(l)$ of a disc of radius $l$ about that point.  We then find two
universalities.  First, for a flat surface, where $A(l)=\pi l^2$,
the finite size scaling series giving corrections to the large $N$
asymptotic behavior of $\langle D_k(N)\rangle$ is independent
of $k$.  Second, for a curved surface, the finite size scaling series for
the average $\int\langle D_k(N)\rangle\,d\mu$ over the surface
is, to $O(1/N)$, a topological invariant.  We discuss the case of higher
dimensions ($d>2$), and also interpret our results using Regge calculus.

\end{abstract}

\clearemptydoublepage
\renewcommand{\baselinestretch}{1.1}\tiny\normalsize
\setcounter{footnote}{0}
\renewcommand{\thefootnote}{\arabic{footnote}}

\section{Introduction}
\label{sect_introduction}
In many physics and computer science
problems, a quantity of interest is the distance between
neighboring sites in a space.  One frequently wishes to calculate
distances to a nearest neighbor, second-nearest neighbor, and
more generally, $k$th-nearest neighbor.  Examples in computational
geometry and optimization abound, ranging from random packing of
spheres to minimum spanning trees.
Such problems also arise naturally in 
physics, both for interacting particle 
systems such as liquids, and for cellular objects such as
foams and random lattices \cite{ItzyksonDrouffe}.

Here we consider the case of $N$ sites placed randomly, with a
uniform distribution, on a 2-D surface of fixed area.  Let the random
variable $D_k(N)$ represent the distance between a given point $\bfx$ and
its $k$th-nearest neighboring site.  Its expectation value $\langle
D_k(N)\rangle$ taken over the ensemble of randomly placed
sites --- and in fact all moments $\langle D^\alpha_k(N)\rangle$ ---
then exhibit some surprising properties.  

When the surface is flat, $\langle D_k(N)\rangle$ at large $N$ is described
by its leading asymptotic behavior times a power series in $1/N$, where all
orders of the series turn out to be
{\it independent\/} of $k$.  In the language of statistical physics, the
finite size scaling series for $\langle D_k(N)\rangle$
is the same at all $k$ to all orders in $1/N$.  Geometrically, this
universality is far from obvious.
Furthermore, we find that the property is not restricted to
two dimensions, and is equally valid for flat spaces of any dimension.

When the surface is curved, this $k$-independence no longer holds,
but one finds on the other hand another universality: the finite size
scaling series, averaged over the entire surface, gives a topological
invariant to $O(1/N)$.  This term thus
depends not on the detailed
shape of the surface but only on the surface's genus.

In this paper we explore these universalities.
First, we express $\langle D_k(N)\rangle$
in terms of the area, $A(l)$, of a disc of radius $l$ on an arbitrary surface.
We observe that in the special case where $A(l)$ consists only of a
power of $l$,
the $1/N$ series for $\langle D_k(N)\rangle$ exhibits the universality
in $k$, i.e., is independent of $k$.  Second, we give the relation between
$A(l)$ and the Gaussian curvature of the surface, and find the leading
correction in the $1/N$ power series for $\langle D_k(N)\rangle$, as functions
of the curvature.  This leads to the topological invariance at $O(1/N)$.  We
then discuss higher order terms, and the case of higher
dimensions.  Finally, we show that a Regge calculus
approach provides a simple means of obtaining this topological invariance
for the case of polyhedral (non-smooth) surfaces.

\section{Preliminaries}
\label{sect_prelim}

Take any point $\bfx$, and consider $P[D_k(N)=l]_\bfx$, the probability
density that the point's $k$th-nearest neighboring site lies at a distance
$l$ from it.  This is equal to the probability density of having $k-1$
(out of $N$) sites {\it within\/} distance $l$, one site (out of $N-k+1$)
{\it at\/} distance $l$, and the remaining $N-k$ sites
{\it beyond\/} distance $l$.  Let us choose units so that our surface has
area 1.  Since sites are distributed uniformly over the surface,
the probability of a single site lying within distance $l$ is then simply
the area $A(l)_\bfx$ of a disc of radius $l$ about point $\bfx$ on the
surface.  Dropping the argument $\bfx$ (in order to simplify the notation),
we may then write
\begin{displaymath}
P[D_k(N)=l] = \left(\begin{array}{c} N \\ k-1 \end{array}\right)\,
\left[ A(l)\right]^{k-1}\times\,
\left(\begin{array}{c} N-k+1 \\ 1 \end{array}\right)\,
\frac{dA(l)}{dl}\,\times\,\left[ 1-A(l)
\right]^{N-k},
\end{displaymath}
giving the expectation value (first moment)
\begin{eqnarray*}
\langle D_k(N)\rangle &=& \int_0^\infty P[D_k(N)=l]\,l\,dl\\
&=& \frac{N!}{(N-k)!\,(k-1)!}\,\int_0^\infty l\,\left[ A(l)\right]^{k-1}
\left[ 1-A(l)\right]^{N-k}\,\frac{dA(l)}{dl}\,dl.
\end{eqnarray*}
Under the variable transformation $w=A(l)$, this may be written in
terms of the inverse function $A^{-1}(w)$ as
\begin{displaymath}
\langle D_k(N)\rangle = \frac{N!}{(N-k)!\,(k-1)!}\,\int_0^1
A^{-1}(w)\,w^{k-1} (1-w)^{N-k}\, dw.
\end{displaymath}

If $A^{-1}(w)$ admits the power series expansion in $w$:
\begin{equation}
\label{eq_series1}
A^{-1}(w)\simeq w^\gamma\,\sum_{j=0}^\infty c_j w^j,\qquad
\mbox{for some }\gamma\in[0,1),
\end{equation}
then
\begin{equation}
\label{eq_integral}
\langle D_k(N)\rangle\simeq\frac{N!}{(N-k)!\,(k-1)!}\,\sum_{j=0}^\infty c_j
\int_0^1 w^{k+j+\gamma-1} (1-w)^{N-k}\, dw.
\end{equation}
Recognizing the integral as the Beta function ${\rm B}(k+j+\gamma,N-k+1)=
(k+j+\gamma-1)!\,(N-k)!/$ $(N+j+\gamma)!$,
\begin{equation}
\label{eq_series2}
\langle D_k(N)\rangle\simeq\sum_{j=0}^\infty c_j\,
\frac{(k+j+\gamma-1)!}{(k-1)!}\,\frac{N!}{(N+j+\gamma)!}.
\end{equation}

Several comments are in order concerning $\langle D_k(N)\rangle$.  First
of all, although we restrict ourselves to discussing the first moment
of $D_k(N)$, we could in fact consider any moment $\langle
D^\alpha_k(N)\rangle$ by taking $\left[ A^{-1}(w)\right]^\alpha$ instead
of $A^{-1}(w)$ in (\ref{eq_series1}).  Doing so would alter $\gamma$ and
the $c_j$'s, but would not change our results qualitatively.  Second of all,
there is no loss of generality in taking our total surface area
to be unity; scaling this area by a constant (or even, as might be more
intuitive to statistical physicists, by $N$) would provide
only a trivial scaling factor in our results.  Third of all, we could
imagine that the point $\bfx$ we consider
is itself an $(N+1)$th site.  This is simply a question of nomenclature.
The problem of finding the expected distance {\it from\/} a point {\it to\/}
its $k$th-nearest site, for a system of $N$ sites, is therefore equivalent
to the problem of finding the expected distance {\it between\/} $k$th-nearest
neighbors, for a system of $N+1$ sites.

Let us now define the reduced variable $\langle\tilde{D}_k(N)\rangle$ by
dividing out the leading asymptotic (large $N$) behavior from $\langle
D_k(N)\rangle$:
\begin{eqnarray}
\langle\tilde{D}_k(N)\rangle &=& \langle D_k(N)\rangle\,\frac{1}{c_0}\,
\frac{(k-1)!}{(k+\gamma-1)!}\,N^\gamma\nonumber\\
&\simeq& N^\gamma\,\sum_{j=0}^\infty\,\frac{c_j}{c_0}\,
\frac{(k+j+\gamma-1)!}{(k+\gamma-1)!}\,
\frac{N!}{(N+j+\gamma)!},
\label{eq_reduced}
\end{eqnarray}
so that $\lim_{N\to\infty}\langle\tilde{D}_k(N)\rangle = 1$ (this is
seen from Stirling's law).  $\langle\tilde{D}_k(N)\rangle$ 
then provides the {\it finite size scaling\/} behavior for $k$th-nearest
neighbor distances.  In what follows, we shall consider the properties of
$A^{-1}(w)$, and its consequences on $\langle\tilde{D}_k(N)\rangle$.

\section{Flat Surfaces}
\label{sect_flat}

On a flat surface, if we could neglect edge effects, the area included
within distance $l$ would simply be $A(l)=\pi l^2$.  In that case,
$A^{-1}(w)=\sqrt{w/\pi}$, and so from (\ref{eq_series1}) and
(\ref{eq_reduced}) we would have
\begin{equation}
\label{eq_noexp}
\langle\tilde{D}_k(N)\rangle = \frac{\sqrt{N}\,N!}{(N+1/2)!}.
\end{equation}
The finite size scaling would thus be completely independent of $k$.

As we are working with a surface of fixed (unit) area, however, we
cannot avoid considering edge effects.  Let us restrict ourselves
to the case where the surface is everywhere locally Euclidean within
some minimum neighborhood of radius $l_0>0$.  (The simplest example of this is
a unit square with periodic boundary conditions, for which $l_0=1/2$.
Obviously, many other constructions are possible.)
Any required modifications to the $A^{-1}(w)$ expression in
(\ref{eq_series1}) then concern only
$w$ greater than $w_0\equiv A(l_0)$.  Correspondingly, (\ref{eq_integral})
remains valid up to remainder terms from the region of integration
$w_0\le w\le 1$.  Since the
$(1-w)^{N-k-1}$ term in the integral is bounded above by $(1-w_0)^{N-k-1}$
within this region, these remainder terms are exponentially small in $N$.
Equation (\ref{eq_noexp}) is therefore still correct {\it to all
orders\/} in $1/N$, and may be written as the expansion
\begin{displaymath}
\langle\tilde{D}_k(N)\rangle = 1 - \frac{3}{8N} +
O\left(\frac{1}{N^2}\right),
\end{displaymath}
where all orders in $1/N$ are independent of $k$.  The finite size scaling
law for
$k$th-nearest neighbor distances on a 2-D flat surface without a boundary
thus exhibits the universality in $k$ to all orders in $1/N$.

The same holds true for flat manifolds in any dimension $d$.  We assume
there is an $l_0$ such that the
volume included within distance $l<l_0$ is simply
the volume of a $d$-dimensional ball:
\begin{eqnarray*}
A(l)&=&\frac{\pi^{d/2}\,l^d}{(d/2)!}, \qquad\mbox{or}\\
A^{-1}(w)&=&\frac{1}{\sqrt{\pi}}\left[ w\left( \frac{d}{2}\right) !
\right]^{1/d}.
\end{eqnarray*}
As before, the boundary conditions allow us to write
$\langle\tilde{D}_k(N)\rangle$ up to remainder terms that are exponentially
small in $N$, so from (\ref{eq_reduced}),
\begin{eqnarray*}
\langle\tilde{D}_k(N)\rangle &\simeq&
\frac{N^{1/d}\,N!}{(N+1/d)!}\,+\,\cdots\\
&=&
1- \frac{1/d + 1/d^2}{2N} + O\left( \frac{1}{N^2}\right).
\end{eqnarray*}
Thus for flat spaces without a boundary, of any dimension $d$,
the universality in $k$ holds to all orders in $1/N$.

It may be interesting to consider a slight variation on the problem,
giving this universality {\it exactly\/} and not only to all
orders.  Take the case of a spherical surface embedded in
3-D Euclidean space, with the usual measure of area
over the sphere, but with a peculiar
sort of ``distance'': rather than the conventional choice of the
arc length (geodesic) metric, use the chord length.  For
a chord of length $l$ originating at a pole of the sphere, the area
of the spherical cap spanned by it is simply $A(l)=\pi l^2$.
The $k$th-nearest neighbor distance properties using chord length
``distance'' on this curved surface then appear analogous to those
on a flat surface.  
There is, however, one important distinction.  The relevant
threshold $w_0$ for edge effects is in this case $w_0=\pi (2R)^2$,
where $R$ is the radius of the sphere.  Since $\pi (2R)^2$ is exactly
equal to the total surface area of the sphere, it is set to 1.  Equation
(\ref{eq_integral}) thus requires no corrections at all, and so the
universality in (\ref{eq_noexp}) is exact.

\section{Curved Surfaces}
\label{sect_curved}

Let us turn to the case of a surface with true curvature, where the
distance is now defined in terms of a metric, i.e., along
geodesics of the surface.  Let us consider again a spherical surface.
With $l$ now representing arc length, the area of the spherical cap
spanned by an arc originating at a pole of the sphere is given by
\begin{eqnarray*}
A(l)_{\rm sphere} &=& 2\pi R^2 \left[ 1-\cos\frac{l}{R}\right] \\
&=& 4\pi R^2 \sin^2\frac{l}{2R}
\end{eqnarray*}
If the total surface area, $4\pi R^2$, is normalized to 1,
\begin{eqnarray}
A(l)_{\rm sphere} &=& \sin^2\sqrt{\pi}l,\qquad\mbox{so}\nonumber\\
A^{-1}(w)_{\rm sphere} &=& \frac{1}{\sqrt{\pi}}\sin^{-1}\sqrt{w}\nonumber\\
&\simeq&\sqrt{\frac{w}{\pi}}\,\sum_{j=0}^\infty\,\frac{w^j}{2j+1}\,
\frac{(2j)!}{2^{2j}(j!)^2}.
\label{eq_area_sphere}
\end{eqnarray}
As in the case of the chord length ``distance'', this $A^{-1}(w)$
expression
is exact everywhere for $0\le w\le 1$.  Equations (\ref{eq_series1}) and
(\ref{eq_reduced}) then require no corrections, and we find
\begin{eqnarray}
\langle\tilde{D}_k(N)\rangle_{\rm sphere} &=&
\sum_{j=0}^\infty\,\frac{1}{2j+1}\,\frac{(2j)!}{2^{2j}(j!)^2}\,
\frac{(k+j-1/2)!}{(k-1/2)!}\,\frac{\sqrt{N}\,N!}{(N+j+1/2)!}\nonumber\\
&=&
1+ \frac{4k-7}{24N} + O\left( \frac{1}{N^2}\right).
\label{eq_nn_sphere}
\end{eqnarray}
Clearly, the $k$ universality does not apply here.

Another sort of universality, however, is found when we turn to the more
general case of an arbitrary closed surface, i.e., an abstract 2-D
manifold with non-constant curvature and no boundary.  For any surface,
we may introduce a system of curvilinear coordinates $u$ and $v$ enabling
us to write (at least piecewise) the differential length element
$ds$ in the conformal, orthogonal form \cite{Eisenhart}:
\begin{equation}
\label{eq_ds}
ds^2 = f(u,v)\,[du^2 + dv^2].
\end{equation}
The Gaussian curvature $K(u,v)$ of the surface may then be expressed
in terms of the function $f(u,v)$ by
\begin{equation}
\label{eq_gausscurv}
K = \frac{1}{2f^3}\left[\left(\frac{\partial f}{\partial u}\right)^2
+ \left(\frac{\partial f}{\partial v}\right)^2
- f\frac{\partial^2 f}{\partial u^2}
- f\frac{\partial^2 f}{\partial v^2}\right].
\end{equation}
What is $A(l)$ on this surface?  In order to know this,
we must first find out what are the manifold's geodesic lines.  For
$ds$ given by (\ref{eq_ds}), we may use the geodesic equation:
\begin{equation}
\label{eq_geodesic}
\frac{d^2 u}{ds^2}+\frac{1}{2f}\,\frac{\partial f}{\partial u}
\left[\left(\frac{du}{ds}\right)^2 - \left(\frac{dv}{ds}\right)^2\right]
+ \frac{1}{f}\,\frac{\partial f}{\partial v}\,\frac{du}{ds}\,\frac{dv}{ds}
=0.
\end{equation}
Let us expand $u$ and $v$ as functions of distance $s$ from an initial
point, along a fixed geodesic:
\begin{eqnarray}
u(s) &=& u_0 + s u_0' + \frac{s^2}{2} u_0'' + \cdots\quad\mbox{and}\nonumber\\
v(s) &=& v_0 + s v_0' + \frac{s^2}{2} v_0'' + \cdots ,
\label{eq_expansion}
\end{eqnarray}
where $u_0\equiv u(0)$, $u_0'\equiv u'(0)$, etc., and likewise for $v$.
Then, expanding $f(u,v)$ in terms of $u$ and $v$ and substituting
(\ref{eq_expansion}),
\begin{eqnarray*}
f(u,v)
&=& f(u_0,v_0) + s \left[ u_0' f_u(u_0,v_0) + v_0' f_u(u_0,v_0)\right]
+ s^2 \left[\frac{u_0''}{2}f_u(u_0,v_0) + \frac{v_0''}{2}f_v(u_0,v_0)\right.\\
&& \qquad\left. + \frac{(u_0')^2}{2}f_{uu}(u_0,v_0) +
u_0' v_0' f_{uv}(u_0,v_0) +
\frac{(v_0')^2}{2}f_{vv}(u_0,v_0)\right] + O(s^3),
\end{eqnarray*}
where subscripts on $f$ denote partial derivatives.

Using (\ref{eq_ds}) and (\ref{eq_geodesic}), we can solve for all
but three of the coefficients in (\ref{eq_expansion}).  Let us choose
$u_0$, $v_0$ and $u_0'$ to be these three.  Now, consider the area $A(l)$
about the point $(u_0,v_0)$.  For $ds$ given in (\ref{eq_ds}), the
differential surface element will be $d\mu=f\,du\,dv$, so:
\begin{eqnarray}
A(l) &=& \int f\,du\,dv\nonumber\\
&=& \int f\,J\left(\frac{u,v}{s,u_0'}\right)ds\,du_0'\nonumber\\
&=& \int f\, \left|\frac{\partial u}{\partial s}
\frac{\partial v}{\partial u_0'} - \frac{\partial u}{\partial u_0'}
\frac{\partial v}{\partial s}\right|\,ds\,du_0'.
\label{eq_area_int}
\end{eqnarray}
The limits of integration over $s$ are $0$ and $l$;
the limits of integration over $u_0'$, which may be found from
(\ref{eq_ds}), are $-\sqrt{1/f(u_0,v_0)}$ and $\sqrt{1/f(u_0,v_0)}$.
In evaluating the Jacobian some care must be taken, as a sign ambiguity
allows two solutions for the coefficients in (\ref{eq_expansion}).
$A(l)$ will be the sum of (\ref{eq_area_int}) evaluated at each of the
two solutions, ultimately causing all odd powers of $l$ to vanish.

The result, after lengthy algebraic manipulations, may be written as
\begin{equation}
\label{eq_area_series1}
A(l)=\pi l^2\left[ 1-\frac{l^2}{24}\,\frac{1}{f^3}\,(f_u^2 + f_v^2
-f f_{uu} -f f_{vv}) + O(l^4)\right],
\end{equation}
where in order to avoid cluttering the notation, we have omitted the
$(u_0,v_0)$ arguments at which all functions are to be evaluated.
For the leading correction term in $A(l)$ given by
(\ref{eq_area_series1}), we recognize
the expression (\ref{eq_gausscurv}) for the Gaussian curvature $K$.
In retrospect, this is not surprising: by symmetry, the correction
series to $\pi l^2$ can contain only even powers of $l$,
and if we consider $A(l)$
as a geometric expansion about a flat space approximation, $K$ will be
the only scalar quantity with dimensions of $l^{-2}$.

With some perseverance,
one may carry (\ref{eq_area_series1}) to higher orders in $l$, and
the coefficients of the expansion turn out to be
\begin{eqnarray}
A(l) &=& \pi l^2\left[ 1-\frac{l^2}{12}\,K + \frac{l^4}{720}\bigl(
2K^2-3\nabla^2K\bigr)\right.\nonumber\\
&& \qquad\left. - \frac{l^6}{161280}\bigl( 8K^3 -
3[10(\nabla K)^2 + 14K\nabla^2K - 5\nabla^4K]\bigr)
+ O(l^8)\right],
\label{eq_area_series2}
\end{eqnarray}
where $\nabla$ is the gradient operator.
We thus have a series expansion giving the area of a disc on any 2-D
surface, in a form that depends only on intrinsic quantities, {\it i.e.\/},
not on the choice of coordinate system.

We may invert (\ref{eq_area_series2}) to obtain the power series
\begin{eqnarray}
\label{eq_area_inverse}
A^{-1}(w) &=& \sqrt{\frac{w}{\pi}}\left[ 1+ \frac{K}{24\pi}\,w +
\frac{9K^2 + 4\nabla^2K}{1920\pi^2}\,w^2 +
\frac{15K^3 + 14K\nabla^2K - 2(\nabla K)^2 + \nabla^4K}
{21504\pi^3}\,w^3\right.\nonumber\\
&& \qquad\left. + O(w^4)\right].
\end{eqnarray}
Take as an example the special case of a spherical surface, where the
Gaussian curvature is a constant $K=1/R^2$, or $K=4\pi$ for a unit
surface.  All derivatives of $K$ then vanish, leaving
\begin{displaymath}
A^{-1}(w)_{\rm sphere}=\sqrt{\frac{w}{\pi}}\left[ 1+ \frac{w}{6} +
\frac{3w^2}{40} + \frac{5w^3}{112} + O(w^4)\right],
\end{displaymath}
from which we recover the first few terms of our earlier result
(\ref{eq_area_sphere}).

Given an expression for $A^{-1}(w)$ on a general 2-D surface, we may find
$\langle\tilde{D}_k(N)\rangle$ using (\ref{eq_series1}) and (\ref{eq_reduced}).
Let us consider the average $\int\langle\tilde{D}_k(N)\rangle\,d\mu$ over
the entire surface, which is found directly from the average of the
series coefficients in (\ref{eq_area_inverse}).  Now examine the $O(w)$
term in the series.
By the Gauss-Bonnet theorem \cite{Eisenhart}, $\int K\,d\mu=2\pi\chi$ on any
closed surface, where $\chi$ is the Euler
characteristic of the surface, a topological invariant.
We therefore obtain, up to leading corrections,
\begin{eqnarray}
\label{eq_chi}
\int A^{-1}(w)\,d\mu &=& \sqrt{\frac{w}{\pi}}\left[ 1+\frac{\chi}{12} w +
O(w^2)\right],\qquad\mbox{giving}\nonumber\\
\int\langle\tilde{D}_k(N)\rangle\,d\mu &=& \frac{\sqrt{N}\,N!}{(N+1/2)!}\,
\left[ 1+ \frac{\chi}{12}\,
\frac{k+1/2}{N+3/2} + O\left(\frac{1}{N^2}\right)\right]\nonumber\\
&=& 1+ \frac{\chi (2k+1)-9}{24N} + O\left( \frac{1}{N^2}\right).
\label{eq_nn_general}
\end{eqnarray}
We thus discover another, very different sort of universality from the one
we had in the case of flat space.  To $O(1/N)$, the finite size scaling law
for $k$th-nearest neighbor distances depends only on the surface's topology,
and not on its detailed properties.

The Euler characteristic $\chi$ for a surface is related to its genus $g$
by $\chi=2(1-g)$.  Taking the torus as one example, $g=1$,
so $\chi=0$ and the
$k$-dependence in (\ref{eq_nn_general}) once again disappears, at least
to $O(1/N)$.  This is to be expected: a flat space with periodic boundary
conditions has, after all, the topology of a torus.  And conversely,
because of the topological invariance, all tori behave like flat space
to $O(1/N)$.  Taking the spherical surface as another example, $g=0$,
so $\chi=2$ and we
recover from (\ref{eq_nn_general}) the result (\ref{eq_nn_sphere}).

The properties of $\int\langle\tilde{D}_k(N)\rangle\,d\mu$ are far less clear
at higher order in $1/N$.
Using the divergence theorem and integration by parts, we may obtain from
(\ref{eq_area_inverse})
\begin{equation}
\label{eq_area_inverse_av}
\int A^{-1}(w)\,d\mu=\int\sqrt{\frac{w}{\pi}}\left[ 1+ \frac{K}{24\pi}\,w +
\frac{3K^2}{640\pi^2}\,w^2 + \frac{15K^3 + 16K\nabla^2K}
{21504\pi^3}\,w^3 + O(w^4)\right]\,d\mu .
\end{equation}
If we looked only at terms up through $O(w^2)$, we might believe that this
series is simply, by analogy with (\ref{eq_area_sphere}), an expansion
of $\int (2/\sqrt{K}) \sin^{-1}\sqrt{Kw/4\pi}\,d\mu$.  Unfortunately,
starting at $O(w^3)$ we see this is not true, since the contributions
of curvature and its gradients
do not all vanish in the average over the surface!  Furthermore,
even for terms in (\ref{eq_area_inverse_av}) of the form $\int K^n\,d\mu$,
at $n>1$ there is no clear equivalent to the Gauss-Bonnet theorem;
the theorem is a direct consequence of the integrand's
linearity.  Thus for a general 2-D surface, there does not appear to be
a simplified form for the terms in $\int\langle\tilde{D}_k(N)\rangle\,d\mu$
beyond $O(1/N)$.  More particularly, the {\it only\/} case in which 
$\int\langle\tilde{D}_k(N)\rangle\,d\mu$ would be independent of $k$ 
beyond $O(1/N)$ is if the curvature
is identically equal to $0$, i.e., a flat surface.

Let us briefly consider the case of curved higher-dimensional manifolds.
The calculation is now far more complicated, as it is no longer possible
to write the metric tensor in a conformal form as we did in
(\ref{eq_ds}).  In addition, whereas in 2-D the only intrinsic
scalar quantity describing curvature is the Gaussian curvature $K$,
for $d>2$ there are $d(d-1)(d-2)(d+3)/12$ different such quantities
\cite{Weinberg}.  However, all of them except $K$ itself have at least
dimensions of $l^{-4}$.  It thus seems reasonable to conjecture that,
as we argued
in 2-D, the $O(l^2)$ correction term in $A(l)$ can only involve $K$.  
(Indeed, we have verified that this is true in 3-D.)  In that case, we
may rely on the example of the spherical surface --- easily generalized to
$d$ dimensions --- to provide us with the initial terms for a general
manifold:
\begin{eqnarray}
A(l) &=& \frac{\pi^{d/2}}{(d/2)!}\,l^d\,\left[ 1-\frac{d(d-1)}{d+2}\,
\frac{K}{6}\,l^2+O(l^4)\right],\qquad\mbox{or}\nonumber\\
A^{-1}(w) &=& \left[\frac{(d/2)!}{\pi^{d/2}}\right]^{1/d} w^{1/d}\,
\left[ 1+\frac{d-1}{d+2}\,\frac{K}{6}\,\left(\frac{(d/2)!}{\pi^{d/2}}
\right)^{2/d}\,w^{2/d}+O(w^{4/d})\right].
\label{eq_area_multid}
\end{eqnarray}
Note that $A^{-1}(w)$ now contains a series in $w^{2/d}$ rather than in
$w$.  Appropriately modifying (\ref{eq_series1}), it may then be shown
that $\langle\tilde{D}_k(N)\rangle$ is in general given by a series in
$1/N^{1/d}$ for odd $d$, and $1/N^{2/d}$ for even $d$.

Consider, finally, the average $\int A^{-1}(w)\,d\mu$ over the manifold.
The higher dimensional generalization of the Gauss-Bonnet theorem
\cite{Nakahara}
involves an integrand of $O(1/l^d)$, or $O(1/w)$.  The leading correction
term from (\ref{eq_area_multid}), $\int K\,d\mu$, therefore cannot be
simplified further for $d>2$; the only term that could possibly give
rise to a topological invariant is the coefficient at $O(l^d)$, or
$O(w)$.  If $d$ is odd, it is rather certain that no topological invariant
will exist in the series.  If $d$ is even, the $O(w)$ term will
first contribute to the $\int\langle\tilde{D}_k(N)\rangle\,d\mu$ series
at $O(1/N)$ --- as in 2-D, although at higher dimensions this will no
longer be the leading correction term.  While we cannot rule out the
possibility
of indeed obtaining a topological invariant at $O(1/N)$, the $O(w)$
term in $A^{-1}(w)$ is in general a complicated one involving
many different curvature scalars, so this is far from obvious.  It
remains an open question.

\section{Regge Calculus}
\label{sect_regge}
We have remarked that from a physical point of view it is natural,
in the 2-D case, for the leading
corrections in $A(l)$ to contain only the
Gaussian curvature, as this represents the leading deviation from
planarity.  Consequently, only the
mean curvature --- or, using the Gauss-Bonnet theorem, the Euler
characteristic $\chi$ --- matters in the $O(1/N)$ term of 
$\int\langle\tilde{D}_k(N)\rangle\,d\mu$.  We have seen using
differential methods (geodesics) that this physical picture is
indeed correct.  These methods apply to a smooth surface.  For
polyhedral surfaces, which are not smooth, we may in fact obtain a
similar result
more easily, using the {\it non-differential\/} method of Regge
calculus.  Consider
a polyhedron with a number of vertices, edges, and faces.
Following the work of Regge \cite{Regge} and others since then
\cite{CheegerMullerSchrader}, we observe that the curvature is
concentrated at the vertices and is measured by a deficit angle: if
$\theta_i$ is the sum of the angles incident on vertex $i$, 
the deficit angle at that vertex is $\Delta_i = 2 \pi - \theta_i$.
It may then be shown that the Gauss-Bonnet theorem, on polyhedra,
reduces to Euler's relation
$2 \pi \chi = \sum_i \Delta_i$.

Let $P$ be a polyhedron with 
a fixed number of vertices, and consider
the problem of finding the finite size scaling
$\int\langle\tilde{D}_k(N)\rangle\,d\mu$ on $P$.
As $N \to \infty$, corrections to the flat space value about a given
point $\bfx$ arise only when $\bfx$ is near one of the vertices,
because only then can curvature (i.e., the deficit angle) enter into
the local calculation of $A(l)$ about $\bfx$.
It is then sufficient to understand the corrections associated with 
one vertex at a time.
Consider a vertex $i$.  $A(l)$ receives a
contribution from $i$ that depends exclusively on the deficit angle
$\Delta_i$.
For small deficit angles (small corrections to $A(l)$), a linear
approximation may be used and this contribution will simply be
proportional to $\Delta_i$.  Correspondingly, the leading corrections
to $A(l)$ --- and thus the $O(1/N)$ term in the finite size scaling
series --- will
be proportional to $\Delta_i$.  
Summing over all the vertices $i = 1,\dots,N$, we find that the leading
correction term in $\int\langle\tilde{D}_k(N)\rangle\,d\mu$
is indeed a linear function of $\chi$, and we recover the topological
invariant derived in the case of a smooth manifold.

A word of caution is necessary, however.  It is tempting at this point
to take the limit where $P$ becomes a smooth manifold, thus recovering
(\ref{eq_chi}).  Unfortunately this will not work; a direct computation
shows that the limit does not commute with the limit $N\to\infty$ taken
above, and the coefficient thus obtained for the $O(1/N)$ term will
not be the correct one.

\section{Conclusions}
\label{sect_conclusions}

We have considered the finite size scaling of mean distances to neighboring
sites, when $N$ sites are distributed randomly and
uniformly on a surface with no boundaries.  When the surface
is flat, we have found that the entire $1/N$ series describing
the mean $k$th-nearest site distance is independent of $k$.
This universality applies equally well to higher moments of the distances,
and to Euclidean manifolds in dimensions greater than 2.
For surfaces with curvature, while this general property is no longer valid,
we have found that the leading correction term in the series, averaged over
the surface, is a topological invariant.  The scaling series thus depends,
to $O(1/N)$, on the genus of the manifold but not on its other properties.

Although we have considered these universalities only for the moments of
point-to-point distances, similar properties hold for higher order
simplices such as areas of triangles associated with nearby points.
The problem is thus a natural one to consider further in the context
of random triangulations, foams or other physical problems
\cite{ItzyksonDrouffe}
tightly connected to geometry.
\section*{Acknowledgments}
\label{sect_ack}
We are grateful to J.~Houdayer and E.~Bogomolny 
for sharing with us their numerous and
valuable insights on this topic, and to O.~Bohigas for having introduced
us to the problem.  OCM acknowledges support from the Institut
Universitaire de France.

\ifx\undefined\BySame
\newcommand{\BySame}{\leavevmode\rule[.5ex]{3em}{.5pt}\ }
\fi
\ifx\undefined\textsc
\newcommand{\textsc}[1]{{\sc #1}}
\fi
\ifx\undefined\emph
\newcommand{\emph}[1]{{\em #1\/}}
\fi

\newpage
\clearemptydoublepage

\renewcommand{\thechapter}{\Alph{chapter}}
\renewcommand{\theequation}{\thechapter.\arabic{equation}}
\setcounter{chapter}{0}

\Appendix{Computational complexity: P vs. NP}
\label{app_pnp}

\noindent
The inherent challenge of the \tsp\ lies in the fact that no known algorithm
can find the optimal tour of an arbitrary $N$-city instance in a number
of steps polynomial in $N$.  In computational complexity theory, this is
what classifies it as a ``hard'' problem.  An ``easy'' problem would be
one that could be solved by a polynomial algorithm.  The word ``easy'',
of course, does not imply ``fast'': an $O(N^3)$ algorithm, at $N=10$,$000$,
could already involve immense running times!  There is nevertheless a
large conceptual difference between a problem that can be solved in
no worse than polynomial time, and a problem that requires, say, exponential
time.

Computer scientists formalize these concepts with the notion of {\sc p}
and \np.  A problem belongs to class {\sc p} if there exists an algorithm
that solves
the problem in a time growing polynomially, or slower, with the size
$N$ of the problem.  A problem belongs to class \np\ if it is merely possible
to test, in polynomial time, whether a certain ``guessed'' solution is
indeed correct.  Note that \np\ stands for ``non-deterministic
polynomial'', and {\it not\/}, as is sometimes mistakenly thought, for
``non-polynomial''.  {\sc p} is in fact a subset of \np, since for any problem
whose solution can be found in polynomial time, one can surely verify the
validity of a potential solution in polynomial time.

Formally speaking, the classes {\sc p} and \np\ apply only to
{\it decision\/} problems, where the solution is simply a ``yes''
or a ``no''.  It is however, generally possible to relate combinatorial
optimization problems to associated decision
problems.  For the case of the \tsp, instead of asking for the optimal
tour length, we may phrase the associated {\it \tsp-decision\/} as follows:
is there a tour whose length is less than a given value $B$?  It is
relatively straightforward to see that the optimal \tsp\ tour length can
be found to arbitrary precision by executing a sequence of \tsp-decisions
(with bound $B$, say, decreasing in discrete steps), where
the length of the sequence is polynomial in $N$
\cite{JohnsonPapadimitriou}.
The \tsp-decision
is thus said to be {\it polynomially reducible\/} to the \tsp; this
implies that the two problems, though not necessarily themselves
equivalent, are at least of equivalent computational complexity.

In order to see that the \tsp-decision is in \np, let us use the
following more careful definition of the \np\ class \cite{CookSA}.
A decision problem belongs to \np\ if any instance calling for a ``yes''
response contains a {\it certificate\/}, itself of size polynomial in
$N$, allowing the ``yes'' answer to be verified in polynomial time.
For the \tsp-decision, this certificate is simply a tour of length
less than $B$: the certificate is of size $N$, and in $O(N)$ steps one
may verify that it is a legal tour and satisfies the length bound.
The certificate is therefore the ``guess'' that is tested.  Of course,
we cannot hope to verify a ``no'' answer in the same way as a ``yes''
answer; the only property characterizing a ``no'' instance is that {\it
no certificate exists\/}.  The fact that a decision problem is in \np\
merely means that we can confirm a ``yes'' instance if we happen to be
given the right certificate.

In the case of the \tsp-decision, one can go further.  An important
subset of the class of \np\ problems is the class of \np-complete problems.
These are problems that are {\it as least as complex\/} as any other
\np\ problem --- in other words, an algorithm capable of solving an
\np-complete problem could be mapped onto an algorithm capable of solving
{\it any\/} \np\ problem, via a polynomial encoding.  (This is not quite the
same thing as the notion of polynomial reducibility, as the encoding
associates a single instance of one problem with a single instance of
another, rather than with a sequence of instances.)
It has been proven that the \tsp-decision is \np-complete
\cite{Papadimitriou}.  Let us consider the implications of this.
We have already noted that {\sc p} $\subseteq$ \np, since a problem
solvable in
polynomial time has by definition a certificate (its solution!)
verifiable in polynomial time.  If it turns out that a polynomial
algorithm could be found for solving the \tsp-decision --- or any
other \np-complete problem --- one would exist for all \np\ problems and
we would have {\sc p} = \np.  No such polynomial algorithm has ever
been found,
and it is conjectured that none exists, so that {\sc p} $\neq$ \np.  Proving
this conjecture, however, has been an open question in complexity theory
since the 1970s.

Note that although it has become standard practice
to refer to the \tsp\ as being \np-complete, the more correct term is
actually {\it \np-hard\/}.  This is because, strictly speaking, the
complexity classes {\sc p}, \np\ and \np-complete are reserved for decision
problems.  \np-hard is the more general designation for problems to
which \np-complete decision problems are polynomially reducible.
The \tsp-decision is \np-complete; the \tsp\ itself, being of equivalent
complexity, is thus \np-hard
\citeaffixed{GareyJohnson,JohnsonPapadimitriou}{see}.

A different issue from that of {\it solving\/} an \np-hard problem in
polynomial time is that of {\it approximating\/} its solution in
polynomial time.  For the Euclidean \tsp, \citeasnoun{Karp} developed
the {\it fixed dissection algorithm\/}, which generates a tour using
the ``divide-and-conquer'' strategy of partitioning space into subparts,
solving the \tsp\ for the cities within each subpart, and joining up
these subtours in such a way as to form one large (non-optimal) tour
through all cities.  Karp proved that this algorithm can with high
probability (prob.\ $\to 1$ as $N\to\infty$) give tours whose
length is within ($1+\epsilon$) times optimality, for arbitrary
$\epsilon$.  The expected execution time of this algorithm is
$O(N^2\log N)$.

Karp's construction has three limitations: first,
the heuristic does not {\it guarantee\/} the $1+\epsilon$ bound at
finite $N$; second, the running time could in a worst-case situation
be exponential; third, the method only applies to instances from the
random ensemble, with a uniform distribution of cities.
Recent work by \citeasnoun{Arora} appears to have resolved these three
difficulties in the case of the Euclidean \tsp.  Using a different
partitioning scheme, Arora's algorithm finds a guaranteed
$(1+\epsilon)$-approximation
in $O\bigl((\log N)^{O(\sqrt{d}/\epsilon)^{d-1}}N\bigr)$ time, for
arbitrary instances in $d$ dimensions.  This sort of method,
unfortunately, cannot easily be
generalized to other problems.  It has in fact been proven \cite{ALMSS}
that if {\sc p} $\neq$ \np, $(1+\epsilon)$-approximation algorithms
cannot exist for all \np-hard problems.  The fact that one has been
found for the
Euclidean \tsp\ is quite remarkable, and suggests that further refinement
of the \np-hard classification may be necessary.

\newpage
\clearemptydoublepage

\Appendix{Outline of a self-averaging proof}
\label{app_selfav}

\noindent
The original proof of self-averaging in the Euclidean optimal tour
length $L_E(N,d)$, by \citeasnoun{BHH}, is quite technical.  There is however 
a more accessible proof, by \citeasnoun{KarpSteele}, using the {\it fixed
dissection algorithm\/} \cite{Karp} mentioned in Appendix \ref{app_pnp}.
This algorithm
generates a (non-optimal) tour by dividing the space into subparts,
finding the optimal subtour within each subpart, and connecting up
all of these subtours (minus one link, in each case) to form one large
tour.  Karp and Steele also took advantage of a lemma (proved by induction),
stating that in the $d$-dimensional unit hypercube there must always
exist a tour of length less than $dN^{1-1/d}+\delta_d N^{1-1/(d-1)}$
($\delta_d$ is a constant depending only on $d$).

They then considered a Poisson process
placing points with unit intensity in the hypercube $[0,t]^d$, and
looked at the expectation value $F(t)$ of the optimal tour length
through these points.  If $\langle L_E(N,d)\rangle$ is the expected
optimal tour length
through $N$ points in the unit hypercube, $t$ times this quantity will be
the expected optimal tour length through $N$ points in $[0,t]^d$.  From
the Poisson distribution, we thus obtain:
\begin{equation}
\label{eq_poisson}
F(t)=\sum_{N=0}^\infty e^{-t^d}\,\frac{t^{dN}}{N!}\,t\,\langle
L_E(N,d)\rangle\mbox{.}
\end{equation}

Now, the fixed dissection algorithm bounds the optimal tour length through
the cities placed by the Poisson process.  This upper bound is equal to 
the lengths of the various subtours plus the length of the large circuit
through space needed to connect the subtours.  Bounding the latter length
using the lemma above, and partitioning space into $m^d$ equal subparts,
we then obtain the expectation value bound:
\begin{eqnarray}
F(t)&\le &m^d F(t/m) + t\bigl( d(m^d)^{1-1/d} + \delta_d 
(m^d)^{1-1/(d-1)}\bigr)\mbox{, or}\\
\frac{F(t)}{t^d}&\le & \frac{F(t/m)}{(t/m)^d} +
\frac{d}{(t/m)^{d-1}}
+ \delta_d\,\frac{m^{-1/(d-1)}}{(t/m)^{d-1}}\mbox{.}
\label{eq_bound2}
\end{eqnarray}
Clearly $F(t)$ is monotone increasing, and because of the bounding lemma,
$F(t)/t^d$ must be bounded.  As $t\to\infty$, then, for fixed $m$,
(\ref{eq_bound2}) implies that
$F(t)/t^d$ approaches a limiting value, which we shall call $\beta_E(d)$.
From (\ref{eq_poisson}), substituting $u=t^d$,
\begin{equation}
\lim_{u\to\infty}\sum_{N=0}^\infty e^{-u}\,\frac{u^N}{N!}\,
\frac{\langle L_E(N,d)\rangle}{u^{1-1/d}}=\beta_E(d)\mbox{.}
\end{equation}
From the Tauberian theorem \citeaffixed[p.~187]{KarpSteele}{see} for the
Poisson distribution, finally, this gives:
\begin{equation}
\lim_{N\to\infty}\frac{\langle L_E(N,d)\rangle}{N^{1-1/d}}=\beta_E(d)\mbox{.}
\end{equation}

The statement is in fact stronger than it appears: Karp and Steele are
also able to bound the variance of $L_E(N,d)$ using an inequality due
to \citeasnoun{EfronStein}, giving Var$[L_E(N,d)/N^{1-1/d}]=O(N^{-1})$.
We now know, therefore, not only that the distribution of
$L_E(N,d)/N^{1-1/d}$ becomes increasingly sharply peaked at large $N$,
but also that its width goes to zero as $1/\sqrt{N}$, {\it i.e.\/}, as
a Gaussian.

\newpage
\clearemptydoublepage

\Appendix{Estimates for a non-uniform distribution of cities in space}
\label{app_nonunif}

\noindent
We have seen that the optimal tour length is self-averaging, in the ensemble
of cities independently and uniformly distributed in Euclidean space.  With
probability 1, the random variable $L_E(N,d)/N^{1-1/d}$ tends to a
constant $\beta_E(d)$ in the $N\to\infty$ limit.  The original proof of this
result,
due to \citeasnoun{BHH}, is however more general and deals with an arbitrary
density $\rho(\bfx)\,d\bfx$ of independently distributed cities in
space.  The general statement of self-averaging is that, with probability 1,
\begin{equation}
\lim_{N\to\infty}\frac{L_E(N,d)}{N^{1-1/d}}=\beta_E(d)
\int[\rho(\bfx)]^{1-1/d}\,d\bfx\mbox{,}
\end{equation}
where $\beta_E(d)$ is independent of the density, and thus equal to
its value for a uniform distribution.
What this means in practice is that the numerical value of $\beta_E(d)$ is of
relevance for cities distributed independently in space with {\it any\/}
distribution.

In this appendix we propose to use this result to give an {\it a priori\/}
estimate of $L_E$ for any instance.
Note, first, that the factor $\int[\rho(\bfx)]^{1-1/d}
\,d\bfx$ applies just as well to quantities such as the distance
$D_1(N,d)$ between nearest neighbors.  For an arbitrary Euclidean instance
at large $N$, then, we may estimate the optimal tour length
$L_E^{\mbox{\scriptsize non-uniform}}$ for that instance as:
\begin{equation}
L_E^{\mbox{\scriptsize non-uniform}}\approx
D_1^{\mbox{\scriptsize non-uniform}}\,
\frac{\langle L_E^{\mbox{\scriptsize
uniform}}\rangle}{\langle D_1^{\mbox{\scriptsize uniform}}\rangle}
\label{eq_nonuniform}
\end{equation}
where $D_1^{\mbox{\scriptsize non-uniform}}$ is the mean nearest neighbor
distance for the instance we are considering.  $\langle L_E^{\mbox{\scriptsize
uniform}}\rangle$ is given by its value in the Euclidean uniform ensemble,
and $\langle D_1^{\mbox{\scriptsize uniform}}\rangle$ is the
ensemble average of nearest neighbor distances.  Both of these quantities,
as functions of $N$ and $d$, have been calculated in Chapter~II.
Note that there is no need to use units where volume equals
1 in the non-uniform instance under consideration; our use of
$D_1^{\mbox{\scriptsize non-uniform}}$ in (\ref{eq_nonuniform}) will take
into account any volume scaling.

\begin{figure}[t]
\begin{center}
\begin{picture}(260,200)
\includegraphics[scale=0.6]{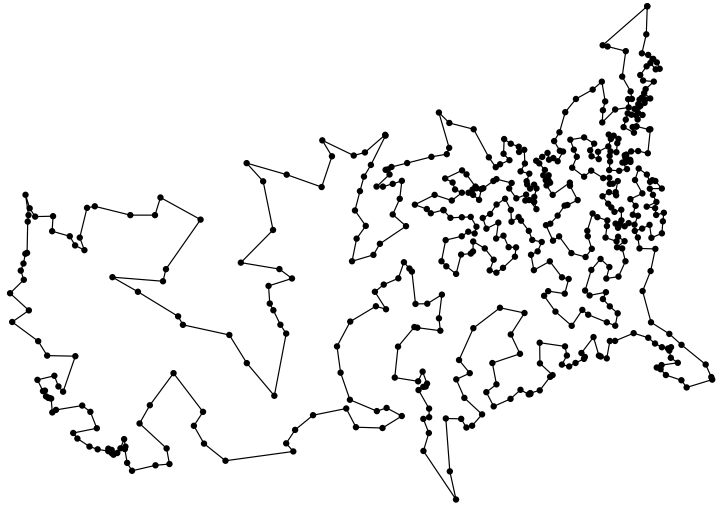}
\end{picture}
\caption{\small Optimal tour in the (non-uniform) {\sc at\&t}-532 instance.}
\label{fig_att532}
\end{center}
\end{figure}

Let us try this out on a \tsp\ instance well known in the operations research
literature, the {\sc at\&t}-532 instance.\footnote{{\sc at\&t}-532
is one of the many instances found in {\sc tsplib}, a library
assembled by \citeasnoun{Reinelt} and available from
{\it http://softlib.rice.edu/softlib/tsplib.\/}}
This consists of 532 sites belonging to the U.S.\ telecommunications
company.  For clear
commercial reasons, the distribution of sites in the instance follows
roughly the population density of the country, and thus is highly
non-uniform.  The instance was first solved to optimality (see Figure
\ref{fig_att532}) by \citeasnoun{PadbergRinaldi}, and in their units,
the optimum was found to be 27,868.  How closely does our approach
predict this result?  In their units, once again, the mean distance
between a city and its nearest neighbor can be measured as 36.205.
Using the expressions
given in Chapter~II, we would then estimate:

\begin{eqnarray}
L_E^{\mbox{\scriptsize non-uniform}}&\approx&
D_1^{\mbox{\scriptsize non-uniform}}\,
\frac{\langle L_E^{\mbox{\scriptsize
uniform}}\rangle}{\langle D_1^{\mbox{\scriptsize uniform}}\rangle} \\
&\approx& D_1^{\mbox{\scriptsize non-uniform}}\times\beta_E(2)
\times\left( 1-\frac{0.0171}{N}-\frac{1.048}{N^2}\right)\times
2N \\
&\approx& 27\mbox{,}427
\end{eqnarray}
which is within 1\% of the actual optimum.

It may seem surprising that our finite size scaling law {\it given periodic
boundary conditions\/} provides such a good estimate for a tour length
calculated given open boundary conditions.  There are two reasons for this
agreement.
First of all, boundary effects are involved in $D_1^{\mbox{\scriptsize
non-uniform}}$.  (\ref{eq_nonuniform}) incorporates these effects into the
$L_E^{\mbox{\scriptsize non-uniform}}$ estimate, and although they may not
be precisely the correct effects for the optimum tour length, they are
undoubtedly close.  Second of all, $\beta_E(2)$ itself does not depend
on the boundary conditions chosen.  For a discussion of why this is so,
the reader is referred to \citeasnoun{Jaillet}, who has proven that open and
periodic boundary conditions give the same $\beta_E(d)$ value.

\newpage
\clearemptydoublepage

\Appendix{Numerical methodology}
\label{app_method}

\noindent
In this appendix we discuss the algorithms used to obtain
our simulation results.  For both the Euclidean and random link {\tsp}s,
we have performed runs at instance sizes $N=12$, 13, 14, 15, 16, 17 using
the \citeasnoun{LinKernighan} heuristic (\lk), and at $N=30$ and $N=100$
using the {\it Chained Local Optimization\/} heuristic (\clo) of
\citeasnoun{MartinOttoFelten_CS}.

\bigskip

\leftline{\bf LK heuristic}

\medskip

\begin{figure}[!b]
\begin{center}
\begin{picture}(231,224)
\epsfig{file=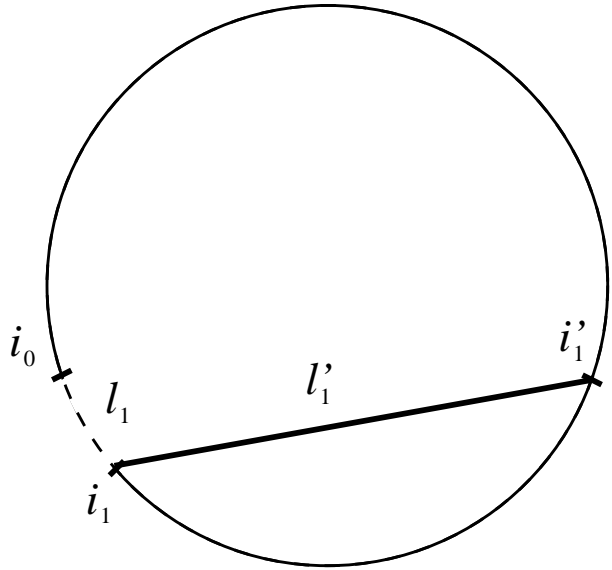}
\end{picture}
\caption{\small One step of \lk-search, showing the removed link $l_1$ (dashed
line) and added link $l_1'$ (bold line).}
\label{fig_lk1}
\end{center}
\end{figure}

\noindent
The \lk\ algorithm may be sketched as follows.  The kernel of the
algorithm is what might be termed {\it \lk-search\/}.  We start off
with an initial (non-optimal) tour.  \lk-search takes an arbitrary
starting city --- for convenience, we generally start with the top
city on a fixed ``list'' in our program.  Call this city $i_0$.  Now
pick a starting direction --- either ``forwards'' or ``backwards'' ---
and call $i_1$ the next city in that direction along the initial tour.
Call $l_1$ the link between $i_0$ and $i_1$, and remove that link.
We will now attempt to reconnect $i_1$ to a new city $i_1'$ that
is somewhere else on the tour, resulting in the situation shown
schematically in Figure \ref{fig_lk1}.  Let $l_1'$ be the link to
that new city $i_1'$.  How do we choose $i_1'$, and thus $l_1'$?
$i_1'$ is the nearest neighbor to $i_1$ that was not already connected
to $i_1$ in the initial tour.  There is another very important
requirement for $i_1'$: it must be such that $l_1'<l_1$.  This it
the {\it gain criterion\/}, and is applied at all stages of \lk-search.
If the gain criterion cannot be met, the search using this $i_0$ and $l_1$
is abandoned.  If the gain criterion is met, however, the search deepens.

\begin{figure}[t]
\begin{center}
\begin{picture}(231,224)
\epsfig{file=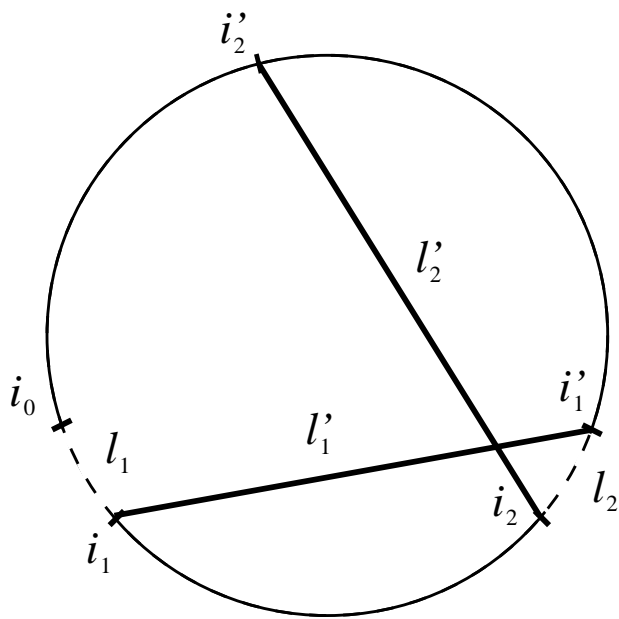}
\end{picture}
\caption{\small \lk-search at step 2, showing tadpole with two links removed
and two links added.}
\label{fig_lk2}
\end{center}
\end{figure}

At this point we have what might be called a ``tadpole graph'' (the more
technical term is a {\it one-tree\/}).  We now pick one of the two
cities next to $i_1'$ along the initial tour, call it $i_2$, and remove
the link $l_2$ joining them.  Which of the two possibilities do we use
for $i_2$, and thus $l_2$?  We pick $i_2$ such that, if $i_2$ were then
to be connected to the dangling end ($i_0$), a single closed
tour would result.  Now continue as before.  We attempt to reconnect
$i_2$ to a new city $i_2'$, resulting in the situation in Figure
\ref{fig_lk2}.  $l_2'$ is the link to the new city, and that city, as
before, is the nearest neighbor that was not an adjoining member of
the original tour.  The gain criteria get stronger as the search gets
deeper: the total gain must remain positive, so we must have
$l_1'+l_2'< l_1+l_2$, or else the search aborts.  \lk-search continues
recursively in this way, with the vertex of the ``tadpole'' hopping
around with the end point $i_0$ staying fixed, until the gain criteria
force it to abort.  There is, however, an important additional part
to the gain criteria.  At each step $m$, whilst $i_m$ is being chosen
in such a way as to make it possible to close up the tour, the length of
that closed tour is recorded.  If at any step of \lk-search the tadpole
length for that step is longer than the best closed tour recorded so
far, the search terminates.

When a search terminates, one of two things can happen.  If the best
recorded tour is shorter than the initial tour, the entire process
begins again but using that tour as the new one.\footnote{This will
happen, notably, if the search is abandoned because of the second part
of the gain criteria (closed-up tour shorter than tadpole), as the
first part of the gain criteria ($\sum_{i} l_i' < \sum_{i} l_i$)
requires that a tadpole always be shorter than the initial tour.}
If, on the other hand, the best recorded tour is not shorter than
the initial tour, we recommence with the alternate choice of $i_1$:
if the starting direction was ``forwards'', now we try ``backwards'',
and vice versa.  When, however, both choices of $i_1$ have been
exhausted, we select a new $i_0$ by advancing to the next city on
the list.  The algorithm, in our implementation, terminates when an
entire ``pass'' of $N$ starting cities $i_0$ results in no
improvement.

\bigskip

\leftline{\bf CLO heuristic}

\medskip

\begin{figure}[t]
\vspace{0.5cm}
\begin{center}
\begin{picture}(260,200)
\includegraphics[scale=0.6]{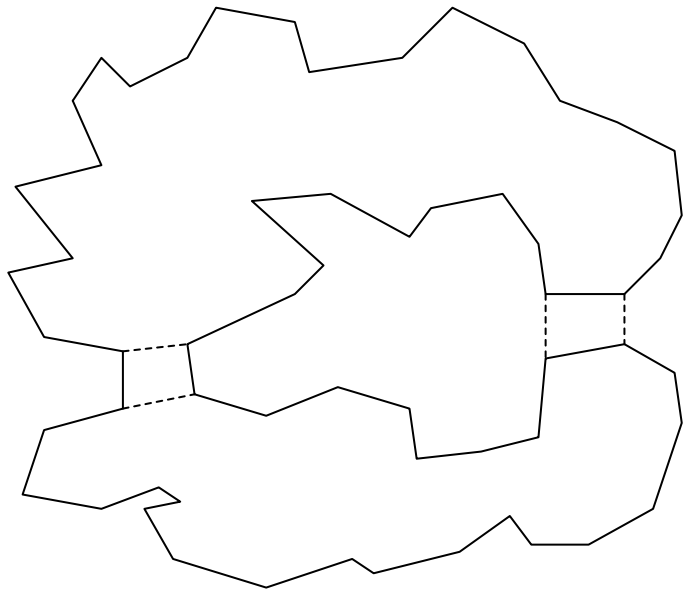}
\end{picture}
\caption{\small {\it Double-bridge\/} change executed by \clo\ heuristic,
generated by removing links shown with dashed lines, and reconnecting them 
differently.}
\label{fig_dbridge}
\end{center}
\end{figure}

\noindent
\clo\ is a stochastic algorithm, combining large Monte Carlo jumps with
embedded \lk\ local search.  In our implementation, it works as follows.
A (non-optimal) initial tour is optimized using \lk.  This brings it
to an \lk\ local minimum.  An attempt is then made to modify the tour by a
random 4-change (4 bonds disconnected and then reconnected differently),
which would not be accessible by the kind of sequential changes performed
by \lk-search.  The 4-change in question is known as a {\it double-bridge\/}
kick, shown in Figure \ref{fig_dbridge}: one 2-change disconnects the
tour, and the other 2-change reconnects the two parts in a different place.
After this 4-change is carried out, \lk\ is again used to optimize the
resulting tour.  If the new local optimum is better than the previous one,
the attempt succeeds, and \clo\ iterates the random 4-change procedure from
the new tour.  If not, the attempts fails and \clo\ tries another random
4-change on the old tour.  This continues for a fixed number of steps.
The idea is that, by directing its search through state space, \clo\ performs
better than simply running \lk\ from random starts an equivalent number of
times.

\bigskip

\leftline{\bf Use of heuristics}

\medskip

\noindent
For instance sizes in the range $12\le N\le 17$, we used the \lk\ heuristic.
Our method consisted of two parts: first, we used a testbed of instances
to estimate the systematic bias in \lk\ arising from the fact that \lk\ does
not always find the true optimum; second, we did our ``production runs''
to determine the actual optimal tour lengths.

In order to determine the systematic bias, we performed the following
procedure on values of $N$ from 12 through 17: (i) we generated a testbed
of 200 random instances; (ii) for each random instance, we generated 100
different random starting tours; (iii) for each random starting tour
we ran the \lk\ heuristic.  Based on the assumption that the best \lk-opt
obtained over 100 random starting tours was indeed the true optimum for
that instance, we calculated the expected bias per instance.  We then
averaged this bias over the testbed of instances, obtaining an estimate
for each value of $N$.

For our production runs, we then performed the following procedure on
values of $N$ from 12 through 17: (i) we generated 100,000 random
instances; (ii) for each random instance, we generated 10 different
random starting tours; (iii) for each random starting tour we ran the
\lk\ heuristic.  We then took, for each instance, the best \lk-opt obtained
over the 10 random starting tours, and averaged over all instances,
obtaining an estimate of the mean optimum length for each value of $N$.

For instance sizes $N=30$ and $N=100$, we used the \clo\ heuristic with
10 steps.  The method and procedures were identical to those for \lk,
apart from the numbers for the production runs: for $N=30$ we generated
8,000 random instances and 5 random starting tours per instance; for
$N=100$ we generated 1,200 random instances and 20 random starting tours
per instance.

\bigskip

\leftline{\bf Other operational issues}

\medskip

\noindent
The choice of number of random starting tours per instance in our
production runs was motivated by the need to keep the systematic bias
to a minimum.  For the \lk\ runs, using 10 random starting tours per
instance allowed us to have in the worst case, at $N=17$, a bias of
1 part in 200,000 for the Euclidean \tsp\ and 1 part in 170,000 for the
random link \tsp.  This bias was therefore negligible at the level
of numerical precision used (4 decimal places).  For the \clo\ runs
at $N=30$, using 5 random starting tours per instance was sufficient
to give us an estimated bias of under 1 part in 400,000 for the
Euclidean \tsp, and under 1 part in 800,000 for the random link \tsp.
We suspect these figures --- particularly the latter --- might be too
good to be true, so it is possible that our estimate is corrupted somewhat
by instances in the testbed where the true optimum was never found.
(Even if our estimate of the bias is off by an order of magnitude, though,
it will still be negligible compared with our statistical error bar,
which at $N=30$ is about 1 in 3,000 for both Euclidean and random link.)
For the \clo\ runs at $N=100$, using 20 random starting tours per instance
gave an estimated bias of 1 part in 120,000 for Euclidean and 1 part
in 60,000 for random link --- again, comfortably below the statistical
error of 1 in 6,000 for Euclidean and 1 in 2,000 for random link.

\begin{table}[!p]
\caption{\small Simulation timings for $d=2$ random link.  Instance sizes
$N=12$ through $N=17$ used \lk, $N=30$ and $N=100$ used 10-step \clo.
\# runs indicates number of instances times number of random starts
per instance.}
\vspace{0.5cm}
\label{table_times}
\begin{center}
\begin{tabular}{ccccc}
$N$&\lk/\clo&\# runs&Machine&{\sc cpu} time (secs)\\
\hline
12&\lk&500,000&Dec Alpha200&7133.4\\
12&\lk&500,000&Sun Ultra1&22001.2\\
13&\lk&500,000&Dec Alpha200&10772.0\\
13&\lk&500,000&Sun Ultra1&26025.2\\
14&\lk&500,000&Sun Sparc100&32755.3\\
14&\lk&500,000&Sun Sparc100&32420.9\\
15&\lk&500,000&Dec Alpha200&10724.9\\
15&\lk&500,000&Dec AS1000&9357.5\\
16&\lk&500,000&Sun Sparc100&41459.0\\
16&\lk&500,000&Sun Sparc100&38795.3\\
17&\lk&500,000&Sun Ultra1&43508.3\\
17&\lk&500,000&Sun Sparc100&46584.2\\
30&\clo&5,000&Sun Ultra1&10801.6\\
30&\clo&5,000&Sun Sparc100&11717.5\\
30&\clo&5,000&Sun Sparc100&11621.3\\
30&\clo&5,000&Sun Ultra1&11163.9\\
30&\clo&5,000&Sun Sparc100&10773.3\\
30&\clo&5,000&Sun Ultra1&10569.0\\
30&\clo&5,000&Sun Sparc100&11558.9\\
30&\clo&5,000&Sun Sparc100&11287.7\\
100&\clo&6,000&Dec Alpha200&19694.6\\
100&\clo&6,000&Dec AS1000&17982.2\\
100&\clo&6,000&Sun Sparc100&71671.1\\
100&\clo&6,000&Sun Sparc100&71137.9\\
\end{tabular}
\end{center}
\end{table}

Finally, let us note a few further details concerning our code.  Timings
for our $d=2$ random link simulations are given in Table \ref{table_times};
as can be seen we ran on four different machines, two of them Dec Alphas
and two of them Sun {\sc sparc}s.  The code itself is written in C, and is a
heavily modified version of the \clo\ package developed by Steve Otto
(otto@cse.ogi.edu) and Robert Prouty (prouty@cse.ogi.edu).  In the code,
cities are placed on a 10,000 $\times$ 10,000 grid, and distances are
rounded to the nearest integer.  We confirmed that the mesh was fine
enough for roundoff error to be negligible to within the 4-decimal-place
precision of our numerical data.  The random number generator used,
from the \clo\ package, is a standard linear congruential algorithm working
as follows:

\bigskip

\leftline{\hskip 4cm\#define MASK ( 0x7fffffff )}
\leftline{\hskip 4cm\#define MULT 1103515245}
\leftline{\hskip 4cm\#define ADD 12345}
\leftline{\hskip 4cm\#define TWOTO31 2147483648.0}

\leftline{\hskip 4cm AAA = MULT \& MASK;}
\leftline{\hskip 4cm BBB = ADD \& MASK;}

\bigskip

\par\noindent
where \& is the binary AND operator.  The routine then updates the
integer quantity randx to be (AAA*randx + BBB) \& MASK, and returns
a double-precision variable randx/TWOTO31.

\newpage
\clearemptydoublepage

\Appendix{Another numerical study of $\beta(d)$}
\label{app_dsj}

\noindent
Subsequent to our work, \citeasnoun{Johnson_HK} have used a variant
of our approach to confirm the values $\beta_E(2)=0.7120\pm 0.0002$
and $\beta_E(3)=0.6979\pm 0.0002$ (to higher precision) and to extend
the results to $d=4$.  Let us briefly summarize their work, and the
differences between our approach and theirs.

Our finite-size scaling analysis, presented in Chapter~II, uses data
points at values of $N$ from $N=12$ to $N=17$, as well as the
points $N=30$ and $N=100$.  Johnson {\it et al.\/}, making use of
our scaling law (\ref{eq_pre_fss}), work with larger instances
sizes --- $N=100$, $N=316$ and $N=1$,000 --- where a better fit may
presumably be obtained.  This requires much more powerful computational
resources than our own, but also good algorithms and efficient methods
for reducing statistical error to a minimum (so as to avoid too large
a number of time-consuming runs).

When one wishes to estimate $\langle L_E(N,d)\rangle$, the most direct
way is to take the numerical average $\overline{L_E(N,d)}$ over a
sample of $K$ randomly chosen instances.  This estimator has an expected
statistical
error $\sigma(K)=\sigma_{L_E}/\sqrt{K}$, where $\sigma_{L_E}$ is the
instance-to-instance standard deviation of $L_E$.  One can improve
on this, however.  Let us instead measure $\overline{L_E(N,d)-L^*(N,d)} +
\overline{L^*(N,d)}$.  If $L^*$ is a quantity closely correlated
with $L_E$, $L_E-L^*$ will have a substantially lower variance than
$L_E$ itself; if in addition $L^*$ has a small variance (or better
yet, if we can measure the ensemble average $\langle L^*\rangle$
exactly) then we will ultimately obtain a measurement of $\langle
L_E\rangle$ to higher accuracy than with the direct estimator
$\overline{L_E}$.  This is the method we have adopted
in our own work.  For $L^*$, we use $\lambda(L_1+L_2)/2$, where
$\lambda$ is a parameter, $L_1$ is the mean distance between nearest
neighboring points in an instance, and $L_2$ is the mean distance
between second-nearest neighboring points in an instance.  The great
advantage of this approach is that an analytical expression can be found
giving $\langle L_1+L_2\rangle /2$ exactly, so the {\it only\/} statistical
error is due to the estimator $\overline{L_E-\lambda(L_1+L_2)/2}$.
$\lambda$ can be chosen so as to minimize the variance of this estimator:
it is relatively simple to show that the value $\lambda^*$ minimizing
$\lambda$ is given by
\begin{equation}
\lambda^*=\frac{\langle L_E (L_1+L_2)/2\rangle - \langle L_E\rangle
\langle (L_1+L_2)/2\rangle}{(\sigma_{(L_1+L_2)/2})^2}
\end{equation}
where $\sigma_{(L_1+L_2)/2}$ is simply the instance-to-instance standard
deviation of $(L_1+L_2)/2$.  Notice that $\lambda^*$ is in fact a measure
of how closely correlated $L_E$ and $(L_1+L_2)/2$ are --- if they are perfectly
correlated, $\lambda^*$ will be equal to 1, and if they are uncorrelated,
$\lambda^*$ will be equal to 0.  In practice, we have found that the
degree to which they are correlated varies little in $N$.  For the
$d=2$ Euclidean case, we found $\lambda^*\approx 0.75$ at $N=15$;
keeping $\lambda^*$ fixed to this value over all $N$, the variance
reduction ranged from $0.38$ at $N=15$ to $0.43$ at $N=100$.  We thus
succeeded in reducing the overall statistical error $\sigma(K)$ by a
factor of between $2.33$ and $2.63$.  (For the random link \tsp, the same
method reduced the error by a factor of between $2.69$ and $3.08$.)
The relative insensitivity of the variance reduction scheme to $N$
suggests that correlations between $L_E$ and $(L_1+L_2)/2$ are
stable at large $N$; a related property has been seen in the random link
case in Chapter~III (see Figure \ref{fig_rl_neighbord2}).

In the work of \citeasnoun{Johnson_HK}, a somewhat different procedure
is used for reduction of variance.  The emphasis of their article is on
analyzing the {\it Held-Karp\/} lower bound, the solution to an integer
programming relaxation of the \tsp\ \cite{HeldKarp_A}.  Thus
instead of measuring $\langle L_E\rangle$
using $\overline{L_E-L^*}+\overline{L^*}$, they use the estimator
$\overline{L_E/L_{HK}}\,\,\overline{L_{HK}}$, where $L_{HK}$ is
the Held-Karp bound.  The advantage of using this
quantity is that it turns out to be correlated {\it very\/} closely
with $L_E$, so that $L_E/L_{HK}$
has extremely low instance-to-instance fluctuations.  (To compare:
at $N=100$ in 2-D, our estimator $L_E-L^*$ leads to a statistical
error for sample size $K$ of about $1.3\%/\sqrt{K}$, whereas theirs
leads to a statistical error of about $0.3\%/\sqrt{K}$.)\footnote{The
estimator of Johnson {\it et al.\/} follows a similar method used
earlier by \citeasnoun{Sourlas}, who measured
$\overline{L_E/L_{th}}\,\,\overline{L_{th}}$, where $L_{th}$ is a
weighted average of $k$th-nearest neighbor distances for $k$ up to
5.  (This is by contrast with our own $L^*$, which is simply an
unweighted average of $k$ up to 2.)  Since Sourlas' results were
for the $d=1$ random link case, it is difficult to compare with
our simulations; let us note, however, that at $N=100$ he found his
method reduced $\sigma(K)$ by a factor of 4.}
Furthermore, in measuring $L_E/L_{HK}$ Johnson {\it et al.\/} used
exact codes for the $N=100$ and
$N=316$ instances, so there is no systematic error in these measurements
(for $N=1$,000 they used the same large $N$ heuristic as we used for
$N=30$ and $N=100$).  For the second part of the estimator,
$\overline{L_{HK}}$ itself, they used a rapid approximate method ---
though they were able to quantify the systematic bias inherent in
this approximation (about $0.005\%$) on the basis of a testbed of
instances where $L_{HK}$ had already been found exactly.  They then
corrected for this systematic bias.

Note that while in our measurements, $\langle L^*\rangle$ is calculated
exactly over the ensemble
and the statistical error is due exclusively to the fluctuations of
$L_E-L^*$, in the measurements of
Johnson {\it et al.\/}, $\overline{L_{HK}}$ involves a significant
statistical error whereas $\overline{L_E/L_{HK}}$ involves an almost
negligible error.  It is instructive to compare our $N=100$ results
with theirs (this is the only 2-D instance size we have in common with
them).  Our data, on the basis of 6000 instances, give $\langle
L_E(100,2) - L^*(100,2)\rangle = 2.4342\pm 0.0012$ and
$\langle L^*(100,2)\rangle = 4.6934$, so $\langle L_E(100,2)\rangle =
7.1276\pm 0.0012$.  Johnson {\it et al.\/} obtain $\langle
L_E(100,2)/L_{HK}(100,2)\rangle = 1.005542\pm 0.000027$ on the
basis of 13,957 instances, and $\langle L_{HK}(100,2)\rangle =
7.0897\pm 0.0006$ on the basis of 98,246 instances, giving
$\langle L_E(100,2)\rangle = 7.1290\pm 0.0008$.  We are unaware of
the running times for the latter results, and thus cannot compare
the relative efficiency of the two approaches.  It is clear, however,
that our resources could not permit instance sizes much in excess of
$N=100$, whereas their tests went up to $N=1$,000.

There is, interestingly, a further bias in the estimator used by
Johnson {\it et al.\/} that they have not noted.  This arises from
the fact that, of course, $\langle L_E\rangle\neq\langle L_E/L_{HK}\rangle
\langle L_{HK}\rangle$.  It is relatively straightforward to calculate
this bias at large $L_E$ and $L_{HK}$:
\begin{equation}
\label{eq_hk}
\left\langle\frac{L_E}{L_ {HK}}\right\rangle\langle L_{HK}\rangle\approx
\langle L_E \rangle
\left( 1 + \frac{\sigma_{L_{HK}}^2}{\langle L_{HK} \rangle ^2}
- \frac{\langle L_E L_{HK}\rangle - \langle L_E\rangle\langle L_{HK}\rangle}
{\langle L_E \rangle \langle L_{HK} \rangle} + \cdots\right)
\end{equation}
where $\sigma_{L_{HK}}$ is the instance-to-instance standard deviation
of $L_{HK}$.  In 2-D, the bias here is of $O(1/N)$.  Without taking
account of this bias, Johnson {\it et al.\/} use our scaling law ---
truncated to subleading order --- to fit data at $N=100$, $N=316$ and
$N=1$,000, obtaining $\beta_E(2)=0.7124\pm 0.0001$.  Fortunately, the
bias affects only the subleading terms ($O(1/N)$
and beyond) in the fit and should not affect the leading term that gives
$\beta_E$
(though care must be taken
when using their $L_E$ results for finite $N$).  What they do not consider,
however, is a test of goodness-of-fit; instead they estimate
the $\beta_E(2)$ error bar ``conservatively'' by adding up the error
bars for the three data points in the fit.\footnote{Furthermore, it
appears that in the notation of Johnson {\it et al.\/}, the error bar
indicates the extremes of the 95\% confidence
interval, hence $\pm 2\sigma$.  In quoting their results, we use the more
standard notation of $\pm\sigma$.}

Let us attempt to perform
the analysis somewhat more carefully, in order to provide a more
meaningful comparison between their conclusions and ours.  As 
$L_E$ and $L_{HK}$ clearly are correlated very closely, we shall
make the assumption that the bias in (\ref{eq_hk}) can in fact be
neglected --- at least at $N=316$ and $N=1$,000 --- and simply combine
their data at these two points with our own data at smaller $N$.
Table \ref{table_dsj} sumarizes the effects of this, comparing
the coefficients in the scaling series 
$\langle L_E(N,2)\rangle = \sqrt{N}\,\beta_E(2)\,[1+A(2)/N+B(2)/N^2]$,
without and with these two new data points.  The two results are consistent
with each other: the coefficients of the fit are relatively stable (even,
surprisingly, to $O(1/N^2)$), and the new $\chi^2$ is consistent with 7
degrees of freedom (10 data points minus 3 fit parameters).  The error
in $\beta_E(2)$ is obtained by the standard procedure of determining the
values of this quantity that make $\chi^2$ increase by 1; those values
are then $\beta_E(2)$ plus or minus one standard deviation.

\begin{table}[h]
\caption{\small Values of coefficients of fit $\langle L_E(N,2)\rangle =$
$\sqrt{N}\,\beta_E(2)\,[1+A(2)/N+B(2)/N^2]$, for previous fit
($N=\{$12,13,14,15,16,17,30,100$\}$ and new fit
($N=\{$12,13,14,15,16,17,30,100,316,1000$\}$).}
\vspace{0.5cm}
\label{table_dsj}
\begin{center}
\begin{tabular}{ccccc}
Result&$\beta_E(2)$&$A(d)$&$B(d)$&$\chi^2$\\
\hline
Previous fit (8 data points)&$0.7120\pm 0.0002$&$0.1088$&$-1.064$&$5.56$\\
New fit (10 data points)&$0.7123\pm 0.0001$&$0.0982$&$-0.9819$&$7.97$\\
\end{tabular}
\end{center}
\end{table}

If we stretch our assumption further and assume that the bias in (\ref{eq_hk})
is negligible even at $N=100$, and therefore combine our own $N=100$ data
with that of Johnson {\it et al.\/} --- obtaining
$\langle L_E(100,2)\rangle = 7.1286\pm 0.0007$ --- the results of ``New fit''
do not change at all, to the precision shown!  Only $\chi^2$ changes,
decreasing now to 7.03.  Another possibility is to fit exclusively to
their 3 data points, though one must be
cautious when using results based on so little information.  Nevertheless,
a two-parameter fit of the form $\sqrt{N}\,\beta_E(2)\,[1+A(2)/N]$ gives
$\beta_E(2)=0.7123\pm 0.0001$ and $A(2)=0.0791$ (with $\chi^2=0.016$,
though this value is purely anecdotal as there is only one degree of
freedom!).  Thus, this $\beta_E(2)$ result indeed appears credible.  It
is unclear why it is quoted in their article as $0.7124$; however, assuming
that the values quoted for the corresponding results at $d=3$ and $d=4$ are
correct, Table \ref{table_betad} compares our results and theirs.

\begin{table}[t]
\caption{\small Comparison of $\beta_E(d)$ from two different numerical
studies.}
\label{table_betad}
\begin{center}
\begin{tabular}{ccc}
$d$&$\beta_E(d)$ from our study&$\beta_E(d)$ from Johnson's study\\
\hline
2&$0.7120\pm 0.0002$&$0.7123\pm 0.0001$\\
3&$0.6978\pm 0.0002$&$0.6980\pm 0.0002$\\
4&N/A&$0.7234\pm 0.0002$\\
\end{tabular}
\end{center}
\end{table}

\begin{figure}[!b]
\begin{center}
\begin{picture}(340,300)
\epsfig{file=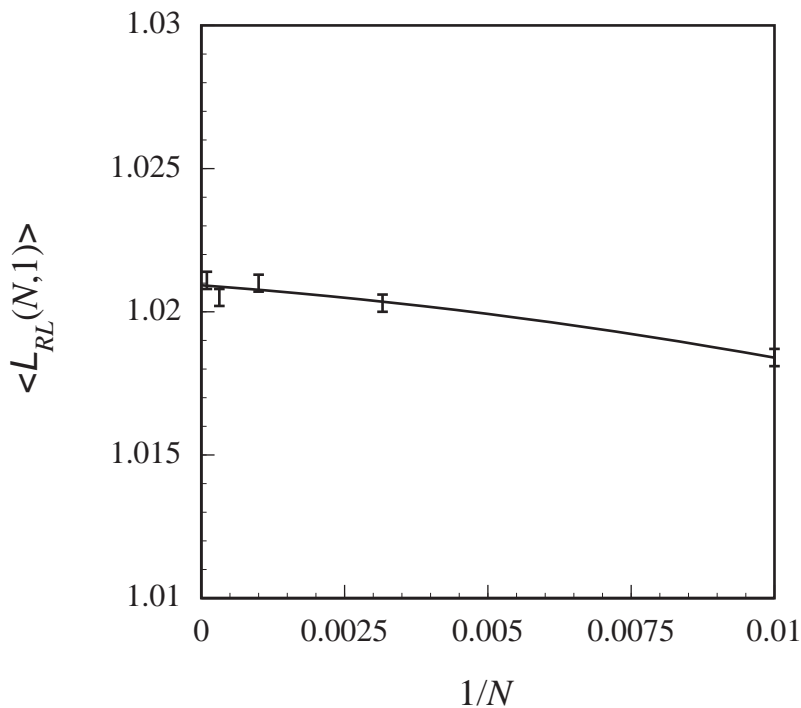}
\end{picture}
\caption{\small Finite size scaling of $d=1$ optimum.  Best fit
($\chi^2=2.62$) is given by:
$\langle L_{RL}(N,1)\rangle = 1.0209
(1 - 0.1437/N - 10.377/N^2)$.  Error bars show one standard deviation
(statistical error).}
\label{fig_dsjrl1}
\end{center}
\end{figure}

Finally, Johnson {\it et al.\/} also performed simulations on the $d=1$ random
link case, although they did not attempt to extrapolate to $\beta_{RL}(1)$,
instead noting simply that their large $N$ data were consistent with
the asymptotic value 1.0208.\footnote{Their notation differs from ours
by a scaling factor of 2; we cite their results following our own notation,
without this factor.}  Let us perform a fit of the usual sort on their
data, using the 5 points that they give: $N=100$, $N=316$, $N=1,$000,
$N=3$,162, and $N=10$,000.  (They do not actually carry out simulations
of $L_{RL}/L_{HK}$ at $N=10,$000, considering that at the level of precision
used this quantity is indistinguishable from 1, based on its
value of $1.000036$ at $N=3,$162.)  The results are plotted in
Figure \ref{fig_dsjrl1}: $\beta_{RL}(1)=1.0209\pm 0.0002$, and
$\chi^2=2.62$ for 2 degrees of freedom (5 data points minus 3 fit parameters).
This provides excellent experimental confirmation of the cavity
predictions at $d=1$.

\newpage
\clearemptydoublepage

\markboth{}{}
\renewcommand{\bibname}{General References}
\addcontentsline{toc}{chapter}{\numberline{}General References}

\ifx\undefined\BySame
\newcommand{\BySame}{\leavevmode\rule[.5ex]{3em}{.5pt}\ }
\fi
\ifx\undefined\textsc
\newcommand{\textsc}[1]{{\sc #1}}
\fi
\ifx\undefined\emph
\newcommand{\emph}[1]{{\em #1\/}}
\fi

\end{document}